\documentclass[traditabstract, longauth]{aa} 
\pdfoutput=1
\usepackage{graphicx}
\usepackage{float}
\usepackage{txfonts}
\usepackage[breaklinks, colorlinks, citecolor=blue]{hyperref}
\usepackage{natbib}
\usepackage[usenames,dvipsnames,svgnames,table]{xcolor}

\bibpunct{(}{)}{;}{a}{}{,} 

\newcommand{\cruler}{{\tt Commander-Ruler}}
\newcommand{\ruler}{{\tt C-R}}
\newcommand{\nilc}{{\tt NILC}}
\newcommand{\sevem}{{\tt SEVEM}}
\newcommand{\smica}{{\tt SMICA}}

\newcommand{\wmap}{\textit{WMAP}}
\newcommand{\lrg}{SDSS-CMASS/LOWZ}
\newcommand{\mg}{SDSS-MphG}

\newcommand{\hatn}{\hat{\vec{n}}}

\newcommand{\gsim}{\; ^{>}_{\sim}\;}

\newcommand{\modif}[1]{\textcolor{Bittersweet}{\textbf{#1}}}

\newcommand{\be}{\begin{equation}}
\newcommand{\ee}{\end{equation}}

\usepackage{multirow}

\def\setsymbol#1#2{\expandafter\def\csname #1\endcsname{#2}}
\def\getsymbol#1{\csname #1\endcsname}

\def\Planck{\textit{Planck}}

\newbox\tablebox    \newdimen\tablewidth
\def\leaderfil{\leaders\hbox to 5pt{\hss.\hss}\hfil}
\def\endPlancktable{\tablewidth=\columnwidth 
    $$\hss\copy\tablebox\hss$$
    \vskip-\lastskip\vskip -2pt}
\def\endPlancktablewide{\tablewidth=\textwidth 
    $$\hss\copy\tablebox\hss$$
    \vskip-\lastskip\vskip -2pt}
\def\tablenote#1 #2\par{\begingroup \parindent=0.8em
    \abovedisplayshortskip=0pt\belowdisplayshortskip=0pt
    \noindent
    $$\hss\vbox{\hsize\tablewidth \hangindent=\parindent \hangafter=1 \noindent
    \hbox to \parindent{$^#1$\hss}\strut#2\strut\par}\hss$$
    \endgroup}
\def\doubleline{\vskip 3pt\hrule \vskip 1.5pt \hrule \vskip 5pt}

\def\L2{\ifmmode L_2\else $L_2$\fi}

\def\DeltaT{\ifmmode \Delta T\else $\Delta T$\fi}
\def\deltat{\ifmmode \Delta t\else $\Delta t$\fi}
\def\fknee{\ifmmode f_{\rm knee}\else $f_{\rm knee}$\fi}
\def\Fmax{\ifmmode F_{\rm max}\else $F_{\rm max}$\fi}
\def\solar{\ifmmode{\rm M}_{\mathord\odot}\else${\rm M}_{\mathord\odot}$\fi}
\def\Msolar{\ifmmode{\rm M}_{\mathord\odot}\else${\rm M}_{\mathord\odot}$\fi}
\def\Lsolar{\ifmmode{\rm L}_{\mathord\odot}\else${\rm L}_{\mathord\odot}$\fi}

\def\inv{\ifmmode^{-1}\else$^{-1}$\fi}
\def\mo{\ifmmode^{-1}\else$^{-1}$\fi}
\def\sup#1{\ifmmode ^{\rm #1}\else $^{\rm #1}$\fi}
\def\expo#1{\ifmmode \times 10^{#1}\else $\times 10^{#1}$\fi}
\def\,{\thinspace}
\def\lsim{\mathrel{\raise .4ex\hbox{\rlap{$<$}\lower 1.2ex\hbox{$\sim$}}}}
\def\gsim{\mathrel{\raise .4ex\hbox{\rlap{$>$}\lower 1.2ex\hbox{$\sim$}}}}

\def\simprop{\mathrel{\raise .4ex\hbox{\rlap{$\propto$}\lower 1.2ex\hbox{$\sim$}}}}
\def\deg{\ifmmode^\circ\else$^\circ$\fi}
\def\pdeg{\ifmmode $\setbox0=\hbox{$^{\circ}$}\rlap{\hskip.11\wd0 .}$^{\circ}
          \else \setbox0=\hbox{$^{\circ}$}\rlap{\hskip.11\wd0 .}$^{\circ}$\fi}
\def\arcs{\ifmmode {^{\scriptstyle\prime\prime}}
          \else $^{\scriptstyle\prime\prime}$\fi}
\def\arcm{\ifmmode {^{\scriptstyle\prime}}
          \else $^{\scriptstyle\prime}$\fi}
\newdimen\sa  \newdimen\sb
\def\parcs{\sa=.07em \sb=.03em
     \ifmmode \hbox{\rlap{.}}^{\scriptstyle\prime\kern -\sb\prime}\hbox{\kern -\sa}
     \else \rlap{.}$^{\scriptstyle\prime\kern -\sb\prime}$\kern -\sa\fi}
\def\parcm{\sa=.08em \sb=.03em
     \ifmmode \hbox{\rlap{.}\kern\sa}^{\scriptstyle\prime}\hbox{\kern-\sb}
     \else \rlap{.}\kern\sa$^{\scriptstyle\prime}$\kern-\sb\fi}
\def\ra[#1 #2 #3.#4]{#1\sup{h}#2\sup{m}#3\sup{s}\llap.#4}
\def\dec[#1 #2 #3.#4]{#1\deg#2\arcm#3\arcs\llap.#4}
\def\deco[#1 #2 #3]{#1\deg#2\arcm#3\arcs}
\def\rra[#1 #2]{#1\sup{h}#2\sup{m}}

\def\dots{\relax\ifmmode \ldots\else $\ldots$\fi}
\def\WHzsr{\ifmmode $W\,Hz\mo\,sr\mo$\else W\,Hz\mo\,sr\mo\fi}
\def\mHz{\ifmmode $\,mHz$\else \,mHz\fi}
\def\GHz{\ifmmode $\,GHz$\else \,GHz\fi}
\def\mKs{\ifmmode $\,mK\,s$^{1/2}\else \,mK\,s$^{1/2}$\fi}
\def\muKs{\ifmmode \,\mu$K\,s$^{1/2}\else \,$\mu$K\,s$^{1/2}$\fi}
\def\muKRJs{\ifmmode \,\mu$K$_{\rm RJ}$\,s$^{1/2}\else \,$\mu$K$_{\rm RJ}$\,s$^{1/2}$\fi}
\def\muKHz{\ifmmode \,\mu$K\,Hz$^{-1/2}\else \,$\mu$K\,Hz$^{-1/2}$\fi}
\def\MJysr{\ifmmode \,$MJy\,sr\mo$\else \,MJy\,sr\mo\fi}
\def\MJysrmK{\ifmmode \,$MJy\,sr\mo$\,mK$_{\rm CMB}\mo\else \,MJy\,sr\mo\,mK$_{\rm CMB}\mo$\fi}
\def\microns{\ifmmode \,\mu$m$\else \,$\mu$m\fi}

\def\muK{\ifmmode \,\mu$K$\else \,$\mu$\hbox{K}\fi}
\def\microK{\ifmmode \,\mu$K$\else \,$\mu$\hbox{K}\fi}
\def\muW{\ifmmode \,\mu$W$\else \,$\mu$\hbox{W}\fi}
\def\kms{\ifmmode $\,km\,s$^{-1}\else \,km\,s$^{-1}$\fi}
\def\kmsMpc{\ifmmode $\,\kms\,Mpc\mo$\else \,\kms\,Mpc\mo\fi}
\setsymbol{LFI:center:frequency:70GHz:units}{70.3\,GHz}
\setsymbol{LFI:center:frequency:44GHz:units}{44.1\,GHz}
\setsymbol{LFI:center:frequency:30GHz:units}{28.5\,GHz}

\setsymbol{LFI:center:frequency:70GHz}{70.3}
\setsymbol{LFI:center:frequency:44GHz}{44.1}
\setsymbol{LFI:center:frequency:30GHz}{28.5}

\setsymbol{LFI:center:frequency:LFI18:Rad:M:units}{71.7\GHz}
\setsymbol{LFI:center:frequency:LFI19:Rad:M:units}{67.5\GHz}
\setsymbol{LFI:center:frequency:LFI20:Rad:M:units}{69.2\GHz}
\setsymbol{LFI:center:frequency:LFI21:Rad:M:units}{70.4\GHz}
\setsymbol{LFI:center:frequency:LFI22:Rad:M:units}{71.5\GHz}
\setsymbol{LFI:center:frequency:LFI23:Rad:M:units}{70.8\GHz}
\setsymbol{LFI:center:frequency:LFI24:Rad:M:units}{44.4\GHz}
\setsymbol{LFI:center:frequency:LFI25:Rad:M:units}{44.0\GHz}
\setsymbol{LFI:center:frequency:LFI26:Rad:M:units}{43.9\GHz}
\setsymbol{LFI:center:frequency:LFI27:Rad:M:units}{28.3\GHz}
\setsymbol{LFI:center:frequency:LFI28:Rad:M:units}{28.8\GHz}
\setsymbol{LFI:center:frequency:LFI18:Rad:S:units}{70.1\GHz}
\setsymbol{LFI:center:frequency:LFI19:Rad:S:units}{69.6\GHz}
\setsymbol{LFI:center:frequency:LFI20:Rad:S:units}{69.5\GHz}
\setsymbol{LFI:center:frequency:LFI21:Rad:S:units}{69.5\GHz}
\setsymbol{LFI:center:frequency:LFI22:Rad:S:units}{72.8\GHz}
\setsymbol{LFI:center:frequency:LFI23:Rad:S:units}{71.3\GHz}
\setsymbol{LFI:center:frequency:LFI24:Rad:S:units}{44.1\GHz}
\setsymbol{LFI:center:frequency:LFI25:Rad:S:units}{44.1\GHz}
\setsymbol{LFI:center:frequency:LFI26:Rad:S:units}{44.1\GHz}
\setsymbol{LFI:center:frequency:LFI27:Rad:S:units}{28.5\GHz}
\setsymbol{LFI:center:frequency:LFI28:Rad:S:units}{28.2\GHz}

\setsymbol{LFI:center:frequency:LFI18:Rad:M}{71.7}
\setsymbol{LFI:center:frequency:LFI19:Rad:M}{67.5}
\setsymbol{LFI:center:frequency:LFI20:Rad:M}{69.2}
\setsymbol{LFI:center:frequency:LFI21:Rad:M}{70.4}
\setsymbol{LFI:center:frequency:LFI22:Rad:M}{71.5}
\setsymbol{LFI:center:frequency:LFI23:Rad:M}{70.8}
\setsymbol{LFI:center:frequency:LFI24:Rad:M}{44.4}
\setsymbol{LFI:center:frequency:LFI25:Rad:M}{44.0}
\setsymbol{LFI:center:frequency:LFI26:Rad:M}{43.9}
\setsymbol{LFI:center:frequency:LFI27:Rad:M}{28.3}
\setsymbol{LFI:center:frequency:LFI28:Rad:M}{28.8}
\setsymbol{LFI:center:frequency:LFI18:Rad:S}{70.1}
\setsymbol{LFI:center:frequency:LFI19:Rad:S}{69.6}
\setsymbol{LFI:center:frequency:LFI20:Rad:S}{69.5}
\setsymbol{LFI:center:frequency:LFI21:Rad:S}{69.5}
\setsymbol{LFI:center:frequency:LFI22:Rad:S}{72.8}
\setsymbol{LFI:center:frequency:LFI23:Rad:S}{71.3}
\setsymbol{LFI:center:frequency:LFI24:Rad:S}{44.1}
\setsymbol{LFI:center:frequency:LFI25:Rad:S}{44.1}
\setsymbol{LFI:center:frequency:LFI26:Rad:S}{44.1}
\setsymbol{LFI:center:frequency:LFI27:Rad:S}{28.5}
\setsymbol{LFI:center:frequency:LFI28:Rad:S}{28.2}

\setsymbol{LFI:white:noise:sensitivity:70GHz:units}{134.7\muKs}
\setsymbol{LFI:white:noise:sensitivity:44GHz:units}{164.7\muKs}
\setsymbol{LFI:white:noise:sensitivity:30GHz:units}{143.4\muKs}

\setsymbol{LFI:white:noise:sensitivity:70GHz}{134.7}
\setsymbol{LFI:white:noise:sensitivity:44GHz}{164.7}
\setsymbol{LFI:white:noise:sensitivity:30GHz}{143.4}

\setsymbol{LFI:white:noise:sensitivity:LFI18:Rad:M:units}{512.0\muKs}
\setsymbol{LFI:white:noise:sensitivity:LFI19:Rad:M:units}{581.4\muKs}
\setsymbol{LFI:white:noise:sensitivity:LFI20:Rad:M:units}{590.8\muKs}
\setsymbol{LFI:white:noise:sensitivity:LFI21:Rad:M:units}{455.2\muKs}
\setsymbol{LFI:white:noise:sensitivity:LFI22:Rad:M:units}{492.0\muKs}
\setsymbol{LFI:white:noise:sensitivity:LFI23:Rad:M:units}{507.7\muKs}
\setsymbol{LFI:white:noise:sensitivity:LFI24:Rad:M:units}{462.2\muKs}
\setsymbol{LFI:white:noise:sensitivity:LFI25:Rad:M:units}{413.6\muKs}
\setsymbol{LFI:white:noise:sensitivity:LFI26:Rad:M:units}{478.6\muKs}
\setsymbol{LFI:white:noise:sensitivity:LFI27:Rad:M:units}{277.7\muKs}
\setsymbol{LFI:white:noise:sensitivity:LFI28:Rad:M:units}{312.3\muKs}
\setsymbol{LFI:white:noise:sensitivity:LFI18:Rad:S:units}{465.7\muKs}
\setsymbol{LFI:white:noise:sensitivity:LFI19:Rad:S:units}{555.6\muKs}
\setsymbol{LFI:white:noise:sensitivity:LFI20:Rad:S:units}{623.2\muKs}
\setsymbol{LFI:white:noise:sensitivity:LFI21:Rad:S:units}{564.1\muKs}
\setsymbol{LFI:white:noise:sensitivity:LFI22:Rad:S:units}{534.4\muKs}
\setsymbol{LFI:white:noise:sensitivity:LFI23:Rad:S:units}{542.4\muKs}
\setsymbol{LFI:white:noise:sensitivity:LFI24:Rad:S:units}{399.2\muKs}
\setsymbol{LFI:white:noise:sensitivity:LFI25:Rad:S:units}{392.6\muKs}
\setsymbol{LFI:white:noise:sensitivity:LFI26:Rad:S:units}{418.6\muKs}
\setsymbol{LFI:white:noise:sensitivity:LFI27:Rad:S:units}{302.9\muKs}
\setsymbol{LFI:white:noise:sensitivity:LFI28:Rad:S:units}{285.3\muKs}

\setsymbol{LFI:white:noise:sensitivity:LFI18:Rad:M}{512.0}
\setsymbol{LFI:white:noise:sensitivity:LFI19:Rad:M}{581.4}
\setsymbol{LFI:white:noise:sensitivity:LFI20:Rad:M}{590.8}
\setsymbol{LFI:white:noise:sensitivity:LFI21:Rad:M}{455.2}
\setsymbol{LFI:white:noise:sensitivity:LFI22:Rad:M}{492.0}
\setsymbol{LFI:white:noise:sensitivity:LFI23:Rad:M}{507.7}
\setsymbol{LFI:white:noise:sensitivity:LFI24:Rad:M}{462.2}
\setsymbol{LFI:white:noise:sensitivity:LFI25:Rad:M}{413.6}
\setsymbol{LFI:white:noise:sensitivity:LFI26:Rad:M}{478.6}
\setsymbol{LFI:white:noise:sensitivity:LFI27:Rad:M}{277.7}
\setsymbol{LFI:white:noise:sensitivity:LFI28:Rad:M}{312.3}
\setsymbol{LFI:white:noise:sensitivity:LFI18:Rad:S}{465.7}
\setsymbol{LFI:white:noise:sensitivity:LFI19:Rad:S}{555.6}
\setsymbol{LFI:white:noise:sensitivity:LFI20:Rad:S}{623.2}
\setsymbol{LFI:white:noise:sensitivity:LFI21:Rad:S}{564.1}
\setsymbol{LFI:white:noise:sensitivity:LFI22:Rad:S}{534.4}
\setsymbol{LFI:white:noise:sensitivity:LFI23:Rad:S}{542.4}
\setsymbol{LFI:white:noise:sensitivity:LFI24:Rad:S}{399.2}
\setsymbol{LFI:white:noise:sensitivity:LFI25:Rad:S}{392.6}
\setsymbol{LFI:white:noise:sensitivity:LFI26:Rad:S}{418.6}
\setsymbol{LFI:white:noise:sensitivity:LFI27:Rad:S}{302.9}
\setsymbol{LFI:white:noise:sensitivity:LFI28:Rad:S}{285.3}

\setsymbol{LFI:knee:frequency:70GHz:units}{29.5\mHz}
\setsymbol{LFI:knee:frequency:44GHz:units}{56.2\mHz}
\setsymbol{LFI:knee:frequency:30GHz:units}{113.7\mHz}

\setsymbol{LFI:knee:frequency:70GHz}{29.5}
\setsymbol{LFI:knee:frequency:44GHz}{56.2}
\setsymbol{LFI:knee:frequency:30GHz}{113.7}

\setsymbol{LFI:knee:frequency:LFI18:Rad:M:units}{16.3\mHz}
\setsymbol{LFI:knee:frequency:LFI19:Rad:M:units}{15.1\mHz}
\setsymbol{LFI:knee:frequency:LFI20:Rad:M:units}{18.7\mHz}
\setsymbol{LFI:knee:frequency:LFI21:Rad:M:units}{37.2\mHz}
\setsymbol{LFI:knee:frequency:LFI22:Rad:M:units}{12.7\mHz}
\setsymbol{LFI:knee:frequency:LFI23:Rad:M:units}{34.6\mHz}
\setsymbol{LFI:knee:frequency:LFI24:Rad:M:units}{46.2\mHz}
\setsymbol{LFI:knee:frequency:LFI25:Rad:M:units}{24.9\mHz}
\setsymbol{LFI:knee:frequency:LFI26:Rad:M:units}{67.6\mHz}
\setsymbol{LFI:knee:frequency:LFI27:Rad:M:units}{187.4\mHz}
\setsymbol{LFI:knee:frequency:LFI28:Rad:M:units}{122.2\mHz}
\setsymbol{LFI:knee:frequency:LFI18:Rad:S:units}{17.7\mHz}
\setsymbol{LFI:knee:frequency:LFI19:Rad:S:units}{22.0\mHz}
\setsymbol{LFI:knee:frequency:LFI20:Rad:S:units}{8.7\mHz}
\setsymbol{LFI:knee:frequency:LFI21:Rad:S:units}{25.9\mHz}
\setsymbol{LFI:knee:frequency:LFI22:Rad:S:units}{15.8\mHz}
\setsymbol{LFI:knee:frequency:LFI23:Rad:S:units}{129.8\mHz}
\setsymbol{LFI:knee:frequency:LFI24:Rad:S:units}{100.9\mHz}
\setsymbol{LFI:knee:frequency:LFI25:Rad:S:units}{38.9\mHz}
\setsymbol{LFI:knee:frequency:LFI26:Rad:S:units}{58.9\mHz}
\setsymbol{LFI:knee:frequency:LFI27:Rad:S:units}{104.4\mHz}
\setsymbol{LFI:knee:frequency:LFI28:Rad:S:units}{40.7\mHz}

\setsymbol{LFI:knee:frequency:LFI18:Rad:M}{16.3}
\setsymbol{LFI:knee:frequency:LFI19:Rad:M}{15.1}
\setsymbol{LFI:knee:frequency:LFI20:Rad:M}{18.7}
\setsymbol{LFI:knee:frequency:LFI21:Rad:M}{37.2}
\setsymbol{LFI:knee:frequency:LFI22:Rad:M}{12.7}
\setsymbol{LFI:knee:frequency:LFI23:Rad:M}{34.6}
\setsymbol{LFI:knee:frequency:LFI24:Rad:M}{46.2}
\setsymbol{LFI:knee:frequency:LFI25:Rad:M}{24.9}
\setsymbol{LFI:knee:frequency:LFI26:Rad:M}{67.6}
\setsymbol{LFI:knee:frequency:LFI27:Rad:M}{187.4}
\setsymbol{LFI:knee:frequency:LFI28:Rad:M}{122.2}
\setsymbol{LFI:knee:frequency:LFI18:Rad:S}{17.7}
\setsymbol{LFI:knee:frequency:LFI19:Rad:S}{22.0}
\setsymbol{LFI:knee:frequency:LFI20:Rad:S}{8.7}
\setsymbol{LFI:knee:frequency:LFI21:Rad:S}{25.9}
\setsymbol{LFI:knee:frequency:LFI22:Rad:S}{15.8}
\setsymbol{LFI:knee:frequency:LFI23:Rad:S}{129.8}
\setsymbol{LFI:knee:frequency:LFI24:Rad:S}{100.9}
\setsymbol{LFI:knee:frequency:LFI25:Rad:S}{38.9}
\setsymbol{LFI:knee:frequency:LFI26:Rad:S}{58.9}
\setsymbol{LFI:knee:frequency:LFI27:Rad:S}{104.4}
\setsymbol{LFI:knee:frequency:LFI28:Rad:S}{40.7}

\setsymbol{LFI:slope:70GHz:units}{$-1.03$\mHz}
\setsymbol{LFI:slope:44GHz:units}{$-0.89$\mHz}
\setsymbol{LFI:slope:30GHz:units}{$-0.87$\mHz}

\setsymbol{LFI:slope:70GHz}{$-1.03$}
\setsymbol{LFI:slope:44GHz}{$-0.89$}
\setsymbol{LFI:slope:30GHz}{$-0.87$}

\setsymbol{LFI:slope:LFI18:Rad:M:units}{$-1.04$\mHz}
\setsymbol{LFI:slope:LFI19:Rad:M:units}{$-1.09$\mHz}
\setsymbol{LFI:slope:LFI20:Rad:M:units}{$-0.69$\mHz}
\setsymbol{LFI:slope:LFI21:Rad:M:units}{$-1.56$\mHz}
\setsymbol{LFI:slope:LFI22:Rad:M:units}{$-1.01$\mHz}
\setsymbol{LFI:slope:LFI23:Rad:M:units}{$-0.96$\mHz}
\setsymbol{LFI:slope:LFI24:Rad:M:units}{$-0.83$\mHz}
\setsymbol{LFI:slope:LFI25:Rad:M:units}{$-0.91$\mHz}
\setsymbol{LFI:slope:LFI26:Rad:M:units}{$-0.95$\mHz}
\setsymbol{LFI:slope:LFI27:Rad:M:units}{$-0.87$\mHz}
\setsymbol{LFI:slope:LFI28:Rad:M:units}{$-0.88$\mHz}
\setsymbol{LFI:slope:LFI18:Rad:S:units}{$-1.15$\mHz}
\setsymbol{LFI:slope:LFI19:Rad:S:units}{$-1.00$\mHz}
\setsymbol{LFI:slope:LFI20:Rad:S:units}{$-0.95$\mHz}
\setsymbol{LFI:slope:LFI21:Rad:S:units}{$-0.92$\mHz}
\setsymbol{LFI:slope:LFI22:Rad:S:units}{$-1.01$\mHz}
\setsymbol{LFI:slope:LFI23:Rad:S:units}{$-0.95$\mHz}
\setsymbol{LFI:slope:LFI24:Rad:S:units}{$-0.73$\mHz}
\setsymbol{LFI:slope:LFI25:Rad:S:units}{$-1.16$\mHz}
\setsymbol{LFI:slope:LFI26:Rad:S:units}{$-0.79$\mHz}
\setsymbol{LFI:slope:LFI27:Rad:S:units}{$-0.82$\mHz}
\setsymbol{LFI:slope:LFI28:Rad:S:units}{$-0.91$\mHz}

\setsymbol{LFI:slope:LFI18:Rad:M}{$-1.04$}
\setsymbol{LFI:slope:LFI19:Rad:M}{$-1.09$}
\setsymbol{LFI:slope:LFI20:Rad:M}{$-0.69$}
\setsymbol{LFI:slope:LFI21:Rad:M}{$-1.56$}
\setsymbol{LFI:slope:LFI22:Rad:M}{$-1.01$}
\setsymbol{LFI:slope:LFI23:Rad:M}{$-0.96$}
\setsymbol{LFI:slope:LFI24:Rad:M}{$-0.83$}
\setsymbol{LFI:slope:LFI25:Rad:M}{$-0.91$}
\setsymbol{LFI:slope:LFI26:Rad:M}{$-0.95$}
\setsymbol{LFI:slope:LFI27:Rad:M}{$-0.87$}
\setsymbol{LFI:slope:LFI28:Rad:M}{$-0.88$}
\setsymbol{LFI:slope:LFI18:Rad:S}{$-1.15$}
\setsymbol{LFI:slope:LFI19:Rad:S}{$-1.00$}
\setsymbol{LFI:slope:LFI20:Rad:S}{$-0.95$}
\setsymbol{LFI:slope:LFI21:Rad:S}{$-0.92$}
\setsymbol{LFI:slope:LFI22:Rad:S}{$-1.01$}
\setsymbol{LFI:slope:LFI23:Rad:S}{$-0.95$}
\setsymbol{LFI:slope:LFI24:Rad:S}{$-0.73$}
\setsymbol{LFI:slope:LFI25:Rad:S}{$-1.16$}
\setsymbol{LFI:slope:LFI26:Rad:S}{$-0.79$}
\setsymbol{LFI:slope:LFI27:Rad:S}{$-0.82$}
\setsymbol{LFI:slope:LFI28:Rad:S}{$-0.91$}

\setsymbol{LFI:FWHM:70GHz:units}{13\parcm01}
\setsymbol{LFI:FWHM:44GHz:units}{27\parcm92}
\setsymbol{LFI:FWHM:30GHz:units}{32\parcm65}

\setsymbol{LFI:FWHM:70GHz}{13.01}
\setsymbol{LFI:FWHM:44GHz}{27.92}
\setsymbol{LFI:FWHM:30GHz}{32.65}

\setsymbol{LFI:FWHM:LFI18:units}{13\parcm39}
\setsymbol{LFI:FWHM:LFI19:units}{13\parcm01}
\setsymbol{LFI:FWHM:LFI20:units}{12\parcm75}
\setsymbol{LFI:FWHM:LFI21:units}{12\parcm74}
\setsymbol{LFI:FWHM:LFI22:units}{12\parcm87}
\setsymbol{LFI:FWHM:LFI23:units}{13\parcm27}
\setsymbol{LFI:FWHM:LFI24:units}{22\parcm98}
\setsymbol{LFI:FWHM:LFI25:units}{30\parcm46}
\setsymbol{LFI:FWHM:LFI26:units}{30\parcm31}
\setsymbol{LFI:FWHM:LFI27:units}{32\parcm65}
\setsymbol{LFI:FWHM:LFI28:units}{32\parcm66}

\setsymbol{LFI:FWHM:LFI18}{13.39}
\setsymbol{LFI:FWHM:LFI19}{13.01}
\setsymbol{LFI:FWHM:LFI20}{12.75}
\setsymbol{LFI:FWHM:LFI21}{12.74}
\setsymbol{LFI:FWHM:LFI22}{12.87}
\setsymbol{LFI:FWHM:LFI23}{13.27}
\setsymbol{LFI:FWHM:LFI24}{22.98}
\setsymbol{LFI:FWHM:LFI25}{30.46}
\setsymbol{LFI:FWHM:LFI26}{30.31}
\setsymbol{LFI:FWHM:LFI27}{32.65}
\setsymbol{LFI:FWHM:LFI28}{32.66}

\setsymbol{LFI:FWHM:uncertainty:LFI18:units}{0.170\arcm}
\setsymbol{LFI:FWHM:uncertainty:LFI19:units}{0.174\arcm}
\setsymbol{LFI:FWHM:uncertainty:LFI20:units}{0.170\arcm}
\setsymbol{LFI:FWHM:uncertainty:LFI21:units}{0.156\arcm}
\setsymbol{LFI:FWHM:uncertainty:LFI22:units}{0.164\arcm}
\setsymbol{LFI:FWHM:uncertainty:LFI23:units}{0.171\arcm}
\setsymbol{LFI:FWHM:uncertainty:LFI24:units}{0.652\arcm}
\setsymbol{LFI:FWHM:uncertainty:LFI25:units}{1.075\arcm}
\setsymbol{LFI:FWHM:uncertainty:LFI26:units}{1.131\arcm}
\setsymbol{LFI:FWHM:uncertainty:LFI27:units}{1.266\arcm}
\setsymbol{LFI:FWHM:uncertainty:LFI28:units}{1.287\arcm}

\setsymbol{LFI:FWHM:uncertainty:LFI18}{0.170}
\setsymbol{LFI:FWHM:uncertainty:LFI19}{0.174}
\setsymbol{LFI:FWHM:uncertainty:LFI20}{0.170}
\setsymbol{LFI:FWHM:uncertainty:LFI21}{0.156}
\setsymbol{LFI:FWHM:uncertainty:LFI22}{0.164}
\setsymbol{LFI:FWHM:uncertainty:LFI23}{0.171}
\setsymbol{LFI:FWHM:uncertainty:LFI24}{0.652}
\setsymbol{LFI:FWHM:uncertainty:LFI25}{1.075}
\setsymbol{LFI:FWHM:uncertainty:LFI26}{1.131}
\setsymbol{LFI:FWHM:uncertainty:LFI27}{1.266}
\setsymbol{LFI:FWHM:uncertainty:LFI28}{1.287}

\setsymbol{HFI:center:frequency:100GHz:units}{100\,GHz}
\setsymbol{HFI:center:frequency:143GHz:units}{143\,GHz}
\setsymbol{HFI:center:frequency:217GHz:units}{217\,GHz}
\setsymbol{HFI:center:frequency:353GHz:units}{353\,GHz}
\setsymbol{HFI:center:frequency:545GHz:units}{545\,GHz}
\setsymbol{HFI:center:frequency:857GHz:units}{857\,GHz}

\setsymbol{HFI:center:frequency:100GHz}{100}
\setsymbol{HFI:center:frequency:143GHz}{143}
\setsymbol{HFI:center:frequency:217GHz}{217}
\setsymbol{HFI:center:frequency:353GHz}{353}
\setsymbol{HFI:center:frequency:545GHz}{545}
\setsymbol{HFI:center:frequency:857GHz}{857}

\setsymbol{HFI:Ndetectors:100GHz}{8}
\setsymbol{HFI:Ndetectors:143GHz}{11}
\setsymbol{HFI:Ndetectors:217GHz}{12}
\setsymbol{HFI:Ndetectors:353GHz}{12}
\setsymbol{HFI:Ndetectors:545GHz}{3}
\setsymbol{HFI:Ndetectors:857GHz}{4}

\setsymbol{HFI:FWHM:Maps:100GHz:units}{9\parcm88}
\setsymbol{HFI:FWHM:Maps:143GHz:units}{7\parcm18}
\setsymbol{HFI:FWHM:Maps:217GHz:units}{4\parcm87}
\setsymbol{HFI:FWHM:Maps:353GHz:units}{4\parcm65}
\setsymbol{HFI:FWHM:Maps:545GHz:units}{4\parcm72}
\setsymbol{HFI:FWHM:Maps:857GHz:units}{4\parcm39}
\setsymbol{HFI:FWHM:Maps:100GHz}{9.88}
\setsymbol{HFI:FWHM:Maps:143GHz}{7.18}
\setsymbol{HFI:FWHM:Maps:217GHz}{4.87}
\setsymbol{HFI:FWHM:Maps:353GHz}{4.65}
\setsymbol{HFI:FWHM:Maps:545GHz}{4.72}
\setsymbol{HFI:FWHM:Maps:857GHz}{4.39}

\setsymbol{HFI:beam:ellipticity:Maps:100GHz}{1.15}
\setsymbol{HFI:beam:ellipticity:Maps:143GHz}{1.01}
\setsymbol{HFI:beam:ellipticity:Maps:217GHz}{1.06}
\setsymbol{HFI:beam:ellipticity:Maps:353GHz}{1.05}
\setsymbol{HFI:beam:ellipticity:Maps:545GHz}{1.14}
\setsymbol{HFI:beam:ellipticity:Maps:857GHz}{1.19}

\setsymbol{HFI:FWHM:Mars:100GHz:units}{9\parcm37}
\setsymbol{HFI:FWHM:Mars:143GHz:units}{7\parcm04}
\setsymbol{HFI:FWHM:Mars:217GHz:units}{4\parcm68}
\setsymbol{HFI:FWHM:Mars:353GHz:units}{4\parcm43}
\setsymbol{HFI:FWHM:Mars:545GHz:units}{3\parcm80}
\setsymbol{HFI:FWHM:Mars:857GHz:units}{3\parcm67}

\setsymbol{HFI:FWHM:Mars:100GHz}{9.37}
\setsymbol{HFI:FWHM:Mars:143GHz}{7.04}
\setsymbol{HFI:FWHM:Mars:217GHz}{4.68}
\setsymbol{HFI:FWHM:Mars:353GHz}{4.43}
\setsymbol{HFI:FWHM:Mars:545GHz}{3.80}
\setsymbol{HFI:FWHM:Mars:857GHz}{3.67}

\setsymbol{HFI:beam:ellipticity:Mars:100GHz}{1.18}
\setsymbol{HFI:beam:ellipticity:Mars:143GHz}{1.03}
\setsymbol{HFI:beam:ellipticity:Mars:217GHz}{1.14}
\setsymbol{HFI:beam:ellipticity:Mars:353GHz}{1.09}
\setsymbol{HFI:beam:ellipticity:Mars:545GHz}{1.25}
\setsymbol{HFI:beam:ellipticity:Mars:857GHz}{1.03}

\setsymbol{HFI:CMB:relative:calibration:100GHz}{$\lsim 1\%$}
\setsymbol{HFI:CMB:relative:calibration:143GHz}{$\lsim 1\%$}
\setsymbol{HFI:CMB:relative:calibration:217GHz}{$\lsim 1\%$}
\setsymbol{HFI:CMB:relative:calibration:353GHz}{$\lsim 1\%$}
\setsymbol{HFI:CMB:relative:calibration:545GHz}{}
\setsymbol{HFI:CMB:relative:calibration:857GHz}{}

\setsymbol{HFI:CMB:absolute:calibration:100GHz}{$\lsim 2\%$}
\setsymbol{HFI:CMB:absolute:calibration:143GHz}{$\lsim 2\%$}
\setsymbol{HFI:CMB:absolute:calibration:217GHz}{$\lsim 2\%$}
\setsymbol{HFI:CMB:absolute:calibration:353GHz}{$\lsim 2\%$}
\setsymbol{HFI:CMB:absolute:calibration:545GHz}{}
\setsymbol{HFI:CMB:absolute:calibration:857GHz}{}

\setsymbol{HFI:FIRAS:gain:calibration:accuracy:statistical:100GHz}{}
\setsymbol{HFI:FIRAS:gain:calibration:accuracy:statistical:143GHz}{}
\setsymbol{HFI:FIRAS:gain:calibration:accuracy:statistical:217GHz}{}
\setsymbol{HFI:FIRAS:gain:calibration:accuracy:statistical:353GHz}{2.5\%}
\setsymbol{HFI:FIRAS:gain:calibration:accuracy:statistical:545GHz}{1\%}
\setsymbol{HFI:FIRAS:gain:calibration:accuracy:statistical:857GHz}{0.5\%}

\setsymbol{HFI:FIRAS:gain:calibration:accuracy:systematic:100GHz}{}
\setsymbol{HFI:FIRAS:gain:calibration:accuracy:systematic:143GHz}{}
\setsymbol{HFI:FIRAS:gain:calibration:accuracy:systematic:217GHz}{}
\setsymbol{HFI:FIRAS:gain:calibration:accuracy:systematic:353GHz}{}
\setsymbol{HFI:FIRAS:gain:calibration:accuracy:systematic:545GHz}{7\%}
\setsymbol{HFI:FIRAS:gain:calibration:accuracy:systematic:857GHz}{7\%}

\setsymbol{HFI:FIRAS:zero:point:accuracy:100GHz:units}{0.8\MJysr}
\setsymbol{HFI:FIRAS:zero:point:accuracy:143GHz:units}{}
\setsymbol{HFI:FIRAS:zero:point:accuracy:217GHz:units}{}
\setsymbol{HFI:FIRAS:zero:point:accuracy:353GHz:units}{1.4\MJysr}
\setsymbol{HFI:FIRAS:zero:point:accuracy:545GHz:units}{2.2\MJysr}
\setsymbol{HFI:FIRAS:zero:point:accuracy:857GHz:units}{1.7\MJysr}

\setsymbol{HFI:FIRAS:zero:point:accuracy:100GHz}{0.8}
\setsymbol{HFI:FIRAS:zero:point:accuracy:143GHz}{}
\setsymbol{HFI:FIRAS:zero:point:accuracy:217GHz}{}
\setsymbol{HFI:FIRAS:zero:point:accuracy:353GHz}{1.4}
\setsymbol{HFI:FIRAS:zero:point:accuracy:545GHz}{2.2}
\setsymbol{HFI:FIRAS:zero:point:accuracy:857GHz}{1.7}

\setsymbol{HFI:unit:conversion:100GHz:units}{0.2415\MJysrmK}
\setsymbol{HFI:unit:conversion:143GHz:units}{0.3694\MJysrmK}
\setsymbol{HFI:unit:conversion:217GHz:units}{0.4811\MJysrmK}
\setsymbol{HFI:unit:conversion:353GHz:units}{0.2883\MJysrmK}
\setsymbol{HFI:unit:conversion:545GHz:units}{0.05826\MJysrmK}
\setsymbol{HFI:unit:conversion:857GHz:units}{0.002238\MJysrmK}

\setsymbol{HFI:unit:conversion:100GHz}{0.2415}
\setsymbol{HFI:unit:conversion:143GHz}{0.3694}
\setsymbol{HFI:unit:conversion:217GHz}{0.4811}
\setsymbol{HFI:unit:conversion:353GHz}{0.2883}
\setsymbol{HFI:unit:conversion:545GHz}{0.05826}
\setsymbol{HFI:unit:conversion:857GHz}{0.002238}

\setsymbol{HFI:colour:correction:alpha=-2:V1.01:100GHz}{0.9893}
\setsymbol{HFI:colour:correction:alpha=-2:V1.01:143GHz}{0.9759}
\setsymbol{HFI:colour:correction:alpha=-2:V1.01:217GHz}{1.0007}
\setsymbol{HFI:colour:correction:alpha=-2:V1.01:353GHz}{1.0028}
\setsymbol{HFI:colour:correction:alpha=-2:V1.01:545GHz}{1.0019}
\setsymbol{HFI:colour:correction:alpha=-2:V1.01:857GHz}{0.9889}

\setsymbol{HFI:colour:correction:alpha=0:V1.01:100GHz}{1.0008}
\setsymbol{HFI:colour:correction:alpha=0:V1.01:143GHz}{1.0148}
\setsymbol{HFI:colour:correction:alpha=0:V1.01:217GHz}{0.9909}
\setsymbol{HFI:colour:correction:alpha=0:V1.01:353GHz}{0.9888}
\setsymbol{HFI:colour:correction:alpha=0:V1.01:545GHz}{0.9878}
\setsymbol{HFI:colour:correction:alpha=0:V1.01:857GHz}{1.0014}

\begin{document}
%
\author{\small
Planck Collaboration:
P.~A.~R.~Ade\inst{89}
\and
N.~Aghanim\inst{62}
\and
C.~Armitage-Caplan\inst{95}
\and
M.~Arnaud\inst{76}
\and
M.~Ashdown\inst{73, 6}
\and
F.~Atrio-Barandela\inst{18}
\and
J.~Aumont\inst{62}
\and
C.~Baccigalupi\inst{88}
\and
A.~J.~Banday\inst{99, 9}
\and
R.~B.~Barreiro\inst{70}
% ...
}
\institute{\small
APC, AstroParticule et Cosmologie, Universit\'{e} Paris Diderot, CNRS/IN2P3, CEA/lrfu, Observatoire de Paris, Sorbonne Paris Cit\'{e}, 10, rue Alice Domon et L\'{e}onie Duquet, 75205 Paris Cedex 13, France\\
\and
Aalto University Mets\"{a}hovi Radio Observatory, Mets\"{a}hovintie 114, FIN-02540 Kylm\"{a}l\"{a}, Finland\\
\and
African Institute for Mathematical Sciences, 6-8 Melrose Road, Muizenberg, Cape Town, South Africa\\
\and
Agenzia Spaziale Italiana Science Data Center, Via del Politecnico snc, 00133, Roma, Italy\\
\and
Agenzia Spaziale Italiana, Viale Liegi 26, Roma, Italy\\
\and
Astrophysics Group, Cavendish Laboratory, University of Cambridge, J J Thomson Avenue, Cambridge CB3 0HE, U.K.\\
\and
Astrophysics \& Cosmology Research Unit, School of Mathematics, Statistics \& Computer Science, University of KwaZulu-Natal, Westville Campus, Private Bag X54001, Durban 4000, South Africa\\
\and
CITA, University of Toronto, 60 St. George St., Toronto, ON M5S 3H8, Canada\\
\and
CNRS, IRAP, 9 Av. colonel Roche, BP 44346, F-31028 Toulouse cedex 4, France\\
\and
California Institute of Technology, Pasadena, California, U.S.A.\\
\and
Centre for Theoretical Cosmology, DAMTP, University of Cambridge, Wilberforce Road, Cambridge CB3 0WA, U.K.\\
\and
Centro de Estudios de F\'{i}sica del Cosmos de Arag\'{o}n (CEFCA), Plaza San Juan, 1, planta 2, E-44001, Teruel, Spain\\
\and
Computational Cosmology Center, Lawrence Berkeley National Laboratory, Berkeley, California, U.S.A.\\
\and
Consejo Superior de Investigaciones Cient\'{\i}ficas (CSIC), Madrid, Spain\\
\and
DSM/Irfu/SPP, CEA-Saclay, F-91191 Gif-sur-Yvette Cedex, France\\
\and
DTU Space, National Space Institute, Technical University of Denmark, Elektrovej 327, DK-2800 Kgs. Lyngby, Denmark\\
\and
D\'{e}partement de Physique Th\'{e}orique, Universit\'{e} de Gen\`{e}ve, 24, Quai E. Ansermet,1211 Gen\`{e}ve 4, Switzerland\\
\and
Departamento de F\'{\i}sica Fundamental, Facultad de Ciencias, Universidad de Salamanca, 37008 Salamanca, Spain\\
\and
Departamento de F\'{\i}sica, Universidad de Oviedo, Avda. Calvo Sotelo s/n, Oviedo, Spain\\
\and
Department of Astronomy and Astrophysics, University of Toronto, 50 Saint George Street, Toronto, Ontario, Canada\\
\and
Department of Astrophysics/IMAPP, Radboud University Nijmegen, P.O. Box 9010, 6500 GL Nijmegen, The Netherlands\\
\and
Department of Electrical Engineering and Computer Sciences, University of California, Berkeley, California, U.S.A.\\
\and
Department of Physics \& Astronomy, University of British Columbia, 6224 Agricultural Road, Vancouver, British Columbia, Canada\\
\and
Department of Physics and Astronomy, Dana and David Dornsife College of Letter, Arts and Sciences, University of Southern California, Los Angeles, CA 90089, U.S.A.\\
\and
Department of Physics and Astronomy, University College London, London WC1E 6BT, U.K.\\
\and
Department of Physics, Carnegie Mellon University, 5000 Forbes Ave, Pittsburgh, PA 15213, U.S.A.\\
\and
Department of Physics, Florida State University, Keen Physics Building, 77 Chieftan Way, Tallahassee, Florida, U.S.A.\\
\and
Department of Physics, Gustaf H\"{a}llstr\"{o}min katu 2a, University of Helsinki, Helsinki, Finland\\
\and
Department of Physics, Princeton University, Princeton, New Jersey, U.S.A.\\
\and
Department of Physics, University of California, Berkeley, California, U.S.A.\\
\and
Department of Physics, University of California, One Shields Avenue, Davis, California, U.S.A.\\
\and
Department of Physics, University of California, Santa Barbara, California, U.S.A.\\
\and
Department of Physics, University of Illinois at Urbana-Champaign, 1110 West Green Street, Urbana, Illinois, U.S.A.\\
\and
Dipartimento di Fisica e Astronomia G. Galilei, Universit\`{a} degli Studi di Padova, via Marzolo 8, 35131 Padova, Italy\\
\and
Dipartimento di Fisica e Scienze della Terra, Universit\`{a} di Ferrara, Via Saragat 1, 44122 Ferrara, Italy\\
\and
Dipartimento di Fisica, Universit\`{a} La Sapienza, P. le A. Moro 2, Roma, Italy\\
\and
Dipartimento di Fisica, Universit\`{a} degli Studi di Milano, Via Celoria, 16, Milano, Italy\\
\and
Dipartimento di Fisica, Universit\`{a} degli Studi di Trieste, via A. Valerio 2, Trieste, Italy\\
\and
Dipartimento di Fisica, Universit\`{a} di Roma Tor Vergata, Via della Ricerca Scientifica, 1, Roma, Italy\\
\and
Discovery Center, Niels Bohr Institute, Blegdamsvej 17, Copenhagen, Denmark\\
\and
Dpto. Astrof\'{i}sica, Universidad de La Laguna (ULL), E-38206 La Laguna, Tenerife, Spain\\
\and
European Space Agency, ESAC, Planck Science Office, Camino bajo del Castillo, s/n, Urbanizaci\'{o}n Villafranca del Castillo, Villanueva de la Ca\~{n}ada, Madrid, Spain\\
\and
European Space Agency, ESTEC, Keplerlaan 1, 2201 AZ Noordwijk, The Netherlands\\
\and
Haverford College Astronomy Department, 370 Lancaster Avenue, Haverford, Pennsylvania, U.S.A.\\
\and
Helsinki Institute of Physics, Gustaf H\"{a}llstr\"{o}min katu 2, University of Helsinki, Helsinki, Finland\\
\and
INAF - Osservatorio Astrofisico di Catania, Via S. Sofia 78, Catania, Italy\\
\and
INAF - Osservatorio Astronomico di Padova, Vicolo dell'Osservatorio 5, Padova, Italy\\
\and
INAF - Osservatorio Astronomico di Roma, via di Frascati 33, Monte Porzio Catone, Italy\\
\and
INAF - Osservatorio Astronomico di Trieste, Via G.B. Tiepolo 11, Trieste, Italy\\
\and
INAF Istituto di Radioastronomia, Via P. Gobetti 101, 40129 Bologna, Italy\\
\and
INAF/IASF Bologna, Via Gobetti 101, Bologna, Italy\\
\and
INAF/IASF Milano, Via E. Bassini 15, Milano, Italy\\
\and
INFN, Sezione di Bologna, Via Irnerio 46, I-40126, Bologna, Italy\\
\and
INFN, Sezione di Roma 1, Universit\`{a} di Roma Sapienza, Piazzale Aldo Moro 2, 00185, Roma, Italy\\
\and
INFN/National Institute for Nuclear Physics, Via Valerio 2, I-34127 Trieste, Italy\\
\and
IPAG: Institut de Plan\'{e}tologie et d'Astrophysique de Grenoble, Universit\'{e} Joseph Fourier, Grenoble 1 / CNRS-INSU, UMR 5274, Grenoble, F-38041, France\\
\and
IUCAA, Post Bag 4, Ganeshkhind, Pune University Campus, Pune 411 007, India\\
\and
Imperial College London, Astrophysics group, Blackett Laboratory, Prince Consort Road, London, SW7 2AZ, U.K.\\
\and
Infrared Processing and Analysis Center, California Institute of Technology, Pasadena, CA 91125, U.S.A.\\
\and
Institut N\'{e}el, CNRS, Universit\'{e} Joseph Fourier Grenoble I, 25 rue des Martyrs, Grenoble, France\\
\and
Institut Universitaire de France, 103, bd Saint-Michel, 75005, Paris, France\\
\and
Institut d'Astrophysique Spatiale, CNRS (UMR8617) Universit\'{e} Paris-Sud 11, B\^{a}timent 121, Orsay, France\\
\and
Institut d'Astrophysique de Paris, CNRS (UMR7095), 98 bis Boulevard Arago, F-75014, Paris, France\\
\and
Institut de Ci\`{e}ncies de l'Espai, CSIC/IEEC, Facultat de Ci\`{e}ncies, Campus UAB, Torre C5 par-2, Bellaterra 08193, Spain\\
\and
Institute for Space Sciences, Bucharest-Magurale, Romania\\
\and
Institute of Astronomy and Astrophysics, Academia Sinica, Taipei, Taiwan\\
\and
Institute of Astronomy, University of Cambridge, Madingley Road, Cambridge CB3 0HA, U.K.\\
\and
Institute of Theoretical Astrophysics, University of Oslo, Blindern, Oslo, Norway\\
\and
Instituto de Astrof\'{\i}sica de Canarias, C/V\'{\i}a L\'{a}ctea s/n, La Laguna, Tenerife, Spain\\
\and
Instituto de F\'{\i}sica de Cantabria (CSIC-Universidad de Cantabria), Avda. de los Castros s/n, Santander, Spain\\
\and
Jet Propulsion Laboratory, California Institute of Technology, 4800 Oak Grove Drive, Pasadena, California, U.S.A.\\
\and
Jodrell Bank Centre for Astrophysics, Alan Turing Building, School of Physics and Astronomy, The University of Manchester, Oxford Road, Manchester, M13 9PL, U.K.\\
\and
Kavli Institute for Cosmology Cambridge, Madingley Road, Cambridge, CB3 0HA, U.K.\\
\and
LAL, Universit\'{e} Paris-Sud, CNRS/IN2P3, Orsay, France\\
\and
LERMA, CNRS, Observatoire de Paris, 61 Avenue de l'Observatoire, Paris, France\\
\and
Laboratoire AIM, IRFU/Service d'Astrophysique - CEA/DSM - CNRS - Universit\'{e} Paris Diderot, B\^{a}t. 709, CEA-Saclay, F-91191 Gif-sur-Yvette Cedex, France\\
\and
Laboratoire Traitement et Communication de l'Information, CNRS (UMR 5141) and T\'{e}l\'{e}com ParisTech, 46 rue Barrault F-75634 Paris Cedex 13, France\\
\and
Laboratoire de Physique Subatomique et de Cosmologie, Universit\'{e} Joseph Fourier Grenoble I, CNRS/IN2P3, Institut National Polytechnique de Grenoble, 53 rue des Martyrs, 38026 Grenoble cedex, France\\
\and
Laboratoire de Physique Th\'{e}orique, Universit\'{e} Paris-Sud 11 \& CNRS, B\^{a}timent 210, 91405 Orsay, France\\
\and
Lawrence Berkeley National Laboratory, Berkeley, California, U.S.A.\\
\and
Max-Planck-Institut f\"{u}r Astrophysik, Karl-Schwarzschild-Str. 1, 85741 Garching, Germany\\
\and
McGill Physics, Ernest Rutherford Physics Building, McGill University, 3600 rue University, Montr\'{e}al, QC, H3A 2T8, Canada\\
\and
MilliLab, VTT Technical Research Centre of Finland, Tietotie 3, Espoo, Finland\\
\and
Niels Bohr Institute, Blegdamsvej 17, Copenhagen, Denmark\\
\and
Observational Cosmology, Mail Stop 367-17, California Institute of Technology, Pasadena, CA, 91125, U.S.A.\\
\and
Optical Science Laboratory, University College London, Gower Street, London, U.K.\\
\and
SB-ITP-LPPC, EPFL, CH-1015, Lausanne, Switzerland\\
\and
SISSA, Astrophysics Sector, via Bonomea 265, 34136, Trieste, Italy\\
\and
School of Physics and Astronomy, Cardiff University, Queens Buildings, The Parade, Cardiff, CF24 3AA, U.K.\\
\and
School of Physics and Astronomy, University of Nottingham, Nottingham NG7 2RD, U.K.\\
\and
Space Research Institute (IKI), Russian Academy of Sciences, Profsoyuznaya Str, 84/32, Moscow, 117997, Russia\\
\and
Space Sciences Laboratory, University of California, Berkeley, California, U.S.A.\\
\and
Special Astrophysical Observatory, Russian Academy of Sciences, Nizhnij Arkhyz, Zelenchukskiy region, Karachai-Cherkessian Republic, 369167, Russia\\
\and
Stanford University, Dept of Physics, Varian Physics Bldg, 382 Via Pueblo Mall, Stanford, California, U.S.A.\\
\and
Sub-Department of Astrophysics, University of Oxford, Keble Road, Oxford OX1 3RH, U.K.\\
\and
Theory Division, PH-TH, CERN, CH-1211, Geneva 23, Switzerland\\
\and
UPMC Univ Paris 06, UMR7095, 98 bis Boulevard Arago, F-75014, Paris, France\\
\and
Universit\"{a}t Heidelberg, Institut f\"{u}r Theoretische Astrophysik, Philosophenweg 12, 69120 Heidelberg, Germany\\
\and
Universit\'{e} de Toulouse, UPS-OMP, IRAP, F-31028 Toulouse cedex 4, France\\
% ...
}

  \title{\Planck\ 2013 results. XIX. The integrated Sachs-Wolfe effect}
\authorrunning{Planck Collaboration}  
\titlerunning{The ISW effect with \Planck}

\date{....}

\abstract{Based on Cosmic Microwave Background (CMB) maps from the 2013
\Planck\ Mission data release, this paper presents the detection of the
Integrated Sachs-Wolfe (ISW) effect, i.e., the correlation between the CMB
and large-scale evolving gravitational potentials. The significance of
detection ranges from 2 to $4\,\sigma$, depending on which method is used.
We investigate three separate approaches, which cover essentially all previous
studies, as well as breaking new ground.
(i) Through correlation of the CMB with the \Planck\ reconstructed
gravitational lensing potential (for the first time). This detection is made
using the lensing-induced bispectrum between the low-$\ell$ and high-$\ell$
temperature anisotropies; the correlation between lensing and the ISW effect
has a significance close to $2.5\,\sigma$.
(ii) Through cross-correlation with tracers of large-scale structure, yielding
around $3\,\sigma$ significance, based on a combination of radio (NVSS) and
optical (SDSS) data.
(iii) Using aperture photometry on stacked CMB fields at the locations of
known large-scale structures, which
yields and confirms, over a broader spectral range, a $4\,\sigma$ signal
when using a previously explored catalogue, but shows strong discrepancies in
amplitude and scale compared to expectations.  More recent catalogues give
more moderate results, ranging from negligible to $2.5\,\sigma$ at most,
but with a more consistent scale and amplitude, the latter being still
slightly above what is expected from numerical simulations within
$\Lambda$CMD.
Where they can be compared, these measurements are compatible with previous
work using data from \wmap, which had already mapped these scales to the
limits of cosmic variance.  \Planck's broader frequency coverage allows for
better foreground cleaning, and confirms that the signal is achromatic,
bolstering the case for ISW detection.  
As a final step we use tracers of large-scale structure to filter the CMB
data, presenting maps of the ISW temperature perturbation.
These results provide complementary and independent evidence for the
existence of a dark energy component that governs the current accelerated
expansion of the Universe.}

\date{Received 2013; accepted 2013}


\authorrunning{Planck Collaboration}

\maketitle

\clearpage
\section{Introduction}

This paper, one of a set associated with the 2013 data release from the
\Planck
\footnote{\Planck\ (\url{http://www.esa.int/Planck}) is a project of the
European Space Agency (ESA) with instruments provided by two scientific
consortia funded by ESA member states (in particular the lead countries France
and Italy), with contributions from NASA (USA) and telescope reflectors
provided by a collaboration between ESA and a scientific consortium led and
funded by Denmark.}
mission \citep{planck2013-p01}, presents the first results on the
integrated Sachs-Wolfe (ISW) effect using \Planck\ data. 
The ISW effect \citep{Sachs1967, Rees1968, Martinez1990b, Hu1994} is a secondary anisotropy
in the cosmic microwave background (CMB), caused by the interaction of CMB
photons with the time-evolving potentials from large-scale structure
(LSS, hereafter). Photons follow a geodesic that is weakly perturbed by the
Newtonian gravitational potential, $\Phi$, and experience a fractional shift
in their temperature given by
\begin{equation}
\Theta = \frac{\Delta T}{T_\mathrm{CMB}} =
 \frac{2}{c^3}\int_{\eta_\ast}^{\eta_0}\mathrm{d}\eta\:
 \frac{\partial\Phi}{\partial\eta},
\end{equation}
where the integral is expressed in terms of the conformal time $\eta$,
defined differentially by $\mathrm{d}\eta/\mathrm{d}a = 1/(a^2H(a))$ with
$H(a)$ the Hubble function and $a$ the scale factor. 
The integration limits here go from the recombination time ($\eta_\ast$)
to the present time ($\eta_0$).

The sensitivity of the ISW effect to gravitational potentials
(that can extend over Gpc scales) 
results in the power of the ISW being concentrated on the largest scales. 
The largest scales for the CMB have been mapped out by the {\it Wilkinson
Microwave Anisotropy Probe\/} (\wmap) to the statistical limit of cosmic
variance.  Some systematics (like foreground removal) can have an impact on
the reconstruction of the CMB especially at the largest scales where our
Galaxy can introduce significant residuals on the reconstructed CMB map. The 
superior sensitivity of \Planck\ together with its better angular resolution
and wider frequency coverage allows
 for a better understanding
(and hence removal) of Galactic and extagalactic foregrounds, therefore 
reducing the possible negative impact of these residuals. \Planck\ allows us to
improve on previous measurements by having a better systematic control,
an improved removal of foregrounds (that permits us to explore the
achromatic nature of the ISW signal on a wider frequency range), and a better 
understanding of systematics affecting tracer catalogues.

For cosmological models where $\Omega_{\rm m}=1$, gravitational potentials
remain constant during linear structure formation, and the ISW signal is
negligible (to first order, although second order nonlinear 
ISW is always expected around smaller over- and under-dense regions). 
In the presence of dark energy, decaying potentials due to the accelerated
expansion rate, result in a net ISW effect which is positive when the CMB
photons cross overdense regions 
and negative when the CMB photons cross underdense regions.  
Therefore, the ISW effect is an indicator of either non-zero curvature
\citep{Kamionkowski1994}\footnote{It is worth mentioning that this was one of the arguments that
suggested a preference for $\Lambda$CDM before the discovery of acceleration using SNe.}, any form of dark energy, such as a
cosmological constant $\Lambda$ \citep{Crittenden1996}, modified gravity
\citep{Hu2002a}, or a combination of these possibilities. 
By measuring the rate at which gravitational potentials
in the LSS decay (up to redshift of around 2), the ISW effect
can be used as an independent probe of cosmology and provides
complementary and independent evidence for dark energy.

Detection of the ISW effect was first made possible with all-sky CMB maps
from \wmap.  Based on these data, many works can be found in the
literature where the authors aim at making, 
and subsequectly improving, the measurement of the ISW effect through 
correlations with tracer catalogues: 2MASS \citep[an infrared
  catalogue out to low redshifts around 0.1,][]{Afshordi2004a, Rassat2006, 
  Francis2009a, Dupe2011}, \textit{HEAO} \citep[an X-ray survey at low
  redshift, with the first positive claim for
  detection,][]{Boughn2004}, Sloan Digital Sky Survey
  \citep[SDSS, an optical survey at intermediate
  redshifts,][]{Fosalba2003, Scranton2003, Fosalba2004, Padmanabhan2005, Cabre2006, 
  Giannantonio2006b, Granett2009a, Xia2009, Bielby2010, Lopez2010,
  Sawangwit2010},
  the NRAO VLA Sky Survey
\citep[NVSS, a radio catalogue with high-redshift sources,]
[]{Boughn2005b,Vielva2006,Pietrobon2006a,Mcewen2007,Raccanelli2008,Hernandez2010,Massardi2010,Schiavon2012},
and combined measurements with multiple tracers 
\citep{Nolta2004,Ho2008,Corasaniti2005,Gaztanaga2006,Giannantonio2008b,Giannantonio2012}. 
The significance of the ISW detections that can be found in the literature
range between $0.9\,\sigma$ and $4.7\,\sigma$. There are a number of
peculiarities related to some of the detection claims, 
as noted by \citet{Hernandez2010} and \citet{Lopez2010}. 
They both found lower significance levels than some previous studies
and pointed out the absence of the signal at low multipoles 
where the ISW effect should be most prominent and the presence of point source emission on small scales for radio surveys.

The main result that is obtained from an ISW detection is a constraint
on the cosmological constant (or dark energy), $\Omega_\Lambda$. 
The general consensus from the variety of ISW analyses is for a
value of $\Omega_\Lambda\simeq0.75$ with an error of about 20\%,
which provides independent evidence for the existence of dark energy 
\citep{Fosalba2003, Fosalba2004, Nolta2004,
  Corasaniti2005, Padmanabhan2005, Cabre2006, Giannantonio2006b,
  Pietrobon2006b, Rassat2006, Vielva2006, Mcewen2007, Ho2008, Schiavon2012}. All tests
on spatial flatness find an upper limit for $\Omega_K$ of a few
percent \citep{Nolta2004,Gaztanaga2006,Ho2008,Li2010}. Using a prior
on spatial flatness, the dark energy
equation of state parameter, $w$, was found to be close to $-1$
\citep{Giannantonio2006b,Vielva2006, Ho2008} and has been excluded from
having a strong time evolution \citep{Giannantonio2008b, Li2010}. 

The ISW effect is achromatic, conserving the Planck spectrum of the
CMB and can be separated from other CMB fluctuations through
cross-correlations with catalogues which trace the LSS gravitational
potentials \citep[see for instance][]{Crittenden1996}. This cross-correlation
can be studied in different ways: angular cross-correlations in real
space between the CMB and the catalogues tracing the LSS;
the corresponding angular cross-power spectrum of the Fourier-transformed maps;
or through the covariance of wavelet-filtered maps as a function of wavelet
scale.  The studies using \wmap\ data mentioned above follow this survey
cross-correlation techique.

An alternative approach, similar to the angular cross-correlation in real
space, consists of stacking CMB fields centred on known supersclusters or 
supervoids \citep{Granett2008a, Granett2008b, Papai2010b}. The advantage of
this technique is that it allows for a detailed study of the profile of the
CMB fluctuations caused by this secondary anisotropy.
 
A novel and powerful approach takes advantage of the fact that the
same potentials that make CMB photons gain or lose energy along their
path (ISW), create lensing distortions that can be measured from the 
CMB map directly~\citep[e.g.,][]{Hu2002b}.
The interplay between weak gravitational lensing and the ISW effect
causes a non-Gaussian contribution to the CMB, which can be measured
through the lensing-induced bispectrum between small and large angular scales.
The measurement of the lensing potential requires a large number of
modes that could not be measured before the arrival of \Planck\ data.

This paper presents new measurements of the ISW effect carried out with
\Planck.  Even although our detections are not in every case as
strong as some previously claimed significance levels,
we believe that our results are an improvement over earlier studies.
This is because we can use the additional power enabled by the frequency
coverage and sensitivity of \Planck.  To establish this we carry out a
comprehensive study of all the main approaches which have previously been
taken to estimate the ISW signal.  We also present new results in
relation to the non-Gaussian structure induced by the ISW effect.

The paper is organized as follows: In Sect.~\ref{sec:data} we describe the
data used in this work (both for the CMB and large-scale structure).
The first ever results on the estimation of the lensing-induced bispectrum
are presented in Sect.~\ref{subsec:iswl}. 
Cross-correlations with external surveys are investigated in
Sect.~\ref{subsec:xcor}, and in Sect.~\ref{subsec:stack} we present the
results for the stacking analysis on the temperature maps,
as well as aperture photometry on super-clusters and super-voids.  
The recovery of the ISW all-sky map is described in Sect.~\ref{subsec:recov}.
Finally, we discuss our main results and their cosmological implications 
in Sect.~\ref{sec:discussion}.


\section{Data description}\label{sec:data}

In this Section we describe the different data sets used in this paper. This
includes \Planck\ data
\citep[the CMB temperature and lensing potential maps, see][]{planck2013-p01,
planck2013-p02,planck2013-p03,planck2013-p06,planck2013-p12}
we refer to the corresponding \Planck\ papers for details) and external
data sets (large-scale structure tracers) used in the ISW determination:
the radio NVSS catalogue; optical luminous galaxies (CMASS/LOWZ) and the
main galaxy sample from the Sloan Digital Sky Survey (SDSS); as well
as several superstructure catalogues.

\subsection{\Planck\ data}

\Planck\ data and products used in this paper are described in the following
sections, in particular the foreground-cleaned CMB maps produced by the
\Planck\ component separation pipelines, and related products, such as
dedicated component-separated frequency maps \citep{planck2013-p06}, as well
as the \Planck\ lensing map~\citep{planck2013-p12}.


\begin{figure*}
\begin{center}
\includegraphics[width=17cm]{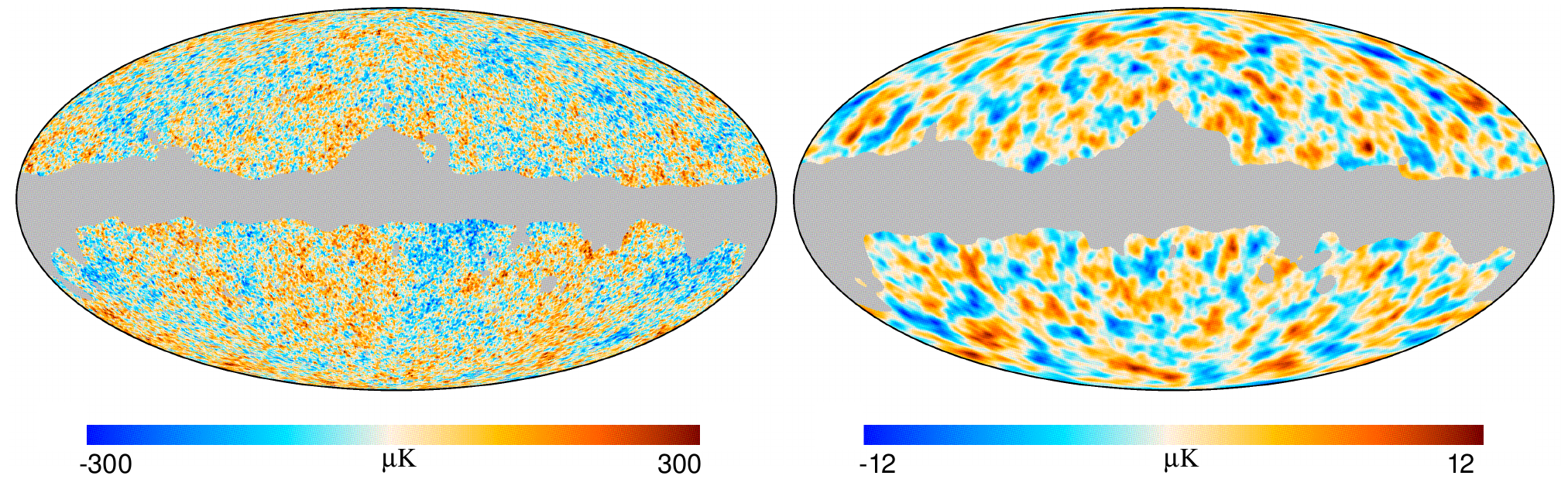}
\end{center}
\caption{\label{fig:cmb_lensing_maps}
{\it Left\/}: one of the CMB maps used in this paper, constructed using \sevem\
(given at $N_\mathrm{side}$ = 64). 
Other \Planck\ CMB maps used in this work are \cruler, \nilc\, and \smica,
in addition to clean \sevem\ maps from 44 to 353\,GHz.
{\it Right\/}: \Planck\ lensing map, optimally filtered to perform the
ISW--lensing cross-correlation (given at $N_\mathrm{side}$ = 1024). See~\cite{planck2013-p06} and~\cite{planck2013-p12}
for a detailed description of these maps.}
\end{figure*}

\subsubsection{CMB maps}
\label{subsubsec:cmbmaps}

For the present work we have made use of the \Planck\ 
foreground-cleaned CMB maps provided by the
data processing centres~\citep[as described in the \Planck\ component
separation paper][]{planck2013-p06}. In particular, to test robustness, some
of the results are presented for different cleaned CMB maps,
which were constructed using four different component separation techniques:
\cruler\ (\ruler, which uses physical parametrization);
\nilc\ (an internal linear combination technique);
\sevem\ (a template fitting method);
and \smica\ (which uses spectral matching). Since the contribution of the ISW
signal is only important on large scales, low resolution
maps, with {\tt HEALPix} \cite{Gorski2005} parameter $N_\mathrm{side} = 64$,
and pixel size of about 55\,arcmin, have been
used for most of the analyses. One exception is the study of the
correlation between the ISW and lensing signals, which requires the
use of full-resolution maps ($N_\mathrm{side}$=2048,
pixel size of $1.7$\,arcmin). The maps are
degraded directly from the original full resolution down to the
corresponding $N_\mathrm{side}$.

In addition, foreground-cleaned maps per frequency (from 44 to 353\,GHz) at
resolution $N_\mathrm{side}=512$ were used for the stacking
analysis presented in Sect.~\ref{subsec:stack}. These maps were
constructed by subtracting a linear combination of internal
templates using \sevem\ \citep[see the \sevem\ Appendix of][for a detailed description of
the method]{planck2013-p06}. 
As an example the \sevem\ CMB map is shown in Fig.~\ref{fig:cmb_lensing_maps} (left panel).

Finally, to minimize the presence of foreground contamination in the
maps, we have used the official mask described in~\cite{planck2013-p06},
which excludes regions with larger Galactic and point-source
contamination (the U73 mask). This mask is given at the full \Planck\ resolution and
is downgraded to the required levels. The downgrading
procedure consists of the following steps:
the mask (originally a map with zero and one values) is convolved with a Gaussian beam of
FWHM three times the characteristic pixel size of the final
$N_\mathrm{side}$ resolution; this convolved map is then degraded to
the required $N_\mathrm{side}$, and, afterwards, a threshold of 0.75 is
imposed (i.e., pixels with a value above this threshold are set to one, whereas
the rest are set to zero).


\subsubsection{Lensing potential map}
\label{subsub:lens}
Weak gravitational lensing distorts the CMB temperature anisotropy pattern.
This effect is sensitive to the projected matter distribution in the
large-scale structure at high redshifts, where structure growth is linear and
the statistics close to Gaussian. Weak lensing causes correlations between
different multipoles which are proportional to the lensing deflection field.
These correlations can be exploited for reconstructing the density field and
for measuring its statistical properties \citep{Hu2002b, Okamoto2003}. The
lensing effect in the CMB can be estimated by this homogeneity breaking, and
in this way individual modes of the lensing potential at multipoles
$\ell<100$ can be reconstructed with a significance of around $0.5\,\sigma$,
showing the necessity of a statistical treatment. Nevertheless the overall
effect of the lensing is measured to better than $25\,\sigma$
\citep{planck2013-p12}.
The additional lensing effect in the temperature power spectrum is
detectable with a significance of about $10\,\sigma$~\citep{planck2013-p08}.

With \Planck\ data, we aim at detecting a correlation between the ISW effect
and the lensing potential, where the latter is a tracer of the large-scale
structure at high redshift. This correlation is restricted to $9\,\sigma$,
even in the ideal case, limited by cosmic variance and the smallness of the
ISW effect in comparison to the primary CMB~\citep{Lewis2011}.
The data products used in this study are the \Planck\ lensing potential
reconstruction, and specific lensing maps obtained from the component
separation pipelines. The lensing potential is available as part of the first
\Planck\ data release. Its detailed development is described in the \Planck\
lensing paper~\citep{planck2013-p12}. In Fig.~\ref{fig:cmb_lensing_maps} we
reproduce (right panel) an optimally filtered version of the \Planck\ lensing
map, suitable for the ISW-lensing cross-correlation.

In addition to a direct correlation between the CMB sky and the reconstructed
lensing map, we measure the bispectrum generated by weak lensing by applying
a range of estimators: the KSW-bispectrum estimator; bispectra binned in
multipole intervals; and a modal decomposition of the bispectrum.
This meassurement is made possible for the first time thanks to the \Planck\
data. In addition, we will use information from the lensing field as a tracer
for an ISW map reconstruction at high redshift (see Sect.~\ref{subsec:recov}).

\subsection{External data sets}
\label{subsec:extdata}

\begin{figure}
\centering
\includegraphics[width=\hsize]{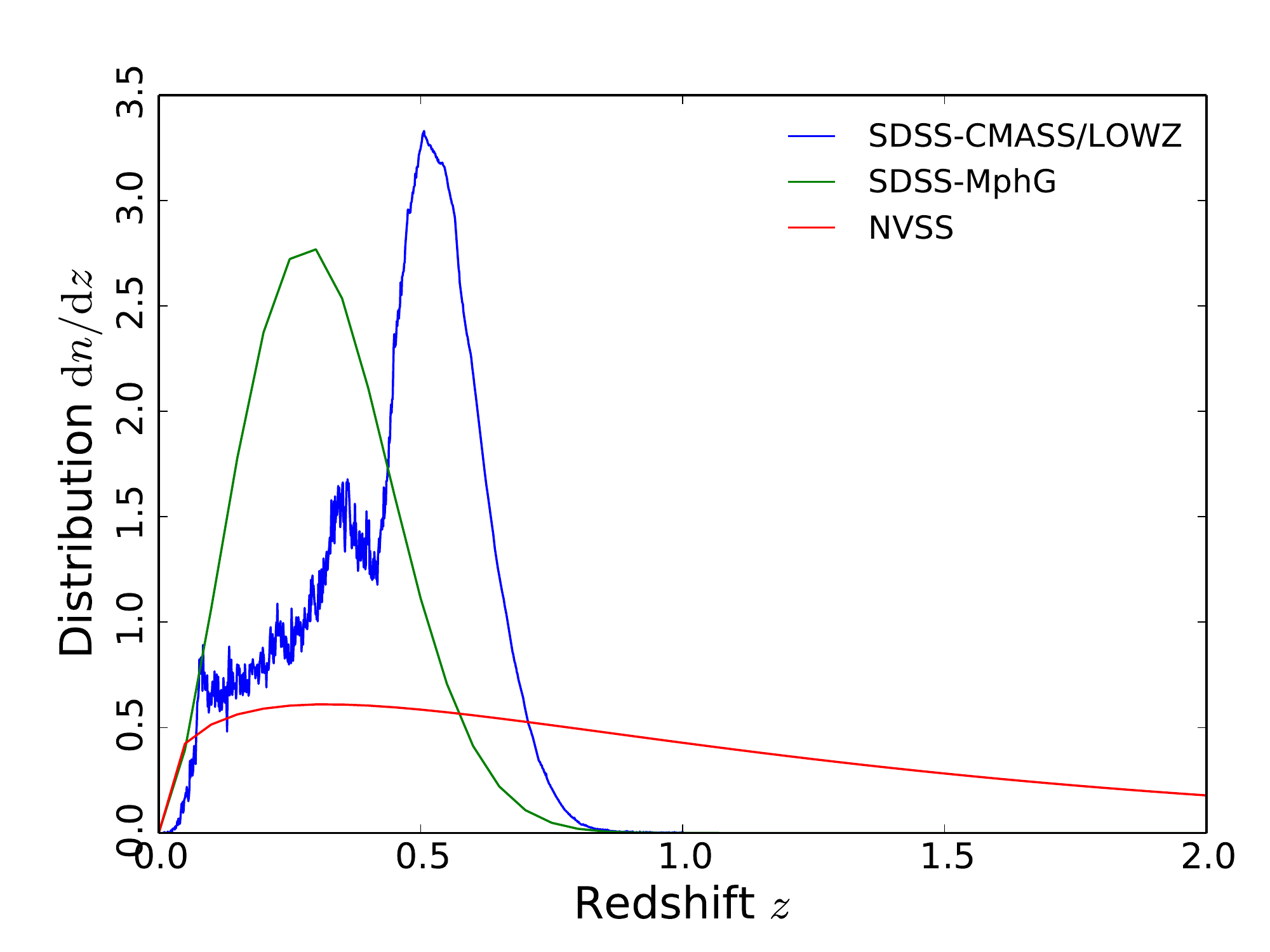}
\caption{
Redshift distributions of the different surveys used in this work as LSS
tracers, to be correlated with the \Planck\ CMB maps. For ease of comparison,
these distributions have been normalised to unity.}
\label{fig:surveys_dndz}
\end{figure}

As described in the introduction, the achromatic nature of the ISW effect
requires a tracer of the gravitational potentials from the large-scale
structure, so that by cross-correlating the CMB temperature map with that
tracer distribution the fluctuations due to the ISW effect are singled out.
The prerequisites for a tracer catalogue to be used in ISW studies are: a
large survey volume; well-understood biasing properties; and low or at least
well-modelled systematics. The radio NVSS catalogue and the optical luminous
galaxies (\lrg) and main photometric galaxy sample (\mg) catalogues
possess these
qualities. Table~\ref{tab:surveys} summarizes some basic properties of these
catalogues. In addition, the redshift distributions of these catalogues are shown
in Fig.~\ref{fig:surveys_dndz}. Notice that NVSS presents the widest redshift
coverage.  The \lrg\ sample is peaked around $z \approx 0.5$, whereas the
\mg\ sample peaks around $z \approx 0.3$. 

\begin{figure}
\centering
\includegraphics[width=\hsize]{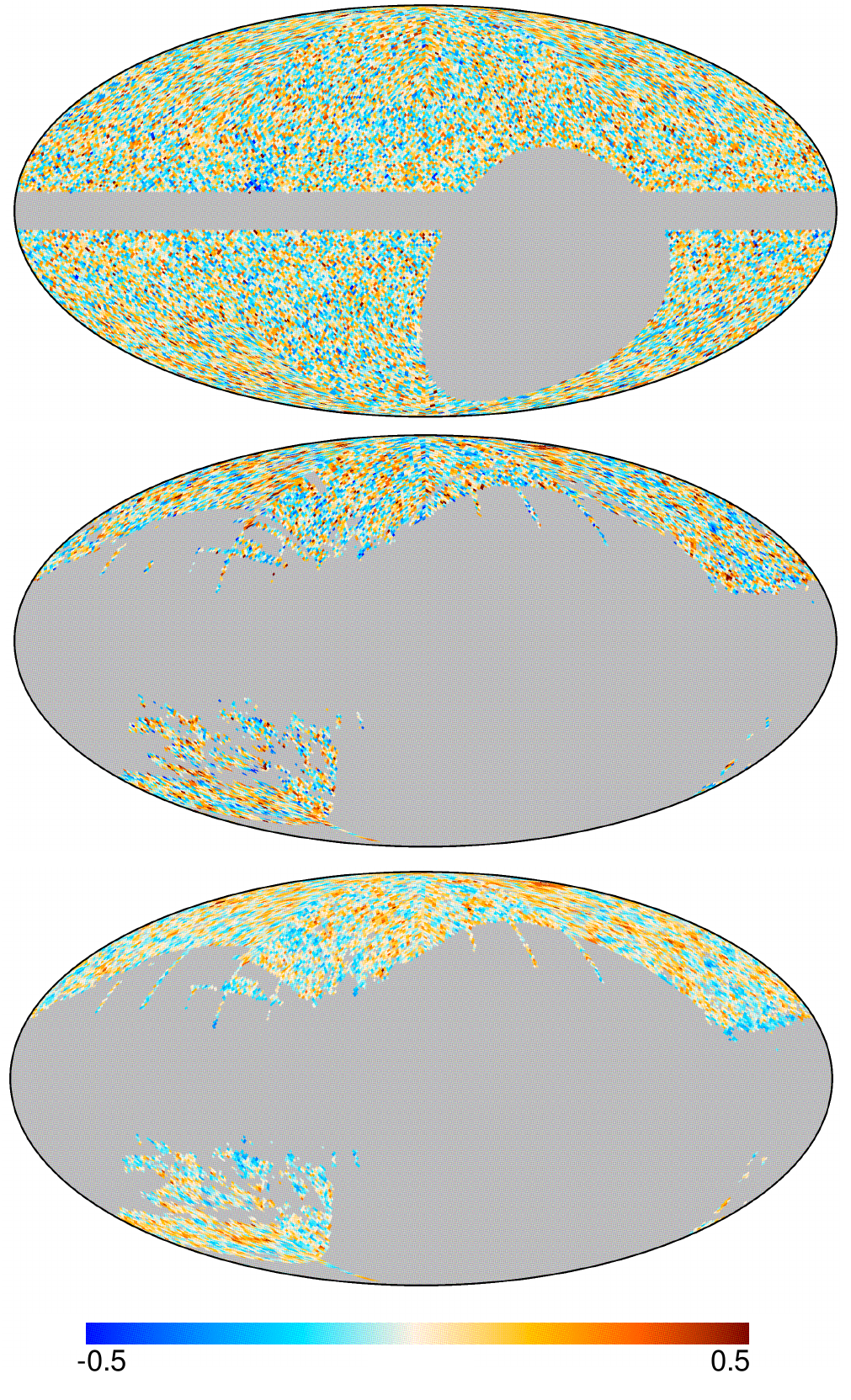}
\caption{
Density contrast maps obtained from the galaxy catalogues at
$N_\mathrm{side}$ = 64. From top to bottom: NVSS; \lrg; and \mg.}
\label{fig:surveys_maps}
\end{figure}

\begin{figure}
\centering
\includegraphics[width=1.0\hsize]{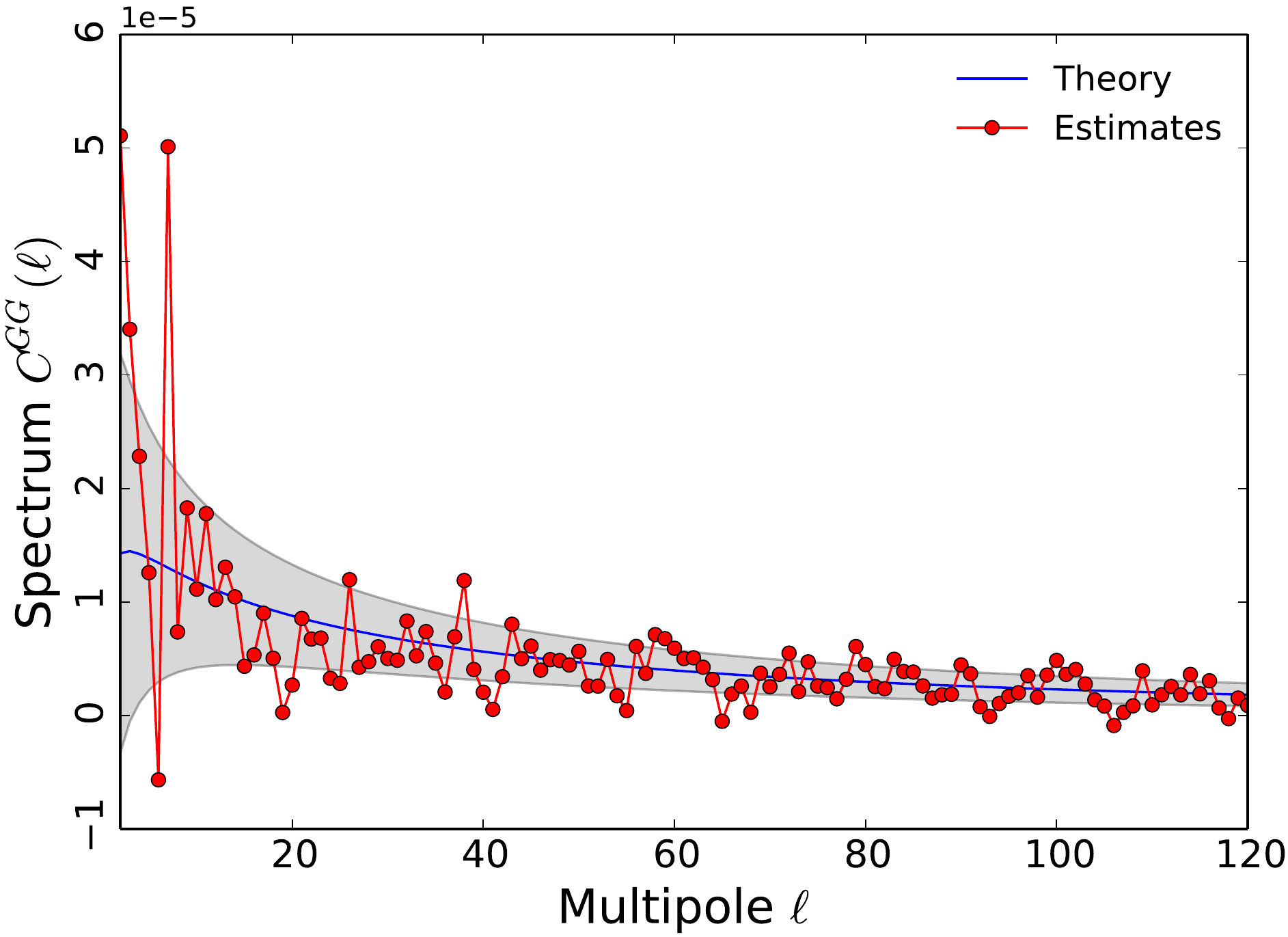}
\includegraphics[width=1.0\hsize]{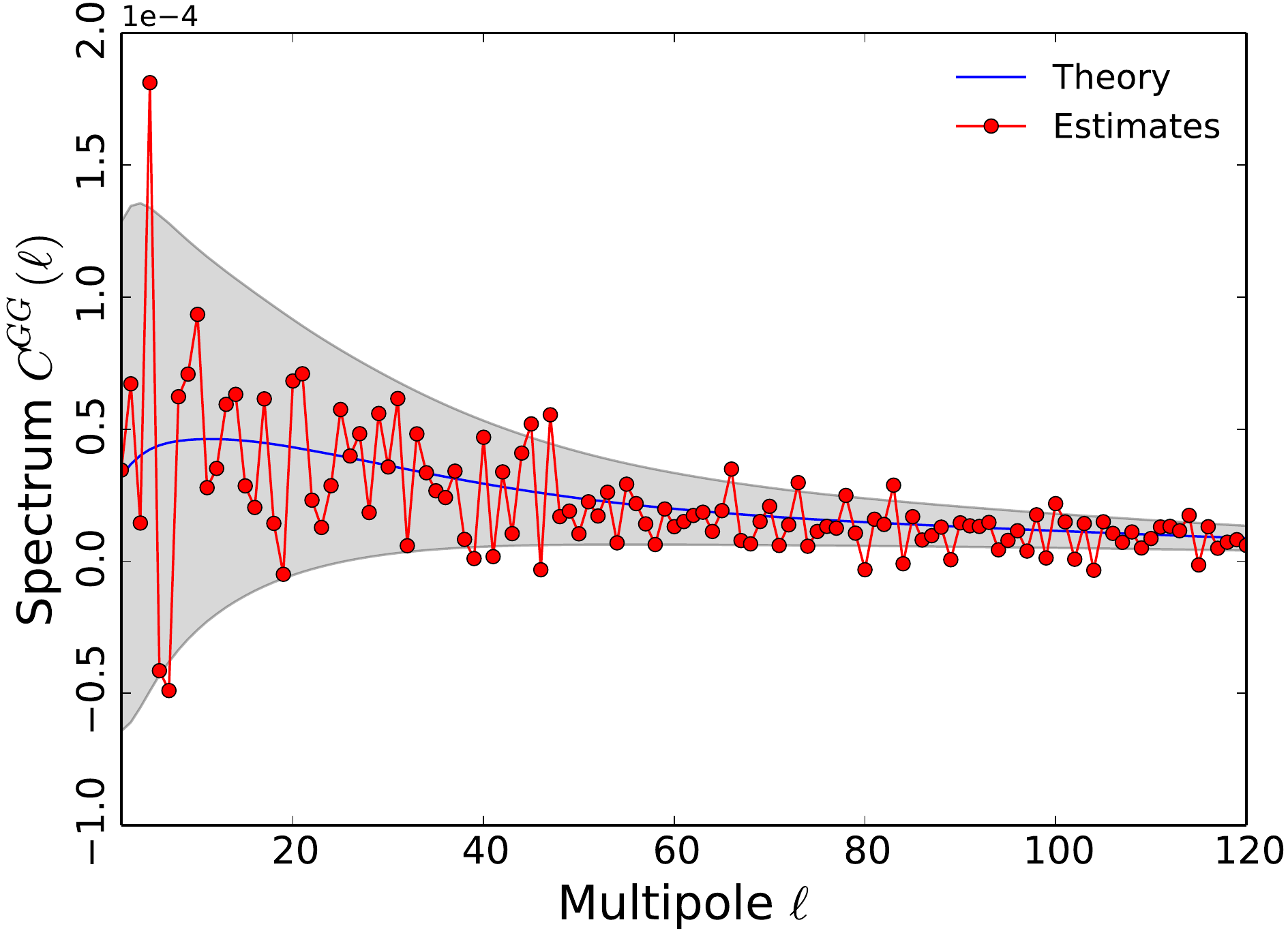}
\includegraphics[width=1.0\hsize]{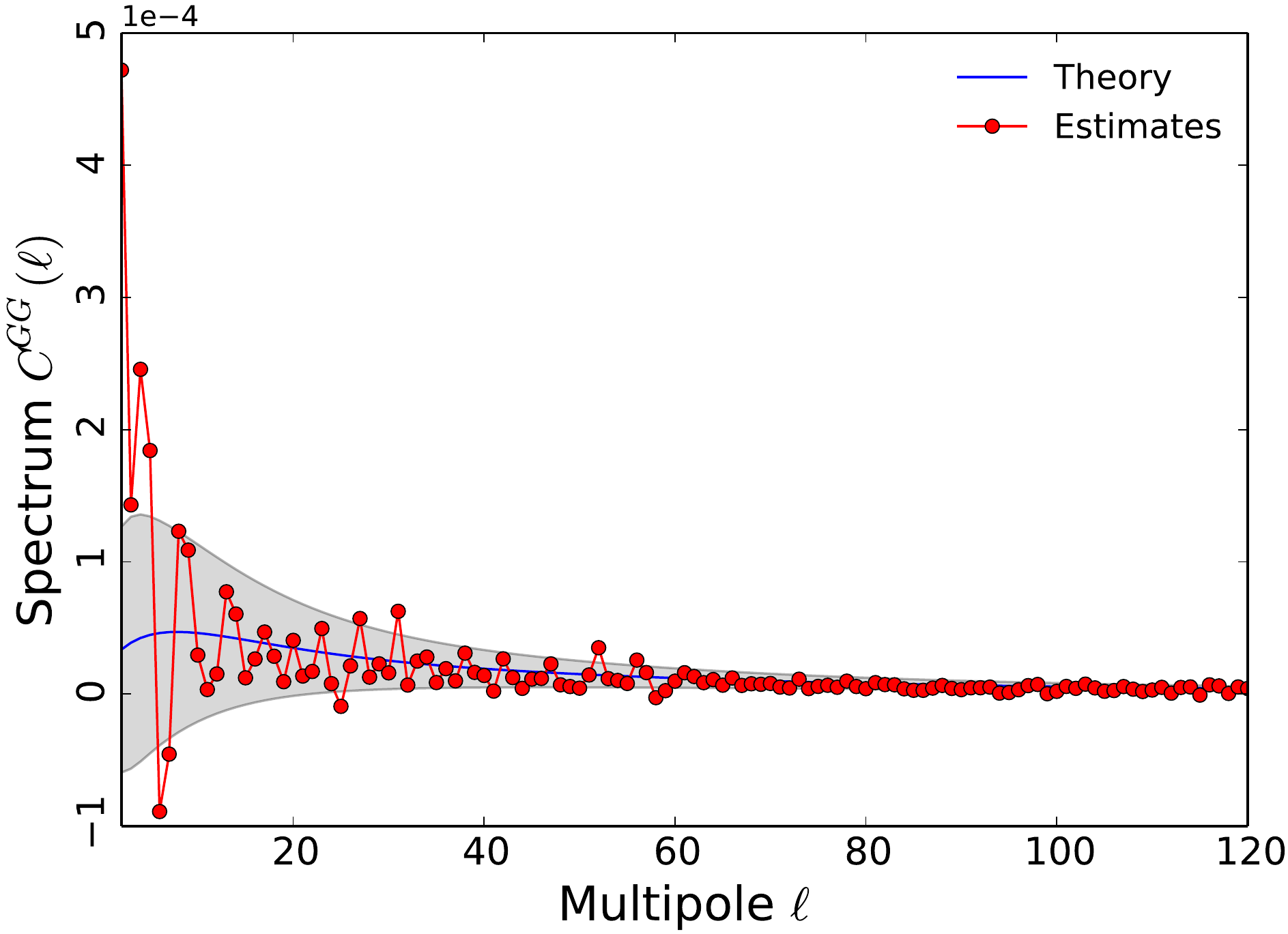}
\caption{
Angular power spectra from the maps in Fig.~\ref{fig:surveys_maps}.
From top to bottom: NVSS; \lrg; and \mg. 
The observed spectra are the red points, whereas the theoretical models are
represented by the black lines (the grey areas correspond to the sampling
variance).}
\label{fig:surveys_cls}
\end{figure}

Figure~\ref{fig:surveys_maps} shows the all-sky density projection for these
maps, where the grey area indicates regions not observed by these surveys
(or discarded for having contamination or low galaxy number density, see next
subsections for details).
In Fig.~\ref{fig:surveys_cls} we give the  angular power spectra
(blue points) of the surveys~\citep[corrected with a procedure similar to
{\tt MASTER}, e.g.,][]{Hivon2002}, as well as the theoretical spectra
(black lines) and their $1\,\sigma$ error bars (grey areas), estimated from
the {\tt MASTER} approach as well. 

\begin{table}[tmb]
\begingroup
\newdimen\tblskip \tblskip=5pt
\caption{Major characteristics of the galaxy catalogues used as tracers of
the gravitational potential. From left to right, the columns indicate: the
number of galaxies per steradian; the fraction of the sky covered by each
survey; the mean bias; and the median redshift. The bias for
NVSS is not provided, since the assumed model has a bias which depends on
redshift (see text for details).
\label{tab:surveys}}
\nointerlineskip
\vskip -3mm
\footnotesize
\setbox\tablebox=\vbox{
   \newdimen\digitwidth 
   \setbox0=\hbox{\rm 0} 
   \digitwidth=\wd0 
   \catcode`*=\active 
   \def*{\kern\digitwidth}
   \newdimen\signwidth 
   \setbox0=\hbox{+} 
   \signwidth=\wd0 
   \catcode`!=\active 
   \def!{\kern\signwidth}
\halign{\hfil#\hfil\tabskip=0.8cm& \hfil#\hfil\tabskip=0.5cm&
 \hfil#\hfil\tabskip=0.5cm& \hfil#\hfil\tabskip=0.5cm& \hfil#\hfil\tabskip=0.cm\cr 
\noalign{\doubleline}
 \noalign{\vskip -2pt}
Galaxy catalogue & $\bar{n}$& $f_{\rm sky}$ & bias & $\bar{z}$\cr 
\noalign{\vskip 3pt\hrule\vskip 5pt}
NVSS& $1.584\times10^5$& 0.73& $\cdots$& 1.17\cr
\lrg& $5.558\times10^5$& 0.22& 2.03& 0.45\cr
\mg& $9.680\times10^6$& 0.22& 1.20& 0.32\cr
\noalign{\vskip 5pt\hrule\vskip 3pt}}}
\endPlancktable                    
\endgroup
\end{table}

Besides the cross-correlation between CMB and LSS tracers
(Sect.~\ref{subsec:xcor}), we will  present results from a different
methodology in Sect.~\ref{subsec:stack}, where we use catalogues of
super-structures to study the ISW through stacking of the CMB fluctuations on
the positions of these super-structures. The relevant catalogues are
described in Sect.~\ref{sssect:catal}.

\subsubsection{NVSS radio catalogues}
\label{subsub:nvss}

Luminous active galactic nuclei (hereafter AGN) are known to be powerful
radio sources, visible out to high redshifts. These sources are hence able to
probe the cosmic density field during the entire
redshift range from matter domination to accelerated expansion due to dark
energy. If AGN are fair tracers of the underlying density field, these
sources should likewise probe the spatial distribution of the large-scale
potential wells that decay at late times after the accelerated expansion
sets in and generates the ISW effect.

We shall focus on a single radio survey, with the level of sensitivity and
sky coverage required for ISW studies, namely the NRAO VLA Sky Survey
\citep[hereafter NVSS,][]{Condon1998}. This survey was
conducted using the Very Large Array (VLA) at $1.4$\,GHz, and covers up to
an equatorial latitude of $b_{\rm E}=-40\deg$, with an average
noise level of $0.45$\,mJy\,beam$^{-1}$.  It results in
roughly $1.4\times10^6$
sources above a flux threshold of $2.5$\,mJy. Fig.~\ref{fig:surveys_maps}
displays the number density map computed from the NVSS survey (top panel).
The AGN population is known to be dominant in radio
catalogues at $1.4$\,GHz in the high flux density regime. \citet{Condon1998}
show that at this frequency, star-forming galaxies (SFGs) contribute
about 30\% of the total number of weighted source counts above
$1$\,mJy, but their presence decreases rapidly as higher flux
thresholds are adopted. The NVSS SFGs are nearby sources ($z<0.01$), and
hence may distort the ability of our radio template to probe the
intermediate and high redshift density field. 

We next address the presence of systematic effects in the NVSS
survey. Two different antenna configurations were used while
conducting the NVSS survey: the D-configuration (for $b_{\rm E}\in
[-10\deg,\ 78\deg]$), and the DnC-configuration for large zenith
angles ($b_{\rm E} < -10\deg,\ b_{\rm E} > 78\deg$).  This change in the
antenna configuration is known to introduce changes in the source number
density above $2.5$\,mJy, as first pointed out by
\citet{Blake2002}. 
The NVSS map at $2.5$\,mJy is corrected for this
declination systematic using the following procedure: the sky is divided into
equatorial strips and the mean number of sources in each strip is
re-normalised to the full sky mean~\citep[see e.g.,][]{Vielva2006}.
With this procedure
the average number of sources in the NVSS map is the same as before the
correction, and hence the shot noise level does not change. The number of
strips into which the map is divided is $70$, 
but the results are independent of this choice.

Regarding the galaxy bias,
in this work we adopt the Gaussian bias evolution model of
\citet{Xia2011}. If $n(M,z)$ is the halo mass function and $b(M,z)$ is
the bias of halos with comoving mass $M$, then the bias of the survey is
given by a mass-weighted integral,
\begin{equation}
b(z) = \frac{\int_{M_{\rm min}}^\infty \mathrm{d}M \ b(M,z) \ M \
n(M,z)}{\int_{M_{\rm min}}^\infty \mathrm{d}M \ M \ n(M,z)} \ .
\label{eq:biasxia}
\end{equation}
This model depends on the minimum mass $M_{\rm min}$ of halos present in
the survey. The upper limit in the mass is taken to be infinity
because the effect of the high mass end on the bias is negligible.
\citet{MarcosCaballero2013} proposed a theoretical
model for the NVSS angular power spectrum, which also takes into account the
information of the redshift distribution given by CENSORS data
\citep{Brookes2008}. The redshift distribution is parametrized by
\begin{equation}
\frac{\mathrm{d}n}{\mathrm{d}z} =
n_0 \left( \frac{z}{z_0} \right)^\alpha e^{-\alpha z / z_0} \ ,
\label{eqn:bias}
\end{equation}
where $z_0=0.33$ and $\alpha=0.37$. The parameter $n_0$ is a constant
in order to have a distribution normalized to unity. This function is
represented in Fig.~\ref{fig:surveys_dndz}. 
The bias follows the prescription of Eq.~\ref{eq:biasxia}, with 
$M_{\rm min}$ equal to $10^{12.67} {\rm M}_{\odot}$, where the
Sheth-Tormen~\citep{Sheth1999} mass function is adopted. 
Hereafter this model will be regarded as our fiducial model for NVSS.

\subsubsection{SDSS luminous galaxies}

For this analysis we use the photometric luminous galaxy (LG) catalogue from
the Baryonic Oscillation Spectroscopic Survey (BOSS) of the SDSS III.
The data used consist of two sub-samples: CMASS; and LOWZ.  Both samples are
combined to form a unique LG map (see Fig.~\ref{fig:surveys_maps},
second panel). Hereafter, these samples will be referred to as SDSS-CMASS,
SDSS-LOWZ, and SDSS-CMASS/LOWZ, for the combination.

\subsubsection*{SDSS-CMASS}

We use the BOSS targets chosen to have roughly constant stellar mass and known
as the photometric ``CMASS'' sample. This sample is mostly contained in the
redshift range $z=0.4$--0.7, with a galaxy number density close to
$110\,{\rm deg}^{-2}$, and is selected after applying the colour cuts
explained in \citet{Ross2011}.

While such color selection yields a catalogue of about $1{,}600{\,},000$
galaxies, further cuts needed to be applied in order to account for dust
extinction (based on the maps by \citealt{Schlegel1998} with the
criterion $E(B-V) < 0.08$), for seeing in the $r$ band (required to be
$<2.0^{\prime\prime}$) and for the presence of bright stars, similar to
\citet{Ho2012}. Finally, we neglected all pixels with a mask value inferred
from the footprint below $0.9$ on a {\tt HEALPix} map of resolution
$N_\mathrm{side}$=64. This procedure left about one million sources
$10{,}500\,{\rm deg}^2$.  Photometric redshifts of this sample are
calibrated using a selection of about 100{,}000 BOSS spectra as a training
sample for the photometric catalogue.
These LGs are among the most luminous galaxies in the Universe and
therefore allow for a good sampling of the largest scales.
Given the large number of such sources included in the sample,
shot-noise does not dominate clustering errors. According to \citet{Ross2011},
about 3.7\,\% of these objects are either stars or quasars, and this
makes further corrections necessary, as explained at the end of this section.

\subsubsection*{SDSS-LOWZ}

The photometric LOWZ sample is one of the two galaxy samples targeted by the
BOSS of Sloan III. It selectd luminous, highly biased, mostly red galaxies,
placed at an average redshift of $\langle z\rangle \sim 0.3$ and below the
redshifts of the CMASS sample ($z<0.4$). Our selection criteria in terms of
the Sloan five model magnitudes $ugriz$ follow those given in Sect.~2 of
\citet{Parejkoetal2012}.
With a total number of sources close to 600{,}000, this photometric sample
contains a higher number density of galaxies in the southern part of the
footprint than in the northern one (by more than 3\,\%), which seems to be
at odds with $\Lambda$CDM predictions. However, most of this effect vanishes
when we subtract the dipole in the effective area under analysis, in such a
way that the low $\ell$ range of the auto power spectrum is consistent
with a $\Lambda$CDM model and a constant bias $b\simeq 2$
\citep{Hernandez2013}.

Both SDSS-CMASS and SDSS-LOWZ samples are further corrected for any
scaling introduced by possible systematics like stars, mask value, seeing,
sky emission, airmass and dust extinction. Following exactly the same
procedure as in \citet{Hernandez2013}, we find that both LG samples are
contaminated by stars, in the sense that the galaxy number density decreases
in areas with higher star density, since the latter tend to ``blind''
galaxy detection algorithms.

\subsubsection{Main photometric SDSS galaxy sample}

We use a sample of photometrically-selected galaxies from the SDSS-DR8
catalogue, which covers a 
total sky area of $14{,}555\,{\rm deg}^2$ \citep{Aihara2011}.
The total number of objects labelled as 
galaxies in this data release is 208 million. From this catalogue, and
following~\cite{Cabre2006}, we 
define a subsample by selecting only objects within the range $18<r<21$,
where this $r$-band model magnitude corrected for extinction.
Following~\cite{Giannantonio2008b}, we also restrict our subsample to objects
with redshifts $0.1 < z < 0.9$, and with measured 
redshift errors such that $\sigma_z < 0.5z$. We rely on the photometric
redshift estimates of the SDSS photo-$z$ primary galaxy table, which
have been obtained through a ``kd-tree'' nearest neighbour technique, 
by fitting the spectroscopic objects observed with similar colour and
inclination angle.  The total number of galaxies in our final sample is 
about 42 million, with redshifts distributed around a median value of
around 0.35, as shown in Fig.~\ref{fig:surveys_dndz}. To avoid possible
errors introduced by singularities in the photometric redshifts estimates,
instead of using the real observed redshift distribution in our analysis 
we resort to the analytical function
\begin{equation}
\frac{dn}{dz}=\frac{\beta}{\Gamma\left(\frac{m+1}{\beta}\right)}\frac{z^m}{z_0^{m+1}}e^{-(z/z_0)^\beta}~,
\label{eq:dndz_xia}
\end{equation}
which is fitted to the data, with parameters $m=1.5$, $\beta=2.3$ and
$z_0=0.34$, which are identical to those found 
by~\cite{Giannantonio2012}. For the galactic bias we use the value
$b=1.2$, which is found by~\cite{Giannantonio2012} by fitting the
$\Lambda$CDM prediction to the observed auto-correlation 
function of the galaxies, and we adopt their proposed mask.

\subsubsection{SDSS, super-structures}\label{sssect:catal}

\citet{Granett2008b} produced a sample\footnote{Available at
\url{http://ifa.hawaii.edu/cosmowave/supervoids/}.} of 50 superclusters and
50 supervoids identified from the Luminous Red Galaxies (LRGs) in the SDSS
(sixth data release, DR6, \citealt{Adelman2008}) that covers an area of
$7500\,{\rm deg}^2$ on the sky. They used publicly available algorithms,
based on the Voronoi tessellation, to find 
2836 superclusters \citep[using VOBOZ, VOronoi BOund Zones,][]{Neyrinck2005}
and
631 supervoids \citep[using ZOBOV, ZOnes Bordering On Voidness,][]{Neyrinck2008}
above a 2$\,\sigma$ significance level (defined as the probability of
obtaining, in a uniform Poisson point sample, the same density contrasts as
those of clusters and voids). 
The 50 superclusters and 50 supervoids they
published in their catalogue correspond to density contrasts of
about $3\,\sigma$ and $3.3\,\sigma$ respectively. They span a redshift range
of $0.4<z<0.75$, with a median of around 0.5, and inhabit a volume of about
5\,$h^{-3}$\,Gpc$^3$. These superstructures can potentially produce measurable
ISW signals, as suggested in \cite{Granett2008a, Granett2008b}. For each
structure, the catalogue provides: the position on the sky of the centre;
the mean and maximum angular distance between the galaxies in the structure
and its centre; the physical volume; and three different measures of the
density contrast (calculated from all its Voronoi cells, from only its
over- or under-dense cells, and from only its most over- under-dense cell).
For the present paper, we concentrate on using the supervoid catalogue by
\citet{Granett2008b}, as they can be compared with two other, more recent
catalogues of voids.

The second catalogue of cosmic voids that we consider here is published by
\cite{Pan2012}\footnote{Available at \url{http://www.physics.drexel.edu/~pan/}.}.
It has been built from the seventh data release (DR7) of the SDSS. Using the
{\tt VoidFinder} algorithm \citep{Hoyle2002}, they identified 1055 voids with
redshifts smaller than $z = 0.1$. Each void is listed with its position on
the sky, its physical radius (defined as the radius of the maximal sphere
enclosing the void), an effective radius defined as the radius of a sphere
of the same volume, its physical distance to us, its volume and mean density
contrast. The filling factor of the voids in the sample volume is 62\,\%.
The largest void is just over 47\,Mpc in effective radius, while the median
effective radius of the void sample is roughly 25\,Mpc. Some of the voids
are both very close to us and relatively large (larger than 30\,Mpc in
radius), which results in large angular sizes of up to $15\deg$.

The third void catalogue that we use has been released by \cite{Sutter2012}
and also made publicly available.\footnote{Available at
\url{http://www.cosmicvoids.net}.} Note that it is being updated regularly,
and the results reported here are based on the 21 February 2013 version of
the catalogue. Using their own modified version of {\tt ZOBOV}, these authors
identified 1495 voids distributed across the 0--0.44 redshift range.
They subdivided their catalogue into six subsamples: \textit{dim1},
\textit{dim2}, \textit{bright1} and \textit{bright2}, constructed from the
main SDSS; and \textit{lrgdim}, \textit{lrgbright} built from the SDSS LRG
sample. For each void, the information provided includes the position of the
centre, the redshift, the volume, the effective radius, and the density
contrast.

\section{ISW-lensing bispectrum}
\label{subsec:iswl}
There is an interesting interplay between gravitational lensing of the CMB
and the ISW effect, which manifests itself as a non-Gaussian feature.
CMB-lensing can be described by a convolution of the CMB-temperature map
$T$ with the weak lensing potential $\phi$ \citep[e.g.,][]{Lewis2006},
\begin{equation}
T(\mathbf{\ell}) \rightarrow 
T(\mathbf{\ell}) - 
\int\frac{\mathrm{d}^2 \mathbf{\ell}^\prime}{2\pi}\:
\mathbf{\ell}^\prime(\mathbf{\ell}-\mathbf{\ell}^\prime)\:
\phi(\mathbf{\ell}-\mathbf{\ell}^\prime)T(\mathbf{\ell}^\prime).
\end{equation}
The CMB lensing can be measured by a direct estimate of the CMB bispectrum,
because the bispectrum acquires first order terms proportional to the product
of two power spectra $\tilde{C}^{\rm TT}_\ell C^{{\rm T}\phi}_\ell$,
where $\tilde{C}^{\rm TT}_\ell$ is the lensed temperature power spectrum
and $C^{{\rm T}\phi}_\ell$ is the temperature-potential cross-power spectrum.
The potential field $\phi$ and the temperature field $T$ are correlated,
because $\phi$, which deflects the CMB photons, also gives rise to the
ISW effect in $T$ \citep{Hu2002b, Seljak1999, Verde2002,Giovi2003}. This
secondary bispectrum contains new information about the cosmological redshift,
because it is generated mainly at redshifts larger than unity, and biases
measurements of the primordial bispectrum. The term $C^{{\rm T}\phi}_\ell$
correlates the CMB temperature on small scales with the lensing potential on
large scales, and causes the bispectrum to assume large amplitudes in the
squeezed triangles configuration \citep[see e.g.,][]{Goldberg1999, Seljak1999,
Hu2000, Giovi2003, Okamoto2003, Giovi2005a, Lewis2006, Serra2008,
Mangilli2009a, Hanson2009a, Hanson2010, Smith2011, Lewis2011}. 

Due to the rotational invariance of the sky, the CMB angular bispectrum
$\langle a_{\ell_1m_1}a_{\ell_2m_2}a_{\ell_3m_3}\rangle$ can be factorized
as follows:
\begin{equation}
\langle a_{\ell_1m_1}a_{\ell_2m_2}a_{\ell_3m_3}\rangle
 = {\cal G}_{\ell_1\ell_2\ell_3}^{m_1m_2m_3}       
 b_{\ell_1\ell_2\ell_3},
\label{the_bispectrum_estimator}
\end{equation}
where ${\cal G}_{\ell_1\ell_2\ell_3}^{m_1m_2m_3}\equiv
 \int \mathrm{d}^\Omega\ Y_{\ell_1}^{m_1}(\hatn)Y_{\ell_2}^{m_2}
 (\hatn)Y_{l_3}^{m_3}(\hatn)$ is the Gaunt-integral and
$b_{\ell_1\ell_2\ell_3}$ is the so-called reduced bispectrum. 
In the case where the bispectral signal on the sky is due to the ISW-lensing 
effect, $b_{\ell_1\ell_2\ell_3} =
 A^{{\rm T} \phi} b_{\ell_1\ell_2\ell_3}^{\rm lens-ISW}$, where 
$A^{{\rm T} \phi}$ parametrizes the amplitude of the effect and 
\begin{eqnarray}
 b_{\ell_1 \ell_2 \ell_3}^{\rm lens-ISW}
 & = &  
\Big \{
\frac{\ell_1 (\ell_1 + 1) - \ell_2 (\ell_2 + 1) + \ell_3 (\ell_3 +
  1)}{2} C_{\ell_1}^{{\rm T}\phi} \tilde{C}_{\ell_3}^{\rm TT}  \\
 & &  
 + (5~\mathrm{permutations}) \Big \}. \nonumber
\label{thebl1l2l3}
\end{eqnarray}
A more general expression for intensity and polarization can be found in
\citet{Lewis2011}. Estimation of the bispectrum then yields a measurement
of $A^{{\rm T}\phi}$. 

We can also define an alternative rotationally-invariant reduced bispectrum 
$B_{\ell_1\ell_2\ell_3}$ as 
$B_{\ell _1 \ell _2 \ell _3} = h^2_{\ell _1 \ell _2 \ell _3}
 b_{\ell _1 \ell _2 \ell _3}$,
where
\begin{eqnarray}
 h^2_{\ell_1\ell_2\ell_3}&\equiv&\sum_{m_1 m_2 m_3}\left({\cal G}_{\ell_1\ell_2\ell_3}^{m_1m_2m_3}\right)^2 \nonumber\\
 &&=\frac{(2\ell_1 +1)(2\ell_2 +1)(2\ell_3 +1)}{4\pi} \left(\begin{array}{ccc} \ell_1 & \ell_2 & \ell_3 \\ 0 & 0 & 0 \end{array}\right)^2.
\end{eqnarray}
The interest in $B_{\ell_1\ell_2\ell_3}$ is that it can be directly
estimated from the observed map using the expression
\begin{equation}
B^\mathrm{obs}_{\ell _1 \ell _2 \ell _3} = \int \mathrm{d}\Omega T_{\ell_1} (\Omega) T_{\ell_2}(\Omega) T_{\ell_3} (\Omega), \label{integral_Bobs}
\end{equation}
where the filtered maps $T_\ell(\Omega)$ are defined as
\begin{equation}
T_{\ell}(\Omega) \equiv \sum_m a_{\ell m} Y_{\ell m}(\Omega). \label{def_filtmaps_Tl}
\end{equation}

By basically combining the single-$\ell$ estimates $B^\mathrm{obs}/B^\mathrm{lens-ISW}$ for $A^{{\rm T}\phi}$ using inverse variance weighting, the ISW-lensing bispectrum estimator can be written as  
\citep[see][for more details]{planck2013-p09a}
\begin{equation}
{\hat A}^{{\rm T}\phi} = \frac{\langle B^{\rm lens-ISW},
 (B^\mathrm{obs} - B^\mathrm{lin}) \rangle}
 {\langle B^{\rm lens-ISW}, B^{\rm lens-ISW} \rangle}, \label{fNL_estim_optim}
\end{equation}
where the inner product is defined by 
\begin{equation}
\langle B^i, B^j \rangle \equiv
\sum_{\ell _1 \leq \ell _2 \leq \ell_3} 
\frac{ 
B^i_{\ell _1 \ell _2 \ell _3} 
B^j_{\ell _1 \ell _2 \ell _3} 
}{ 
V_{\ell _1 \ell _2 \ell _3}
}. \label{Binnerprod}
\end{equation}
Here,  $B^\mathrm{lin}$ is a linear 
correction that has zero average but reduces the variance of the estimator
in the presence of anisotropic noise and a mask. 
Furthermore, $V_{\ell _1 \ell _2 \ell _3} = g_{\ell _1 \ell _2 \ell _3} 
h^2_{\ell _1 \ell _2 \ell _3} C_{\ell_1} C_{\ell_2} C_{\ell_3}$,
with $g$ being a simple permutation factor ($g=6$ when all $\ell$ are equal,
$g=2$ when two $\ell$ are equal and $g=1$ otherwise).  
As in all expressions in this section,
we have implicitly taken the beam 
and noise of the experiment into accout, e.g., $C_\ell$ should actually be 
$b_\ell^2 C_\ell + N_\ell$ with $b_\ell$ the beam transfer function
and $N_\ell$ the noise power spectrum. 

In Eq.~\ref{fNL_estim_optim} we have also used the fact that, as discussed in 
detail in \citet{planck2013-p09a}, the full inverse covariance weighting can be
replaced by a diagonal covariance term, 
$(C^{-1} a)_{\ell m} \rightarrow {a_{\ell m}/C_\ell}$, without loss of
optimality, if the masked regions of the map
are filled in with a simple diffusive inpainting scheme.

The normalization of the lensing-ISW estimator in the denominator of
Eq.~\ref{fNL_estim_optim} can be replaced by \citep[see e.g.][]{Lewis2011}
\begin{equation}
F = \sum_{\ell} \bigg( F_{\ell}^{-1} + \frac{1+r_{\ell}^{-2}}{2 \ell + 1}
 \bigg)^{-1},
\label{iswlensing_fisher_T}
\end{equation}
where $r_{\ell} \equiv C_{\ell}^{{\rm T}\phi}/
 \sqrt{\tilde{C}^{\rm TT}_{\ell}\:C^{\phi\phi}_{\ell}}$ parameterizes
the deviation from the Cauchy-Schwarz relation and $F_{\ell}$ is given in
terms of the ISW-lensing bispectrum \citep[see for example][]{Lewis2011}.
The first term in Eq.~\ref{iswlensing_fisher_T} corresponds to the Fisher
errors assuming Gaussian $a_{\ell m}$. However, contrary to the null
hypothesis that is assumed, for example, in the primordial bispectra
(Gaussianity), there is an actual non-Gaussian signal already present in
the ISW-lensing bispectrum. This guarantees a larger variance for the
estimators than are included in the additional terms present in the previous
equations.

An important issue is the impact of the ISW-lensing bispectrum on estimates
of the primordial non-Gaussianity. Assuming weak levels of non-Gaussianity
and considering both the primordial bispectrum
$B^{\rm prim}_{\ell_1 \ell_2 \ell_3}$ and the ISW-lensing bispectrum
$B^{\rm lens-ISW}_{\ell_1 \ell_2 \ell_3}$, one can compute the expected bias $\Delta$ induced in the primordial bispectrum using the formula:
\begin{equation}
\Delta^{\rm prim} = 
\frac{\langle B^\mathrm{lens-ISW}, B^\mathrm{\rm prim} \rangle}
{\langle B^\mathrm{\rm prim}, B^\mathrm{\rm prim} \rangle}
\label{fnl_bias_iswl}
\end{equation}
with the inner product defined in Eq.~\ref{Binnerprod}.
Predictions of this bias on the primordial $f_\mathrm{NL}$ for \Planck\
resolution can be seen for example in \citet{Hanson2009a},
\citet{Mangilli2009a}, \citet{Smith2011}, and \citet{Lewis2011}. The most
important bias is introduced to the local shape and, considering
$\ell_\mathrm{max}\sim 2000$, is expected to be
$\Delta^{\rm local} \sim 7$ \citep{planck2013-p09a}.


\subsection{ISW-lensing estimators}
\label{subsub:iswl_est}

There are several implementations of the optimal estimator given in Eq. \ref{fNL_estim_optim}. For
their 
detailed description in the context of \Planck\ see \citet{planck2013-p09a,planck2013-p12}. 
We have applied four of these implementations to \Planck\ data in order to constrain the ISW-lensing bispectrum. Three of them represent a direct bispectrum estimation: the KSW estimator \citep{Komatsu2005, Creminelli2006}, the binned bispectrum \citep{Bucher2010}, and the modal decomposition \citep{Fergusson2010}.The remaining approach is based on a previous estimation of the gravitational lensing potential field \cite{Lewis2011}. 
These estimators differ in the implementation and approximations that are used in order to compute the expression given in Eq.~\ref{fNL_estim_optim}, the direct computation of which is out of reach of current computing facilities.
They will be reviewed in the next subsections.

\subsubsection{Lensing potential reconstruction}

\label{subsubsec:theory_lensing}

The estimator given in Eq.~\ref{the_bispectrum_estimator} can be written in
terms of the lensing potential amplitude reconstruction $\hat{\phi}$ as
\begin{equation}
\hat{A}^{{\rm T}\phi} \equiv \hat{S} = \frac{1}{N}
 \sum_{\ell m}C_{\ell}^{{\rm T}\phi}
 \frac{\tilde{T}_{\ell m}}{\tilde{C}^{\rm TT}_{\ell}}
 \frac{\hat{\phi}^*_{\ell m}}{N_{\ell}^{\phi\phi}},
\label{eqn:cltp_estimator}
\end{equation}
where $\hat{\phi}^*_{\ell m}$ can be estimated using a quadratic estimator
\citep{Okamoto2003} and $N_{\ell}^{(0)}$ is given in terms of the
ISW-lensing bispectrum \citep{Lewis2011}. Therefore, this estimator
quantifes the amount of cross-correlation between the temperature map and
the reconstruction of the lensing signal, and most of the correlation is
found at multipoles below 100.

\subsubsection{KSW-estimator}
\label{subsubsec:ksw_ext}

The KSW bispectrum estimator~\citep{Komatsu2003} for the
ISW-lensing signal can be written as
\begin{equation}
\label{eq:KSW}
\hat{A}^{{\rm T}\phi}=(F^{-1}) \hat{S},
\end{equation}
where $ \hat{S}$ can be computed from data as
\begin{eqnarray}
 \hat{S} &\equiv &
\frac16\sum_{\ell_1 m_1}\sum_{\ell_2 m_2}\sum_{\ell_3 m_3}
{\cal G}_{\ell_1\ell_2\ell_3}^{m_1m_2m_3}b_{\ell_1\ell_2\ell_3}^{\rm lens-ISW}
 \times \\
& & \big[ (C^{-1}a)_{\ell_1m_1}(C^{-1}a)_{\ell_2m_2}(C^{-1}a)_{\ell_3m_3}
 - \nonumber \\
& & 3(C^{-1})_{\ell_1m_1,\ell_2m_2}(C^{-1}a)_{\ell_3m_3} \big].\nonumber
\end{eqnarray}
and $(F^{-1})$ is the inverse of the ISW-lensing Fisher matrix
$F$ of Eq.~\ref{iswlensing_fisher_T}.
Details on the implementation of the KSW estimator for the ISW-lensing
signal can be found in~\citet{Mangilli2013}.
In particular, Eq.~\ref{eq:KSW} takes the form
\begin{equation}
\hat{A}^{{\rm T}\phi}=(F^{-1}) (\hat{S}_{\rm cubic}+\hat{S}_{\rm linear}),
\end{equation}
where $\hat{S}_{\rm cubic}$ is the term that extracts the amplitude
information from the data contained in the bispectrum, while
$\hat{S}_{\rm linear}$ 
is a zero-mean term that reduces estimator variance when the experimental
setup breaks rotational invariance, i.e., in the presence of sky cut and
anisotropic noise.
To estimate $\hat{A}^{{\rm T}\phi}$ we used the KSW estimator with an
implementation of the linear term truncated at $\ell_{\rm max}$ as described 
in~\cite{Munshi2009} and~\cite{planck2013-p09a}.

\subsubsection{Binned bispectrum}
\label{subsubsec:binned_ext}

The binned bispectrum estimator \citep{Bucher2010} achieves the required
computational reduction in determining $A^{{\rm T}\phi}$ by binning 
Eq.~\ref{fNL_estim_optim}. In particular, the maximally filtered 
maps in Eq.~\ref{def_filtmaps_Tl} are replaced by 
\begin{equation}
T_i(\Omega) = \sum_{\ell\in\Delta_i} \sum_{m=-\ell}^{+\ell}
a_{\ell m} Y_{\ell m}(\Omega),
\end{equation}
where the $\Delta_i$ are suitably chosen intervals (bins) of multipole values
(chosen in such a way as to minimize the variance of the quantities to
be estimated).
These maps are then used in Eq.~\ref{integral_Bobs} to obtain the binned
observed bispectrum, and analogously for $B^\mathrm{lin}$. The bispectrum
template $B^\mathrm{lens-ISW}$ and inverse-variance weights $V$ are also binned
by summing them over all $\ell$ values in the bin. Finally these binned
quantities are inserted in the general expression for $A^{{\rm T}\phi}$
(Eq.~\ref{fNL_estim_optim}), with the sum over $\ell$ replaced by a sum over 
bin indices $i$. Since most bispectrum shapes change rather slowly
(with features on the scale of the acoustic peaks, like the power spectrum),
the binned estimator works very well, increasing the variance only
slightly, while achieving an enormous computational reduction (from
about 2000 multipoles in each of the three directions to only about 50
bins).


\subsubsection{Modal bispectra}
\label{subsubsec:theory_modal}
Modal decomposition of bispectra has been introduced by \citet{Fergusson2010}
as a way to compute reduced bispectra that uses a diagonalization ansatz such
that the shape function in Fourier space can be separated, which reduces the
dimensionality of the integration. At the same time it greatly reduces the
complexity of estimating bispectra from data. The separation of the
bispectrum shape function into coefficients
$q_p^\ell(x)$ allows the derivation of a filtered map $M_p(\hat{n},x)$,
\begin{equation}
M_p(\hat{n},x) = \sum_{\ell m}
 \frac{q_p^\ell(x)a_{\ell m}}{C_\ell}Y_{\ell m}(\hat{n}),
\end{equation}
from the coefficients $a_{\ell m}$ of the temperature map. With that
expression, one can obtain a mode expansion coefficient $\beta$,
\begin{equation}
\beta_{prs} = \int\mathrm{d}\Omega\:\int x^2\mathrm{d} x\:
 M_p(\hat{n},x)M_r(\hat{n},x)M_s(\hat{n},x).
\end{equation}
With that decomposition, the estimator of the bispectrum assumes a
particularly simple diagonal shape,
\begin{equation}
\hat{S} = \frac{6}{N}\Delta_\Phi^2\sum_{prs}\alpha_{prs}\beta_{prs},
\end{equation}
where the $\alpha_{prs}$ are the equivalent coefficients obtained by
performing the modal decomposition of the theoretical bispectrum shape
function. The relation between modal bispectra and wavelet bispectra is
derived by \citet{Regan2013}.


\subsection{Results}
\label{subsubsec:iswl_res}
\begin{table*}[tmb]
\begingroup
\newdimen\tblskip \tblskip=5pt
\caption{Amplitudes $A^{{\rm T}\phi}$, errors $\sigma_{A}$ and significance
levels (SNR) of the non-Gaussianity due to the ISW effect, for all component
separation algorithms (\ruler, \nilc, \sevem, and \smica) and all the
estimators (potential reconstruction, KSW, binned, and modal). For the
potential reconstruction case, an additional minimum variance (MV) map has
been considered (see \citealt{planck2013-p12} for details).
\label{table:isw-lensing}}
\nointerlineskip
\vskip -3mm
\footnotesize
\setbox\tablebox=\vbox{
   \newdimen\digitwidth 
   \setbox0=\hbox{\rm 0} 
   \digitwidth=\wd0 
   \catcode`*=\active 
   \def*{\kern\digitwidth}
   \newdimen\signwidth 
   \setbox0=\hbox{+} 
   \signwidth=\wd0 
   \catcode`!=\active 
   \def!{\kern\signwidth}
\halign{#\hfil\tabskip=0.4cm & 
 \hfil#\hfil\tabskip=0.4cm& \hfil#\hfil\tabskip=0.4cm& \hfil#\hfil\tabskip=0.4cm& \hfil#\hfil\tabskip=0.4cm& \hfil#\hfil\tabskip=0.4cm& \hfil#\hfil\tabskip=0.4cm& \hfil#\hfil\tabskip=0.4cm& \hfil#\hfil\tabskip=0.4cm& \hfil#\hfil\tabskip=0.4cm& \hfil#\hfil\tabskip=0cm\cr  
\noalign{\doubleline}
\noalign{\vskip -2pt}
Estimator & {\ruler}& & {\nilc}& & {\sevem}& & {\smica}& & MV\cr  
\noalign{\vskip 2pt\hrule\vskip 3pt} 
& $A^{{\rm T}\phi}$ & SNR & $A^{{\rm T}\phi}$ & SNR & $A^{{\rm T}\phi}$ & SNR & $A^{{\rm T}\phi}$ & SNR & $A^{{\rm T}\phi}$ & SNR\cr
\noalign{\vskip 3pt\hrule\vskip 5pt}                                    
${\rm T}\phi \quad \ell \geq 10$ & 0.52 $\pm$ 0.33& 1.5& 0.72 $\pm$ 0.30& 2.4& 0.58 $\pm$ 0.31& 1.9& 0.68 $\pm$ 0.30& 2.3& 0.78 $\pm$ 0.32& 2.4\cr
$\rm{T }\phi \quad \ell \geq 2$ & 0.52 $\pm$ 0.32& 1.6& 0.75 $\pm$ 0.28& 2.7& 0.62 $\pm$ 0.29& 2.1& 0.70 $\pm$ 0.28& 2.5& & \cr
\noalign{\vskip 3pt\hrule\vskip 5pt}KSW & 0.75 $\pm$ 0.32& 2.3& 0.85 $\pm$ 0.32& 2.7& 0.68 $\pm$ 0.32& 2.1& 0.81 $\pm$ 0.31& 2.6& & \cr
\noalign{\vskip 3pt\hrule\vskip 5pt}Binned & 0.80 $\pm$ 0.40& 2.0& 1.03 $\pm$ 0.37& 2.8& 0.83 $\pm$ 0.39& 2.1& 0.91 $\pm$ 0.37& 2.5& & \cr
\noalign{\vskip 3pt\hrule\vskip 5pt}Modal & 0.68 $\pm$ 0.39& 1.7& 0.93 $\pm$ 0.37& 2.5& 0.60 $\pm$ 0.37& 1.6& 0.77 $\pm$ 0.37& 2.1& & \cr
\noalign{\vskip 3pt\hrule\vskip 3pt}}}
\endPlancktablewide                 
\endgroup
\end{table*}                        

The detection of the ISW effect via the non-Gaussian signal induced by the
gravitational lensing potential is summarized in Table~\ref{table:isw-lensing}.
We provide the estimates of the ISW-lensing amplitude $A^{{\rm T}\phi}$, its
uncertainty $\sigma_{A}$ and the signal-to-noise obtained with the different
estimator pipelines  described in Sec.~\ref{subsub:iswl_est}. The estimators
have been applied to the official \Planck\ CMB maps made using
\ruler, \nilc, \sevem, and \smica~\citep{planck2013-p06}.
The quantity $\sigma_{A}$ is obtained from 200 simulations representative of
the analysed CMB data maps. These Monte Carlo
simulations~\citep[FFP-6, see][]{planck2013-p01} account for the expected
non-Gaussian ISW-lensing signal, according to the \Planck\ best-fit model,
and have been passed through the different component separation pipelines, as
described in~\cite{planck2013-p06}. 
lensed simulations can be found in \cite{planck2013-p12}.  The mask
used in the analysis is the combined Galactic and point source common
mask~\citep[U73,][]{planck2013-p01} with sky fraction $f_{\rm sky}=0.73$.

The KSW and the ${\rm T}\phi$ estimators show similar sensitivity, finding,
respectively, $A^{{\rm T}\phi}=0.81 \pm 0.31$
and $A^{{\rm T}\phi}=0.70 \pm 0.28$ from the \smica\ CMB map, which
corresponds to a significance at about the $2.5\,\sigma$ level. The modal and
binned estimators are slightly less optimal, but give consistent results,
which is consistent with the imperfect overlap of the modal estimator
templates with the ISW-lensing signal; the ISW-lensing bispectrum has a
rapidly oscillating shape in the squeezed limit and both, binned and modal
estimates, are better suited (and originally implemented) to deal with smooth
bispectra of the kind predicted by primordial inflationary theories. Since
the correlation coefficient of the binned and modal ISW-lensing
templates relative to the actual ISW-lensing bispectrum (Eq.~\ref{thebl1l2l3})
is generally $0.8 < r < 0.9$~\citep[to be compared with $r=0.99$, achieved by
both estimators for local, equilateral and orthogonal primordial
templates,][]{planck2013-p09a}, the corresponding estimator's weights are
expected to be
about 20 \% suboptimal, consistent with observations.

The ${\rm T}\phi$ estimator has also been applied to the specific \Planck\ lensing
baseline, i.e., the MV map, which is a noise-weighted combination of the
217\,GHz and 143\,GHz channel maps, previously cleaned from infrared
contamination through subtraction of the 857\,GHz map, taken as
a dust template. From this map the lensing potential is recovered and then
correlated with that potential field in order to estimate the amplitude
$A^{{\rm T}\phi}$. The official baseline adopts a more conservative high-pass
filtering, such that as only multipoles $\ell \geq 10$ are considered, and the
mask with $f_{\rm sky}=0.7$ is used. In this case, the ISW-lensing estimate
is $0.78 \pm 0.32$ (a $2.4\,\sigma$ detection, where the 
error bars are obtained from 1000 simulations), as reported in the first
sub-row for ${\rm T}\phi$ in Table~\ref{table:isw-lensing}. 
The full multipole range is considered in the second sub-row, obtaining about
7\% better sensitivity. 

Notice that, according to all the estimators, the \ruler\ CMB map provides
lower significance for ISW-lensing, since its resolution is slightly lower
than that of the other maps. \nilc\ and \smica\ exhibit a somewhat
larger detection of the ISW signal, since they are the least noisy maps. 

In order to explore the agreement among the different estimators, we performed
a validation test based on 200 lensed simulations processed through the
\smica\ pipeline. 
The results are summarized in Table~\ref{tab:Aisw_compared}.  For each pair
of statistics, we provide the difference in amplitudes
estimated for the data ($\Delta A^{{\rm T}\phi}$), the dispersion of the
difference of amplitudes obtained from the simulations ($s_{A}$),
the ratio between this dispersion and the largest of the corresponding
sensitivities ($\eta$, according to Table~\ref{table:isw-lensing}), and the
correlation coefficient ($\rho$). As can be seen from the Table, the
agreement among estimators is good and the discrepancies are only
around $0.5\,\sigma$, which is the expected scatter, given the correlation
between the weights of different
estimators discussed above.
Overall, the bispectrum estimators provide a larger value of the amplitude
$A^{{\rm T}\phi}$, as compared to the ${\rm T}\phi$ estimator.
 
We have also explored the joint estimation of the two bispectra that are
expected to be found in the data: the ISW-lensing; and the residual point
sources. A detailed description of the non-Gaussian signal coming from
point sources can be found in \cite{planck2013-p09a}. The joint analysis of
these two signals performed with the KSW estimator, and the binned, and modal
estimators has shown that the ISW-lensing amplitude estimation can be
considered almost completely independent of the non-Gaussian signal induced by
the residual sources, and that the two bispectra are nearly perfectly
uncorrelated.

There is not a unique way of extracting a single signal-to-noise value from
Table~\ref{table:isw-lensing}.  However, all the estimators show evidence
of ISW-lensing at about the $2.5\,\sigma$ level.

Finally, we estimate that the bias introduced by the ISW-lensing signal on
the estimation of the primordial local shape bispectrum
(Eq.~\ref{fnl_bias_iswl}) is $\Delta^{\rm prim}\simeq 7$, corresponding to
the theoretical expectation, as described in detail in~\cite{planck2013-p09a}. 

\begin{table}[tmb]
\begingroup
\newdimen\tblskip \tblskip=5pt
\caption{For each pair of estimators, we give the mean difference among
the amplitudes estimated from the data ($\Delta A^{{\rm T}\phi}$), the
dispersion of the differences between the amplitudes estimated from the
simulations ($s_{A}$), the ratio of this dispersion to the larger of the
corresponding sensitivities ($\eta$), and the correlation coefficient ($\rho$).
\label{tab:Aisw_compared}}
\nointerlineskip
\vskip -3mm
\footnotesize
\setbox\tablebox=\vbox{
   \newdimen\digitwidth 
   \setbox0=\hbox{\rm 0} 
   \digitwidth=\wd0 
   \catcode`*=\active 
   \def*{\kern\digitwidth}
   \newdimen\signwidth 
   \setbox0=\hbox{+} 
   \signwidth=\wd0 
   \catcode`!=\active 
   \def!{\kern\signwidth}
\halign{#\hfil\tabskip=0.2cm& \hfil#\hfil\tabskip=0.2cm&
 \hfil#\hfil\tabskip=0.2cm& \hfil#\hfil\tabskip=0.2cm& \hfil#\hfil\tabskip=0.cm\cr 
\noalign{\doubleline}
 \noalign{\vskip -2pt}
&&KSW&Binned&Modal\cr 
\noalign{\vskip 3pt\hrule\vskip 5pt}
& $\Delta A\pm s_A$& $-0.11\pm0.10$& $-0.21\pm0.21$&  $-0.07\pm0.21$\cr
${\rm T}\phi$	& $\eta$& $0.32$& $0.56$& $0.56$\cr
& $\rho$& $0.95$&$0.84$& $0.84$\cr
\noalign{\vskip 3pt\hrule\vskip 5pt}
& $\Delta A\pm s_A$&	&$-0.10\pm0.19$&!$0.04\pm0.19$\cr
KSW& $\eta$& &$0.52$& $0.51$\cr
& $\rho$& &$0.86$& $0.87$\cr
\noalign{\vskip 3pt\hrule\vskip 5pt}
& $\Delta A\pm s_A$&& &!$0.14\pm0.15$\cr
Binned& $\eta$&	& &$0.41$\cr
& $\rho$& & & $0.92$\cr
\noalign{\vskip 1pt\hrule\vskip 3pt}}}
\endPlancktable                    
\endgroup
\end{table}

\section{Cross-correlation with surveys}
\label{subsec:xcor}

The ISW effect can be probed through several different approaches. Among the
ones already explored in the literature, the classical test is to study the
cross-correlation of the CMB temperature fluctuations with a tracer of the
matter distribution, typically a galaxy or cluster catalogue. 
As mentioned in the introduction, the correlation of the CMB with LSS tracers
was first proposed by~\cite{Crittenden1996} as 
a natural way to amplify the ISW signal, otherwise very much subdominant with
respect to the primordial CMB fluctuations.
Indeed, this technique led to the first reported detection of the ISW
effect \citep{Boughn2004}.

Several methods have been proposed in the literature to study statistically
the cross-correlation of the CMB fluctuations with LSS tracers, and, they
can be divided into: real space statistics (e.g., the cross-correlation
function, hereinafter CCF); harmonic space statistics (e.g., the cross-angular
power spectrum, hereinafter CAPS); and wavelet space statistics (e.g., the
covariance of the Spherical Mexican Hat Wavelet coefficients, or SMHWcov from
now on). These statistics are equivalent (in the sense of the significance of
the ISW detection) under ideal conditions. However, ISW data analysis presents
several problematic issues (incomplete sky coverage, selection biases in the
LSS catalogues, foreground residuals in the CMB map, etc.).
Hence, the use of several different statistical approaches provides a more
robust framework for studying the ISW-LSS cross-correlation, since different
statistics may have different sensitivity to these systematic effects,
The individual methods are described in more detail in
Sect.~\ref{subsubsec:methods}.

Besides the choice of specific statistical tool, the ISW cross-correlation can
be studied from two different (and complementary) perspectives. On the one
hand, we can determine the amplitude of the ISW signal, as well as the
corresponding signal-to-noise  ratio, by comparing the observed
cross-correlation to the expected one. On the other hand, we can postulate a
null hypothesis (i.e., that there is no correlation between the CMB and the
LSS tracer) and study the probability of obtaining the observed
cross-correlation. Whereas the former answers a question regarding the
compatibility of the data with the ISW hypothesis (and provides an estimation
of the signal-to-noise associated with the observed signal), the latter tells
us how incompatible the measured signal is with the no-correlation hypothesis,
i.e., against the presence of dark energy (assuming that the Universe is
spatially flat). Obviously, both approaches can be extended to account for
the cross-correlation signal obtained from several surveys at the same time.
These two complementary tests are described in detail in
Sect.~\ref{subsubsec:xcorrdef}, with the results presented in
Sect.~\ref{subsubsec:results}.


\subsection{Cross-correlation statistics}
\label{subsubsec:methods}

Let us denote the expected cross-correlation of two signals ($x$ and $y$) by $\xi_a^{xy}$, where $a$ stands for a distance measure (e.g., the angular distance $\theta$ between two points in the sky, the multipole $\ell$ of the harmonic transformation, or the wavelet scale $R$). For simplicity, we assume that the two signals are given in terms of a fluctuation field (i.e., with zero mean and dimensionless). 

This cross-correlation could represent either the CCF, the CAPS or the SMHWcov. It has to be understood as a vector of $a_\mathrm{max}$ components, where $a_\mathrm{max}$ is the maximum number of considered \emph{distances}. Obviously, when $x \equiv y$, $\xi_a^{xy}$ represents an auto-correlation. 
The specific forms for $\xi_a^{xy}$ and $\mathbf{C}_{\xi^{xy}}$ for the different cross-correlation statistics (CAPS, CCF, and SMHWcov) are given below.

\subsubsection{Angular cross-power spectrum}
\label{subsub:caps}

The angular cross-power spectrum (CAPS) is a natural tool for studying
the cross-correlation of the CMB fluctuations and tracers of the
LSS. Under certain conditions, it provides a statistical tool with
uncorrelated (full-sky coverage) or nearly uncorrelated (binned
spectrum for incomplete sky coverage) components.  Even the unbinned
CAPS, estimated on incomplete signals, can be easily worked out, since
the correlations are mostly related to the geometry of the mask. This
is the case for the CAPS obtained through {\tt MASTER}
approach~\citep[e.g.,][]{Hivon2002,Hinshaw2003}. 
Another approach is to work in the map domain, making use of a quadratic 
maximum likelihood (QML henceforth) estimator \citep{Tegmark1997c} 
for the CAPS \citep{Padmanabhan2005,Schiavon2012}. Such approach 
is optimal, i.e., leads to unbiased estimates for the CAPS with  
minimum error bars.

\subsubsection*{Pseudo angular power spectrum}

Let us denote the CAPS between the CMB field $T\left(p\right)$ and an LSS
tracer $G\left(p\right)$ map (where $p =\left(\theta, \phi \right)$
represents a given pixel) as: $C_{\ell}^{\mathrm{TG}}$ (i.e., $\xi_a^{xy}
\equiv C_{\ell}^{\mathrm{TG}}$ for this cross-correlation estimator). In the
full-sky case, an optimal estimator of the CAPS is given by:
\begin{equation}
\hat{C}_{\ell}^{\mathrm{TG}} = \frac{1}{2\ell + 1}
 \sum_{m=-\ell}^{+\ell} t_{\ell m} g_{\ell m}^*,
\label{eq:opt_caps}
\end{equation}
where $t_{\ell m}$ and $g_{\ell m}$ are the spherical harmonic
coefficients of the CMB and the LSS maps, respectively. This CAPS can
be seen as a vector with $\ell_\mathrm{max}$ components, where
$\ell_\mathrm{max}$ is the maximum multipole considered in the
analysis. Here we adopt $3N_{\rm side} - 1$,
which suffices for ISW analysis, since it is know that most of the ISW signal
is contained within $\ell \lesssim 80$ \citep{Afshordi2004b,Hernandez2008}.
When a mask $\Pi\left(p\right)$ is applied to
the maps, it acts as a weight that modifies the underlying harmonic
coefficients. Now, we have $\tilde{t}_{\ell m}$ and $\tilde{g}_{\ell
  m}$, where
\begin{eqnarray}
\tilde{t}_{\ell m} & = & \int \int \mathrm{d}\left(\cos{\theta} \right)
 \mathrm{d}\phi\, T\left(\theta, \phi \right)\Pi\left(\theta, \phi \right)
 Y_{\ell m}^* \left(\theta, \phi \right), \\ \nonumber
\tilde{g}_{\ell m} & = & \int \int \mathrm{d}\left( \cos{\theta} \right)
 \mathrm{d}\phi\, G\left(\theta, \phi \right)\Pi\left(\theta, \phi \right)
 Y_{\ell m}^* \left(\theta, \phi \right), 
\label{eq:masked_harm}
\end{eqnarray}
and $Y_{\ell m} \left(\theta, \phi \right)$ are the spherical harmonic
functions. In these circumstances, the estimator in Eq.~\ref{eq:opt_caps} is
not longer optimal, and is referred to as pseudo-CAPS. A nearly optimal
estimator is given by decoupling the masked CAPS (denoted by
$\tilde{C}_{\ell}^{\mathrm{TG}}$) through the masking kernel
$\tens{B}$~\citep[e.g.,][]{Xia2011}:
\begin{equation}
\label{eq:pcaps}
\hat{C}_{\ell}^{\mathrm{TG}} = \tens{B}^{-1}\tilde{C}_{\ell}^{\mathrm{TG}},
\end{equation}
where
\begin{equation}
B_{\ell \ell', {\rm G}} = \frac{2\ell+1}{4\pi}\sum_{\ell''} J_{\ell''}^{\rm G}
\left(\begin{array}{ccc}
      \ell & \ell' & \ell'' \\
      0 & 0 & 0
      \end{array}
\right)^2,
\label{eq:master1}
\end{equation}
with $J_{\ell''}^G$ the cross-angular power spectrum of the $T$ and $G$ masks.

The estimator in Eq.~\ref{eq:pcaps} is nearly optimal because
$\tilde{C}_{\ell}^{\mathrm{TG}}$ has to be understood as the mean value over
an ensemble average of skies.
Let us point out that when more than a single CAPS is considered, for instance
when one is interested in the cross-correlation of the \Planck\ CMB map with
more than one LSS tracer map, the CAPS estimator can be seen as a single vector
with $N\ell_{\rm max}$ components, with $N$ being the number of surveys.

It can be shown that the element $C_{\ell \ell', i j}$ of the covariance
matrix of the CAPS estimator in Eq.~\ref{eq:pcaps} (for the case of a masked
sky and for $N$ surveys) is given by
\begin{equation}
C_{\ell \ell', i j} = K_{\ell, i j} K_{\ell', i j}
 \frac{\left(M_{ij}\right)_{\ell \ell'}^{-1}}{(2\ell' + 1)} ,
\label{eq:cov_caps1}
\end{equation}
where
\begin{equation}
K_{\ell, i j} = \left[ C_{\ell}^{\mathrm{TG}_i}C_{\ell}^{\mathrm{TG}_j}
 + C _{\ell}^\mathrm{T}\left(C_{\ell}^{\mathrm{G}_i \mathrm{G}_j}
 + N_{\ell}^{\mathrm{G}_i \mathrm{G}_j}\delta_{ij} \right) \right]^{1/2},
\label{eq:cov_caps2}
\end{equation}
and $\left(M_{ij}\right)_{\ell \ell'}^{-1}$ is the
$\left(\ell,\ell'\right)$ element of the inverse matrix of $\tens{M}$
for surveys $i$ and $j$ fixed, such as
\begin{equation}
M_{\ell \ell', ij} = \frac{2\ell+1}{4\pi}\sum_{\ell''} H_{\ell''}^{ij}
\left(\begin{array}{ccc}
      \ell & \ell' & \ell'' \\
      0 & 0 & 0
      \end{array}.
\right)^2,\label{eq:master2}
\end{equation}
Here $H_{\ell''}^{ij}$ is the angular cross-power spectrum of the two joint
masks, i.e., the masks resulting from the multiplication of the 
$T$ with $G_i$ and $G_j$, respectively.
The quantities $C_{\ell}^{xy}$ are expected or fiducial spectra,
$N_{\ell}^{yy}$ is the Poissonian noise of the $y$ survey
(deconvolved by any beam or pixel filter), and $\delta_{ij}$ is the Kronecker
delta. In Eq.~\ref{eq:cov_caps2}, the instrumental noise associated with
the CMB data has been
ignored, since the \Planck\ sensitivity is such that the noise contribution on
the scales of interest is negligible.
When more than one survey has poor sky coverage, then the complexity of the
correlations is not well reflected by the previous expression. Therefore, in
this paper, we will compute $C_{\ell \ell', i j}$ from coherent CMB and LSS
Monte Carlo simulations. 
For each simulation, we generate four independent, Gaussian, and white
realizations (at $N_\mathrm{nside}=64$), which are afterwards properly
correlated using the expected auto- and cross-correlations of the signals.
Corresponding Poissonian shot noise realizations are added to each survey map.
The resulting four maps are masked with the corresponding masks (i.e., one for
the CMB and one for mask for each survey).

The computation of the CAPS in Eq.~\ref{eq:pcaps} is
extremely fast (especially for the resolutions used in the study of the ISW).
However, as stated above, it is only a {\it nearly\/} optimal estimator of the
underlying CAPS. Moreover, its departure from
optimality is largest at the smallest
multipoles (largest scales), where the ISW signal is more important. 

\subsubsection*{The QML angular power spectrum}

The QML method for the power spectrum estimation 
of temperature CMB anisotropies was introduced 
by~\cite{Tegmark1997c} and later extended to polarization by
\cite{Tegmark2001}. For an application to temperature and polarization to
{\it WMAP\/} data see~\cite{Gruppuso2009} and \cite{Paci2013}.
The same method was employed to measure the cross-correlation between 
the CMB and LSS in~\cite{Padmanabhan2005},
\cite{Ho2008}, and~\cite{Schiavon2012}.
The QML method is usually stated to be optimal, since it provides unbiased
estimates and the smallest error bars allowed by the Fisher-Cram{\'e}r-Rao
inequality.  As a drawback, from the computational point of view, 
the QML is a very expensive approach. Let us denote 
the QML estimator of the CAPS between the CMB map $T$ 
and an LSS tracer $G$ (at multipole $\ell$) by
$Q_{\ell}^{\mathrm{TG}}$ (i.e., $\xi_a^{xy} \equiv Q_{\ell}^{\mathrm{TG}}$
for this cross-correlation estimator). 

A detailed description of the algebra of the QML is given 
in~\cite{Schiavon2012}. We briefly recall here the
basics of the CAPS estimator, which is given by:
\begin{equation}
\label{eq:qmlcaps}
\hat{Q}_\ell^{\mathrm{TG}} = \sum_{\ell' X'}
 (F^{-1})^{{\mathrm{TG}} ~ X'}_{\ell\ell'}
 \left[ \vec{x}^{\rm T} \tens{E}_{\ell'}^{X'} \vec{x}
 -{\rm Tr}(\tens{N}\tens{E}_{\ell'}^{X'}) \right],
\end{equation}
where $X$ represents any of the following spectra:
$X = \left\lbrace \mathrm{T}, \mathrm{TG}, \mathrm{G} \right\rbrace$.
The vector $\vec{x}$ has $2N_{\mathrm{pix}}$ elements (with $N_{\mathrm{pix}}$
being the total number of pixels allowed by the joint CMB and LSS mask): the
first set of $N_{\mathrm{pix}}$ corresponds to the CMB map, and the second one
accounts for the LSS map. The $F^{{\mathrm{TG}} ~ X'}_{\ell \ell '}$ is the
Fisher matrix defined as
\begin{equation}
\label{eq:fisher}
F_{\ell\ell'}^{\mathrm{TG}~ X'}=\frac{1}{2}{\rm Tr}
 \Big[\tens{C}^{-1}\frac{\partial {\tens{C}}}{\partial
 C_\ell^{\mathrm{TG}}}\tens{C}^{-1}\frac{\partial
 \tens{C}}{\partial C_{\ell'}^{X'}}\Big],
\end{equation}
and the $\tens{E}$ matrix is given by
\begin{equation}
\label{eq:Elle}
\tens{E}_\ell^X=\frac{1}{2}\tens{C}^{-1}\frac{\partial \tens{C}}{\partial
  C_\ell^X}\tens{C}^{-1}.
\end{equation}
The object $\tens{C} =\tens{S}(C_{\ell}^{X})+\tens{N}$ is the global
covariance matrix, including the signal $\tens{S}$ and noise $\tens{N}$
contributions, with $C_\ell^X$ being the fiducial theoretical angular power
spectrum. 
The uncertainties in the QML estimates are given by the inverse of the 
Fisher matrix, which includes the correlation among different multipoles.
The error associated with the shot noise of the galaxy surveys is modelled
in the galaxy submatrix of $\tens{N}$.

The results presented on this paper are based on
$\hat{C}_{\ell}^{\mathrm{TG}}$, whereas $\hat{Q}_{\ell}^{\mathrm{TG}}$
is used as a cross-check, applied to a lower resolution version of the maps of $N_\mathrm{side} = 32$.
In addition the maximum multipole considered in this case is $\ell_\mathrm{max} = 2N_\mathrm{side}$, which
has been already verified as a conservative limit for the QML.

\subsubsection{Cross-correlation function}
\label{subsub:ccf}

The cross-correlation function (CCF) is a suitable tool for studying the
ISW effect via cross-correlation of the
CMB fluctuations and tracers of the LSS, and it has been one of the most
extensively used in this
context~\citep[e.g.,][]{Boughn2002b,Giannantonio2008b,Xia2009b}. On the one
hand, the signal only appears at very large scales and, therefore,
it is sufficient to work at resolutions at which the low performance of the
CCF (in terms of computational time) is not
a serious handicap. On the other hand, neither the CMB nor the LSS data are
available with full sky coverage and, in some cases,
the geometry of the masks is very complicated: the CCF adapts perfectly to the
effects of partial sky coverage, since it is defined in real
space. As a drawback, the Poissonian noise of the galaxy tracer appears at
the smallest angular scales, where the signal-to-noise of the ISW effect is
higher for this estimator. Therefore, a proper characterization of the shot
noise is especially important for the CCF, in order to obtain a good
estimation of the uncertainties.

Let us denote the CCF between the CMB map $T$ and an LSS tracer $G$
(at an angular distance of $\theta$) as
$C^{\mathrm{TG}}(\theta)$ (i.e., $\xi_a^{xy} \equiv C^{\mathrm{TG}}(\theta)$
for this cross-correlation statistic).
The CCF estimator is defined as
\begin{equation}
\hat{C}^{\mathrm{TG}}(\theta) = \frac{1}{N_\theta}\sum_{i,j}T_i G_j,
\label{eq:ccf}
\end{equation}
where the sum runs over all pixels with a given angular separation.
For each angular bin centred around $\theta$, $N_\theta$ is the
number of pixel pairs separated by an angle within the bin. Only the pixels
allowed by the joint CMB and LSS mask are considered.
The angular bins used in this work are:
$\theta_1 \in \left[0,1\right]^\circ$; $\theta_2 \in \left(1,3\right]^\circ$;
$\theta_3 \in \left(3,5\right]^\circ$; \ldots; and $\theta_{61} \in
\left(119,121\right]^\circ$. The choice of binning does not
affect the results significantly.

The covariance of the CCF estimator can be easily derived from the one
already computed for the CAPS in Eq.~\ref{eq:cov_caps1}. It is sufficient
to know that the CCF can be expressed in terms of the CAPS as
\begin{equation}
C^{\mathrm{TG}}\left(\theta\right) =
 \sum_{\ell = 0}^{\ell_\mathrm{max}}
 \frac{2\ell + 1}{4\pi}C_{\ell}^{\mathrm{TG}}P_\ell\left(\cos{\theta}\right),
\label{eq:ccf_in_caps}
\end{equation}
where $P_\ell\left(\cos{\theta}\right)$ are the Legendre polynomials.
Hence, it is straightforward to prove that the covariance between the
$\theta$ and $\theta'$ components of the CCF for the surveys $i$ and $j$,
respectively, is given by
\begin{equation}
C_{\theta \theta', i j} = \sum_{\ell} \sum_{\ell'}
 \frac{\left(2\ell + 1\right)}{4\pi}
 \frac{\left(2\ell' + 1\right)}{4\pi}
 P_\ell\left(\cos{\theta}\right) P_{\ell'}\left(\cos{\theta}'\right)
 C_{\ell \ell', i j}.
\label{eq:ccf_cov}
\end{equation}
%


\subsubsection{Wavelet covariance}

Wavelets provide an interesting alternative to more traditional tools
(e.g., CCF or CAPS) for studying the CMB-LSS correlation. They exploit the
fact that the ISW signal is mostly concentrated at scales of a few
degrees~\citep[e.g.,][]{Afshordi2004b}. Wavelets are ideal kernels to enhance
features with a characteristic size, since the wavelet analysis at an
appropriate scale $R$ amplifies those features over the background. Therefore,
wavelets could provide most of the signal-to-noise of the ISW effect by just
analysing a narrow range of scales.
They were first proposed for the ISW detection by~\cite{Vielva2006}, where the
Spherical Mexican Hat Wavelet~\citep[SMHW,][]{Martinez2002} was proposed as
the filtering kernel.
The basic idea of this approach is to estimate the covariance of the SMHW
coefficients (SMHWcov) as a function of the wavelet
scale~\citep[see e.g.,][for details]{Vielva2006}. Other wavelet kernels can
be considered, such as needlets \citep{Pietrobon2006a}, directional
wavelets \citep{Mcewen2007}, or steerable wavelets \citep{McEwen2008b}.

Let us denote the SMHWcov between the CMB map $T$ and a LSS tracer $G$
(at a wavelet scale $R$) as
$\Omega^{\mathrm{TG}}\left(R\right)$, i.e.,
$\xi_a^{xy} \equiv \Omega^{\mathrm{TG}}\left(R\right)$
for this cross-correlation statistic).
The SMHWcov estimator is defined as
\begin{equation}
\hat{\Omega}^{\mathrm{TG}}\left(R\right) =
 \frac{1}{N_\mathrm{pix}}\sum_{i}\omega_{{\rm T}_i}\left(R\right)
 \omega_{{\rm G}_i}\left(R\right),
\label{eq:smhwcov}
\end{equation}
where $\omega_{T}\left(R\right)$ and $\omega_{G}\left(R\right)$ are the SMHW
coefficients for the CMB and the LSS at scale $R$, respectively
(note that wavelet coefficients are forced to have zero mean on the observed
sky). The scales considered in our study are
$R = \left\lbrace  60, 90, 120, 150, 200, 250, 300, 350, 400, 500, 600
 \right\rbrace$ in arcminutes.

As for the CCF, the covariance of the SMHWcov estimator can be easily derived
from the one already computed for the CAPS in Eq.~\ref{eq:cov_caps1}.
It is sufficient to know that the SMHWcov can be expressed in terms of the
CAPS as
\begin{equation}
\Omega^{\mathrm{TG}}\left(R\right)
 = \sum_{\ell = 0}^{\ell_\mathrm{max}}\frac{2\ell + 1}{4\pi}
 C_{\ell}^{\mathrm{TG}}\omega_{\ell}^2\left(R\right),
\label{eq:ccf_in_caps}
\end{equation}
where $\omega_\ell\left(R\right)$ is the SMHW window function at the scale
$R$. Hence, it is straightforward to prove that the covariance between the
$R$ and $R'$ components of the SMWHcov for surveys $i$ and $j$, respectively,
is given by:
\begin{equation}
C_{R R', i j} = \sum_{\ell} \sum_{\ell'} \frac{\left(2\ell + 1\right)}{4\pi}
 \frac{\left(2\ell' + 1\right)}{4\pi} \omega_{\ell}^2\left(R\right)
 \omega_{\ell'}^2\left(R'\right) C_{\ell \ell', i j}.
\label{eq:ccf_cov}
\end{equation}
%


\subsection{Cross-correlation tests}
\label{subsubsec:xcorrdef}

For any of the cross-correlation estimators described above, we aim two
different statistical tests.  First,
if the observed cross-correlation is given by $\hat{\xi}_a^{xy}$, then,
a simple $\chi^2$ can be proposed to estimate the amplitude $A$, such that
$A\xi_a^{xy}$ is the closest solution to $\hat{\xi}_a^{xy}$:
\begin{equation}
\chi^2\left(A\right) = \left[\hat{\xi}_a^{xy} - A\xi_a^{xy} \right]^{\rm T}
 \tens{C}_{\xi^{xy}}^{-1} \left[\hat{\xi}_a^{xy} - A\xi_a^{xy}\right],
\label{eq:xcorr_fit}
\end{equation}
where $\tens{C}_{\xi^{xy}}$ is the covariance matrix (of dimension
$a_\mathrm{max} \times a_\mathrm{max}$) of the expected cross-correlation
$\xi_a^{xy}$, i.e.,
$\mathbf{C}_{{\xi^{xy}}_{i,j}} \equiv
 \left\langle \xi_{a_i}^{xy} \xi_{a_j}^{xy} \right\rangle$.
It is straightforward to show that the best-fit amplitude $A$ and its
dispersion are given by
\begin{eqnarray}
\label{eq:fit}
A&=&\left[\hat{\xi}_a^{xy}\right]^t \tens{C}_{\xi^{xy}}^{-1} \xi_a^{xy}
 \left[\left[\xi_a^{xy}\right]^t
 \tens{C}_{\xi^{xy}}^{-1} \xi_a^{xy} \right]^{-1}, \\
 \sigma_A&=&\left[\left[\xi_a^{xy}\right]^t
 \tens{C}_{\xi^{xy}}^{-1} \xi_a^{xy} \right]^{-1/2} \nonumber.
\end{eqnarray}
An analogy with Eq.~\ref{eq:xcorr_fit} can be defined for the null hypothesis
case:
\begin{equation}
{\chi^2}_{\mathrm{null}} = \left[\hat{\xi}_a^{xy}\right]^{\rm T}
 \tens{D}_{\xi^{xy}}^{-1} \hat{\xi}_a^{xy},
\label{eq:xcorr_null}
\end{equation}
where $\tens{D}_{\xi^{xy}}$ is the covariance of the cross-correlation of the
two signals, in the absence of an intrinsic dependence, i.e.,
when $\xi_a^{xy} \equiv 0$. The ISW signal is very weak and, therefore it
is a good approximation to assume that that
$\tens{D}_{\xi^{xy}} \approx \tens{C}_{\xi^{xy}}$. 

For Gaussian statistics, ${\chi^2}_{\mathrm{null}}$ already provides the direct
probability of the observed cross-correlation $\hat{\xi}_a^{xy}$ under the
null hypothesis. However, several non-idealities (sky coverage, systematics,
foregrounds residuals, etc.) forces is to use alternative approaches to
estimate the probability. One of the most common options is to perform the
cross-correlation of survey signal $y$ with realistic simulations of the CMB,
$x$, and compute a joint statistics (e.g., ${\chi^2}_{\mathrm{null}}$) for
each simulation. The probabiltiy value associated with the data will come then
then be the fraction of simulations having a value of
${\chi^2}_{\mathrm{null}}$ equal to or larger than the one obtained for
the data.
Both, $\tens{C}_{\xi^{xy}}$ and $\tens{D}_{\xi^{xy}}$ can be derived either
analytically or numerically (via simulations). 

The latter approach is computationally expensive, but, in some cases,
could provide a more accurate defence against certain systematics, in
particular the incomplete sky coverage. There are several options to perform
such kind of simulations. The standard approach is the one
mentioned above, i.e., cross-correlation of the LSS map with CMB simulations.
This is a very robust approach, since it is usually
hard to reproduce the systematics present in the LSS tracers,
but incomplete because the LSS is fixed, resulting in a lack of randomness.
An alternative method is to use a jack-knife test, which unfortunately tends
to underestimate the errors.
Finally, one can produce simulations of both the CMB and the LSS, assuming
perfect knowledge of the properties of both signals, in particular of the
LSS field (which, as mentioned above, is almost never the case).
Comprehensive discussions of these approaches are given
in~\cite{Cabre2007} and~\cite{Giannantonio2008b}.

\subsection{Results}
\label{subsubsec:results}

\begin{table}[tmb]
\begingroup
\newdimen\tblskip \tblskip=5pt
\caption{Expected significance $A/\,\sigma_A$ of the CMB-LSS
cross-correlation. Values obtained from each survey independently, as well as
jointly, are given for all the estimators (CAPS, CCF, and
SMHWcov).\label{tab:s2n_expected}
\label{tab:surveys}}
\nointerlineskip
\vskip -3mm
\footnotesize
\setbox\tablebox=\vbox{
   \newdimen\digitwidth 
   \setbox0=\hbox{\rm 0} 
   \digitwidth=\wd0 
   \catcode`*=\active 
   \def*{\kern\digitwidth}
   \newdimen\signwidth 
   \setbox0=\hbox{+} 
   \signwidth=\wd0 
   \catcode`!=\active 
   \def!{\kern\signwidth}
\halign{\hfil#\hfil\tabskip=0.4cm& \hfil#\hfil\tabskip=0.4cm&
 \hfil#\hfil\tabskip=0.4cm& \hfil#\hfil\tabskip=0.4cm& \hfil#\hfil\tabskip=0cm\cr 
\noalign{\doubleline}
 \noalign{\vskip -2pt}
$\hat{\xi}_a^{xy}$&NVSS&SDSS&SDSS&All\cr 
& &CMASS/LOWZ&MphG&\cr 
\noalign{\vskip 3pt\hrule\vskip 5pt}
CAPS		& 3.0& 1.9& 0.6& 3.2\cr
CCF		& 3.0& 1.9& 0.6& 3.1\cr
SMMHWcov	& 3.0& 1.9& 0.5& 3.1\cr
\noalign{\vskip 3pt\hrule\vskip 3pt}}}
\endPlancktable                    
\endgroup
\end{table}

In this section we present the results obtained from the cross-correlation
of the galaxy catalogues described in Sect.~\ref{subsec:extdata}
(NVSS, \lrg and \mg) with the four \Planck\ CMB maps presented in 
Sect.~\ref{subsubsec:cmbmaps} (\ruler, \nilc, \sevem, and \smica).
All these maps are analysed at a {\tt HEALPix} resolution of
$N_\mathrm{side} = 64$. The cross-correlation estimators described in the
previous section are applied to all cases. This comprehensive analysis
will help to achieve a robust estimation of the ISW.

As already mentioned the covariance among all the components of the estimators
are obtained from coherent Gaussian simulations of the CMB and the three
galaxy catalogues. Since we are only considering large-scale effects (above
about $1^\circ$), the same set of CMB simulations are equally valid for the
four CMB maps, since they are nearly identical on such
scales~\citep[see][for details]{planck2013-p06}.
We have used 70{,}000 coherent Monte Carlo simulation sets (as described in
Sect.~\ref{subsub:caps}) to compute the correlations; this is enough to
characterize the covariance.

The expected signal-to-noise ratio for the ISW effect detection is summarized
in Table~\ref{tab:s2n_expected}. Values for all the cross-correlation
estimators are given. We consider the case of a survey-by-survey detection,
as well as the joint analysis of all the surveys.
A signal-to-noise of about $3\,\sigma$ is expected for the joint analysis,
which is actually dominated by the NVSS cross-correlation. This is expected,
since, firstly, NVSS covers a much larger fraction of the sky compared to
other surveys, and secondly, it extends over a redshift interval ideal for
the detection of the ISW signal~\citep[e.g.][]{Afshordi2004b}.

The differences among estimators are not significant, indicating that none of
them is clearly optimal compared with the others.
To explore this agreement further, we have analysed an extra set of
1{,}000 CMB and LSS clustered simulations, and have compared, simulation by
simulation, the ISW amplitude estimation derived for each cross-correlation
estimator ($C_\ell^{\rm TG}$, $C^{\rm TG}\left(\theta\right)$ and
$\Omega^{\rm TG}\left(R\right)$). 
In Table~\ref{tab:A_compared} we summarize the comparison. We only report
values for the joint fit to the ISW amplitude for the three surveys.
Similar results are found survey by survey. For each pair of estimators,
we provide the mean difference among the amplitude estimations ($\Delta A$),
the dispersion of these differences ($s_A$), the ratio ($\eta$) of this
dispersion to the expected sensitivity (i.e., the inverse of the
signal-to-noise numbers given in the last column of
Table~\ref{tab:s2n_expected}), and the correlation coefficient ($\rho$).
It is clear that the agreement between estimators is very high and that
differences are, on average, lower than half the statistical uncertainty
imposed by the sampling variance. 

\begin{table}[tmb]
\begingroup
\newdimen\tblskip \tblskip=5pt
\caption{For each pair of estimators, we give the mean difference among
the amplitude estimations ($\Delta A$), the dispersion of these differences
($s_A$), the ratio ($\eta$) of this dispersion to the expected sensitivity
(i.e., the inverse of the signal-to-noise numbers given in the last column
of Table~\ref{tab:s2n_expected}), and the correlation coefficient
($\rho$).\label{tab:A_compared}}
\nointerlineskip
\vskip -3mm
\footnotesize
\setbox\tablebox=\vbox{
   \newdimen\digitwidth 
   \setbox0=\hbox{\rm 0} 
   \digitwidth=\wd0 
   \catcode`*=\active 
   \def*{\kern\digitwidth}
   \newdimen\signwidth 
   \setbox0=\hbox{+} 
   \signwidth=\wd0 
   \catcode`!=\active 
   \def!{\kern\signwidth}
\halign{\hfil#\hfil\tabskip=0.2cm& \hfil#\hfil\tabskip=0.2cm&
 \hfil#\hfil\tabskip=0.2cm& \hfil#\hfil\tabskip=0.cm\cr 
\noalign{\doubleline}
 \noalign{\vskip -2pt}
&&CCF&SMHWcov\cr 
\noalign{\vskip 3pt\hrule\vskip 5pt}
& $\Delta A\pm s_A$& $-0.01\pm0.12$& $0.06\pm0.07$\cr
CAPS& $\eta$& $0.36$& $0.21$\cr
& $\rho$& $0.93$& $0.98$\cr
\noalign{\vskip 3pt\hrule\vskip 5pt}
& $\Delta A\pm s_A$&& $0.08\pm0.14$\cr
CAPS& $\eta$& & $0.42$\cr
& $\rho$& & $0.92$\cr
\noalign{\vskip 3pt\hrule\vskip 3pt}}}
\endPlancktable                    
\endgroup
\end{table}

\begin{table*}[tmb]
\begingroup
\newdimen\tblskip \tblskip=5pt
\caption{Amplitudes $A$, errors $\sigma_A$, and significance
levels $A/\,\sigma_A$ of the CMB-LSS cross-correlation (survey by survey and
all together) due to the ISW effect, for all component separation algorithms
for the different estimators.\label{tab:s2n_data}}
\nointerlineskip
\vskip -3mm
\footnotesize
\setbox\tablebox=\vbox{
   \newdimen\digitwidth 
   \setbox0=\hbox{\rm 0} 
   \digitwidth=\wd0 
   \catcode`*=\active 
   \def*{\kern\digitwidth}
   \newdimen\signwidth 
   \setbox0=\hbox{+} 
   \signwidth=\wd0 
   \catcode`!=\active 
   \def!{\kern\signwidth}
\halign{#\hfil\tabskip=0.3cm& \hfil#\hfil\tabskip=0.3cm&
 \hfil#\hfil\tabskip=0.3cm& \hfil#\hfil\tabskip=0.3cm&
 \hfil#\hfil\tabskip=0.3cm& \hfil#\hfil\tabskip=0.3cm&
 \hfil#\hfil\tabskip=0.3cm& \hfil#\hfil\tabskip=0.3cm&
 \hfil#\hfil\tabskip=0.3cm& \hfil#\hfil\tabskip=0.cm\cr 
\noalign{\doubleline}
 \noalign{\vskip -2pt}
LSS data& $\hat{\xi}_a^{xy}$& \ruler&& \nilc\hfil&& \sevem&& \smica\hfil&\cr 
\noalign{\vskip 2pt\hrule\vskip 3pt} 
& & $A^{{\rm T}\phi}$ & SNR & $A^{{\rm T}\phi}$ & SNR & $A^{{\rm T}\phi}$ & SNR & $A^{{\rm T}\phi}$ & SNR\cr
\noalign{\vskip 3pt\hrule\vskip 5pt}
NVSS		& CAPS 		& $0.86\pm0.33$& 2.6& $0.91\pm0.33$& 2.8& $0.90\pm0.33$& 2.7& $0.91\pm0.33$& 2.7\cr 
			& CCF 		& $0.80\pm0.33$& 2.4& $0.84\pm0.33$& 2.5& $0.83\pm0.33$& 2.5& $0.84\pm0.33$& 2.5\cr 
			& SMHWcov 	& $0.89\pm0.34$& 2.6& $0.93\pm0.34$& 2.8& $0.89\pm0.34$& 2.6& $0.92\pm0.34$& 2.7\cr 
\noalign{\vskip 3pt\hrule\vskip 5pt}
\lrg			& CAPS 		& $0.98\pm0.52$& 1.9& $1.09\pm0.52$& 2.1& $1.06\pm0.52$& 2.0& $1.09\pm0.52$& 2.1\cr 
		& CCF 		& $0.81\pm0.52$& 1.6& $0.91\pm0.52$& 1.8& $0.89\pm0.52$& 1.7& $0.90\pm0.52$& 1.7\cr 
			& SMHWcov 	& $0.80\pm0.53$& 1.5& $0.89\pm0.53$& 1.9& $0.87\pm0.53$& 1.6& $0.88\pm0.53$& 1.7\cr 
\noalign{\vskip 3pt\hrule\vskip 5pt}
\mg			& CAPS 		& $1.31\pm0.57$& 2.3& $1.43\pm0.57$& 2.5& $1.35\pm0.57$& 2.4& $1.42\pm0.57$& 2.5\cr 
			& CCF 		& $1.00\pm0.57$& 1.8& $1.11\pm0.57$& 2.0& $1.10\pm0.57$& 1.9& $1.10\pm0.57$& 1.9\cr 
			& SMHWcov 	& $1.03\pm0.59$& 1.8& $1.18\pm0.59$& 2.0& $1.15\pm0.59$& 2.0& $1.17\pm0.59$& 2.0\cr 
\noalign{\vskip 3pt\hrule\vskip 5pt}
All				& CAPS 		& $0.84\pm0.31$& 2.7& $0.91\pm0.31$& 2.9& $0.88\pm0.31$& 2.0& $0.90\pm0.31$& 2.9\cr 
		& CCF 		& $0.77\pm0.31$& 2.5& $0.83\pm0.31$& 2.7& $0.82\pm0.31$& 2.6& $0.82\pm0.31$& 2.7\cr 
			& SMHWcov 	& $0.86\pm0.32$& 2.7& $0.92\pm0.32$& 2.9& $0.89\pm0.32$& 2.8& $0.91\pm0.32$& 2.9\cr 
\noalign{\vskip 5pt\hrule\vskip 3pt}}}
\endPlancktablewide                    
\endgroup
\end{table*}

\begin{figure*}[!ht]
\begin{center}
\includegraphics[width=6cm]{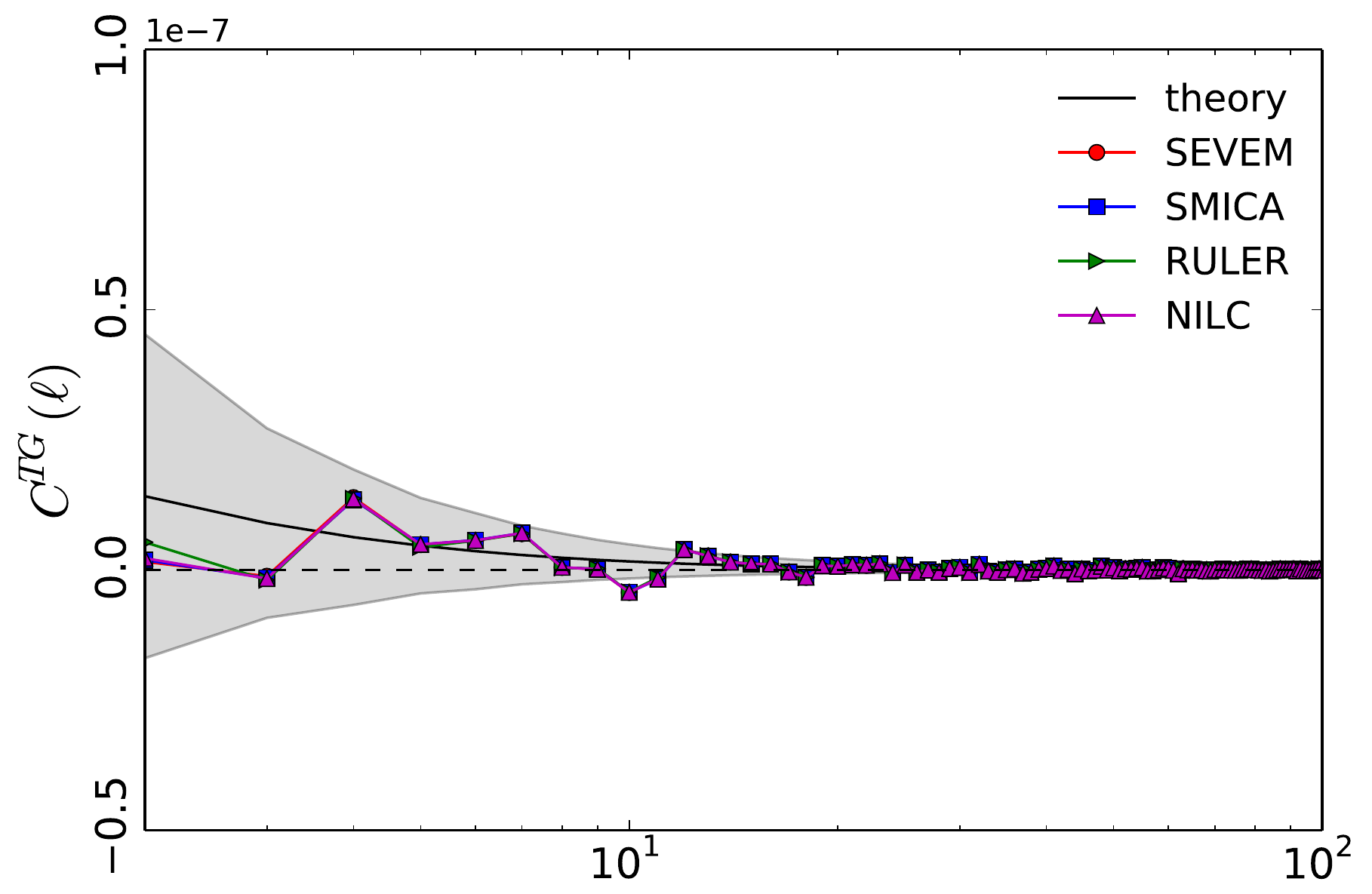}
\includegraphics[width=6cm]{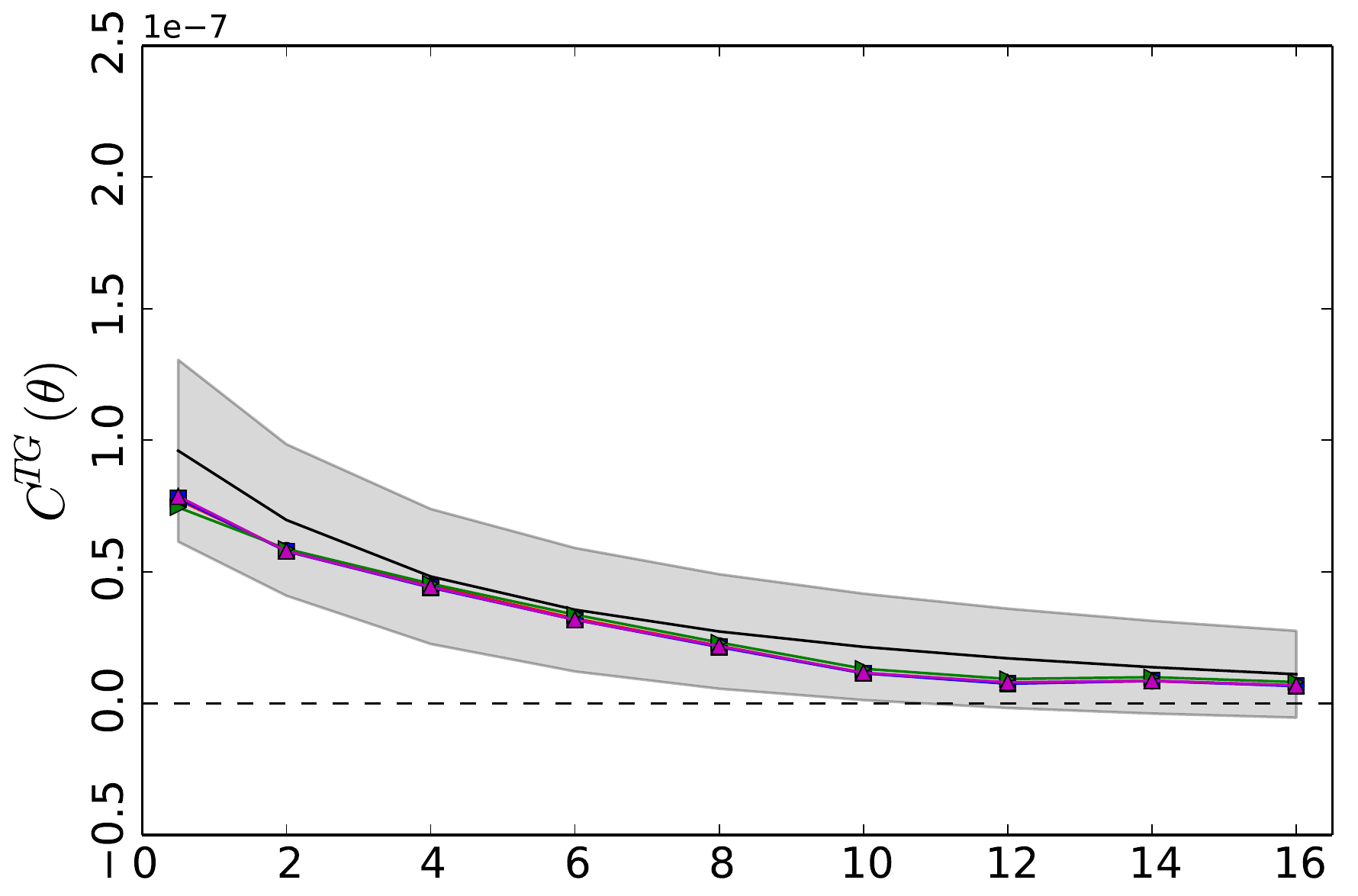}
\includegraphics[width=6cm]{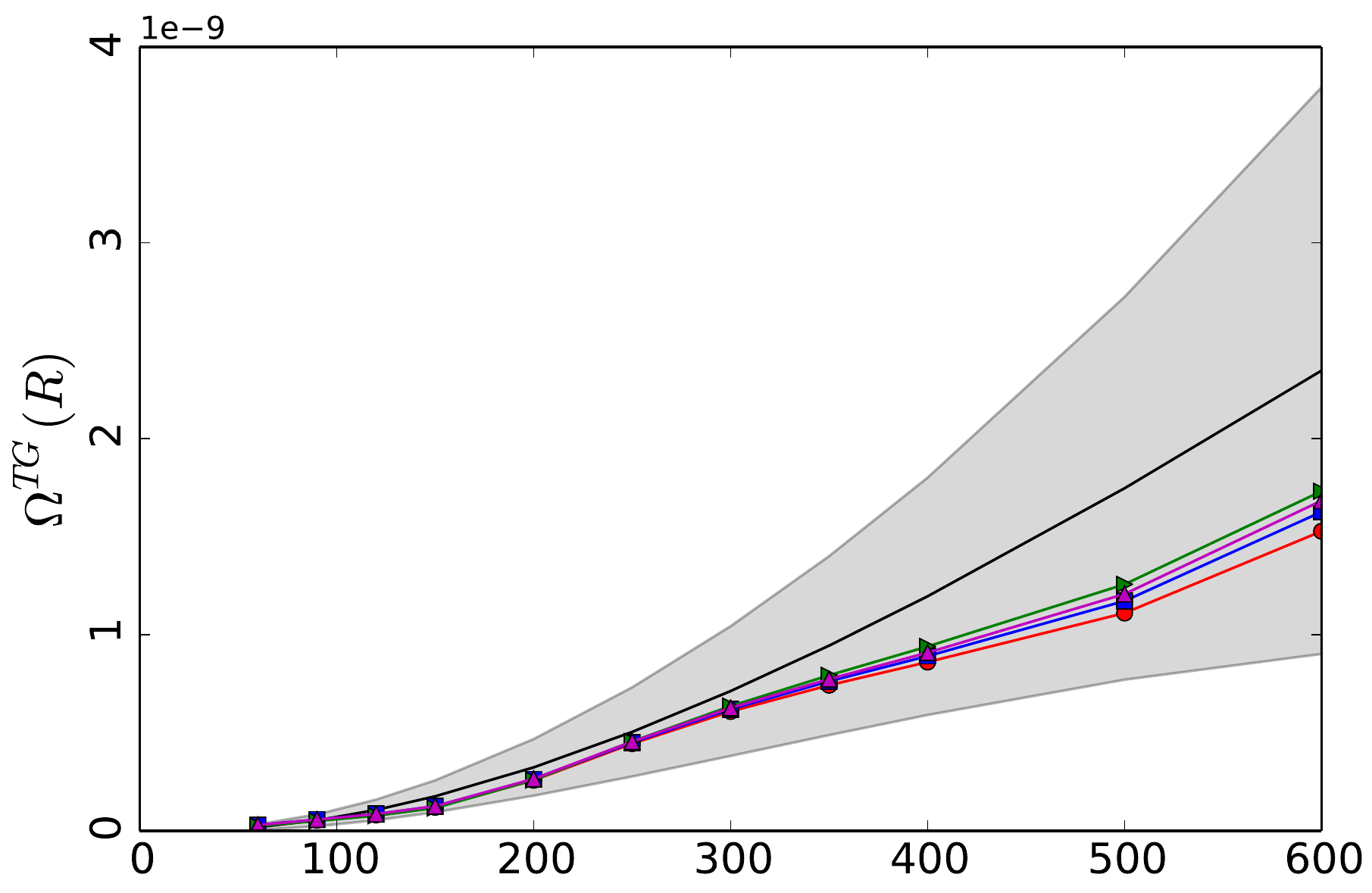}
\includegraphics[width=6cm]{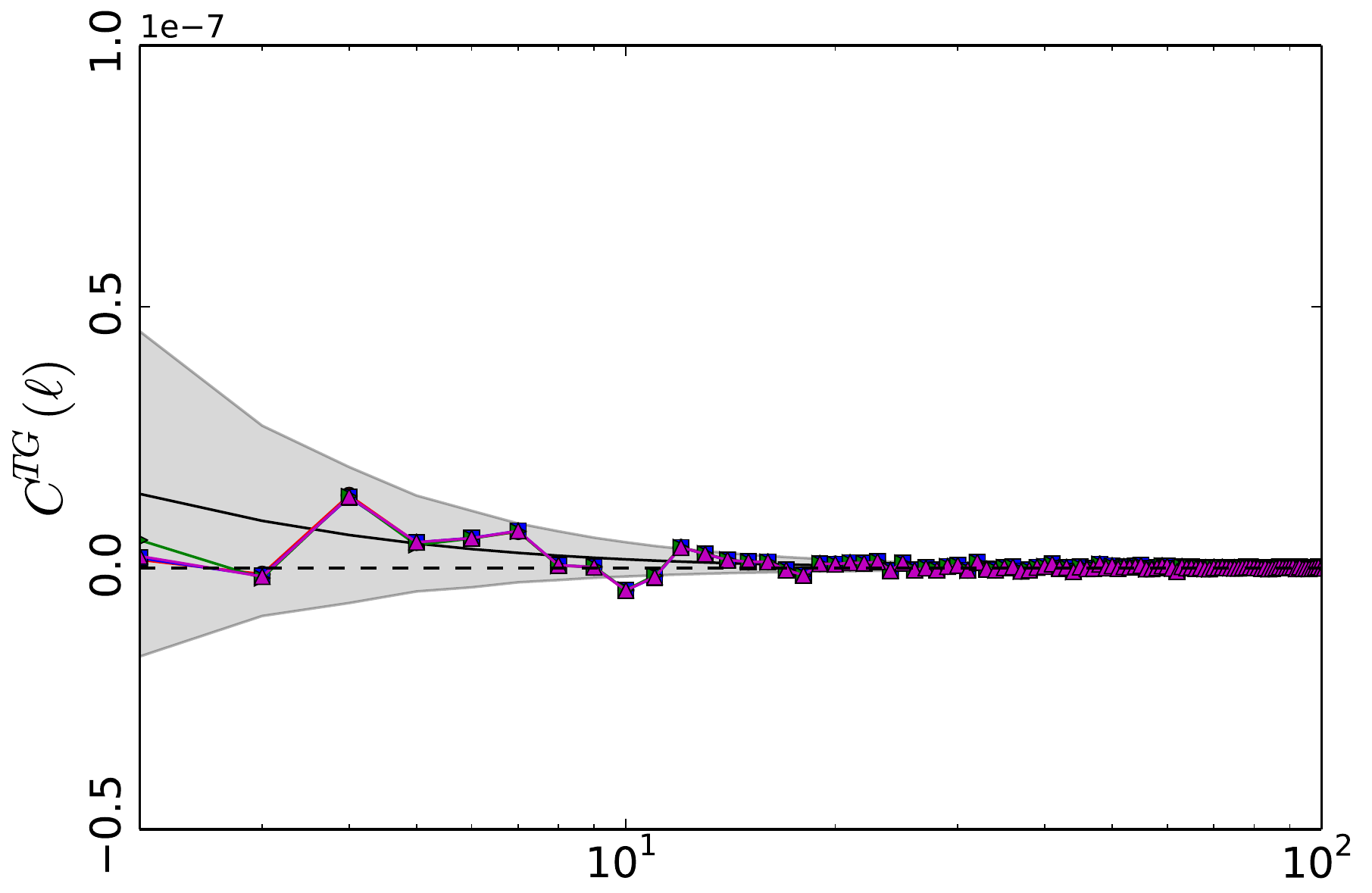}
\includegraphics[width=6cm]{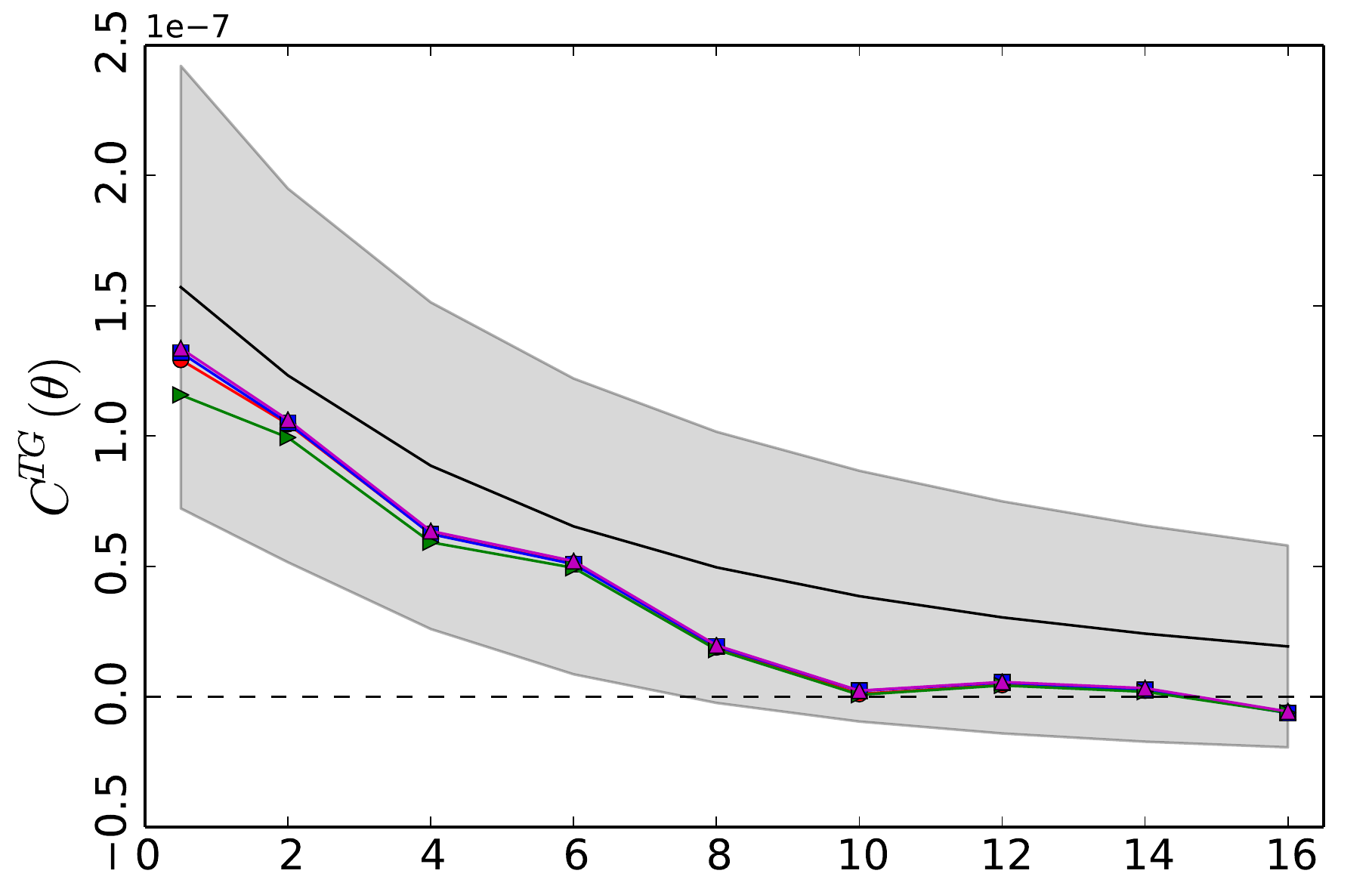}
\includegraphics[width=6cm]{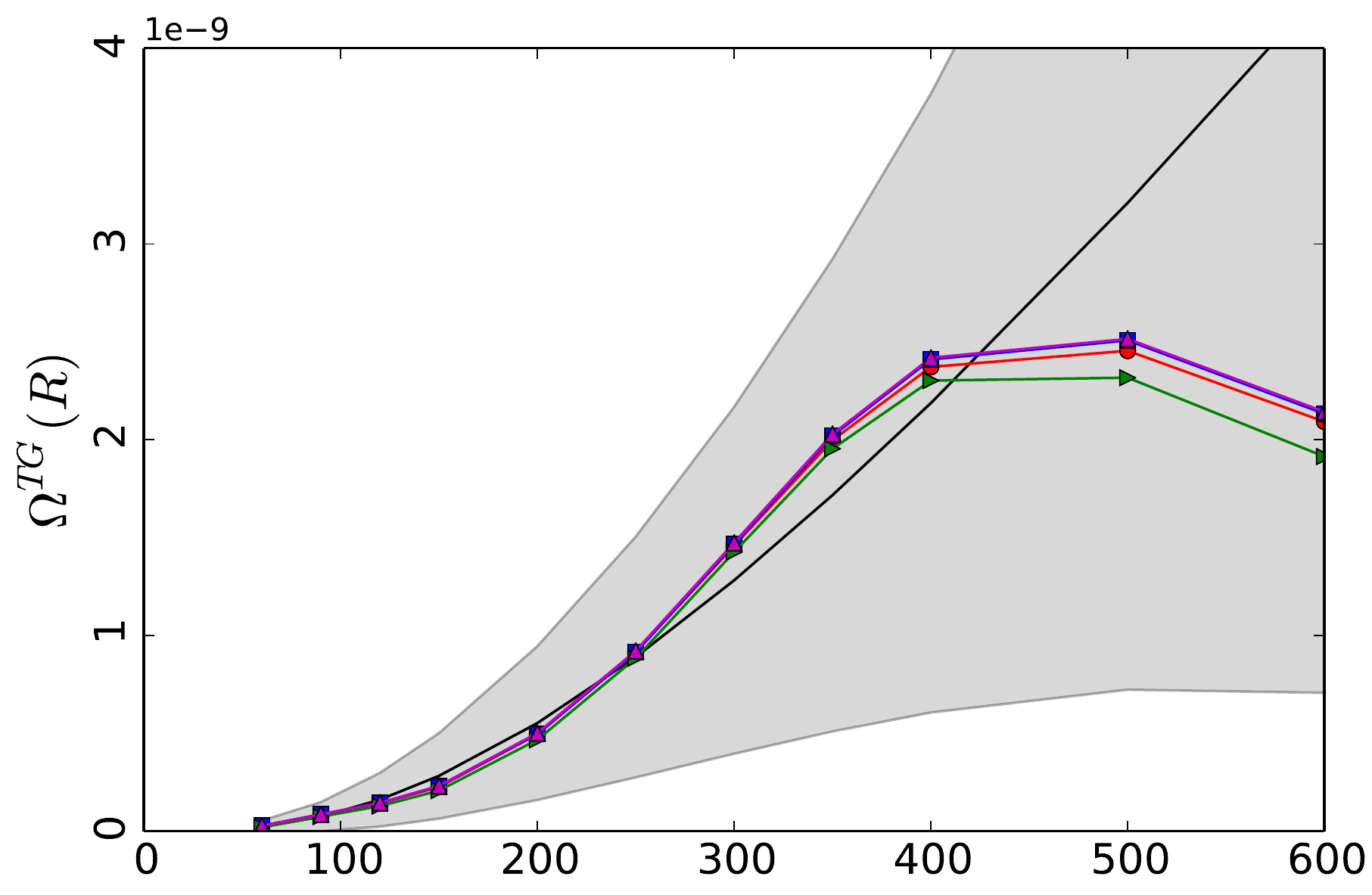}
\includegraphics[width=6cm]{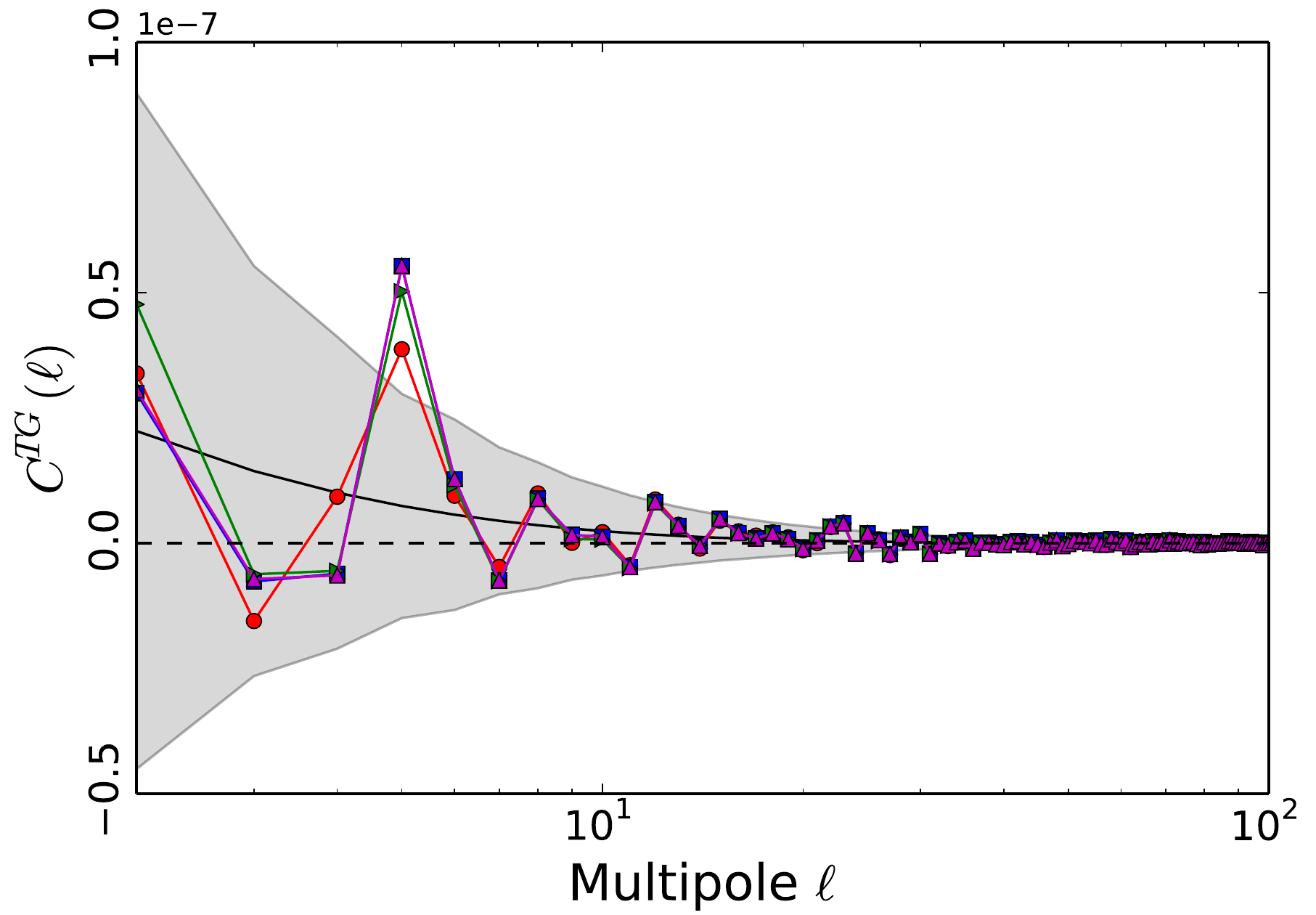}
\includegraphics[width=6cm]{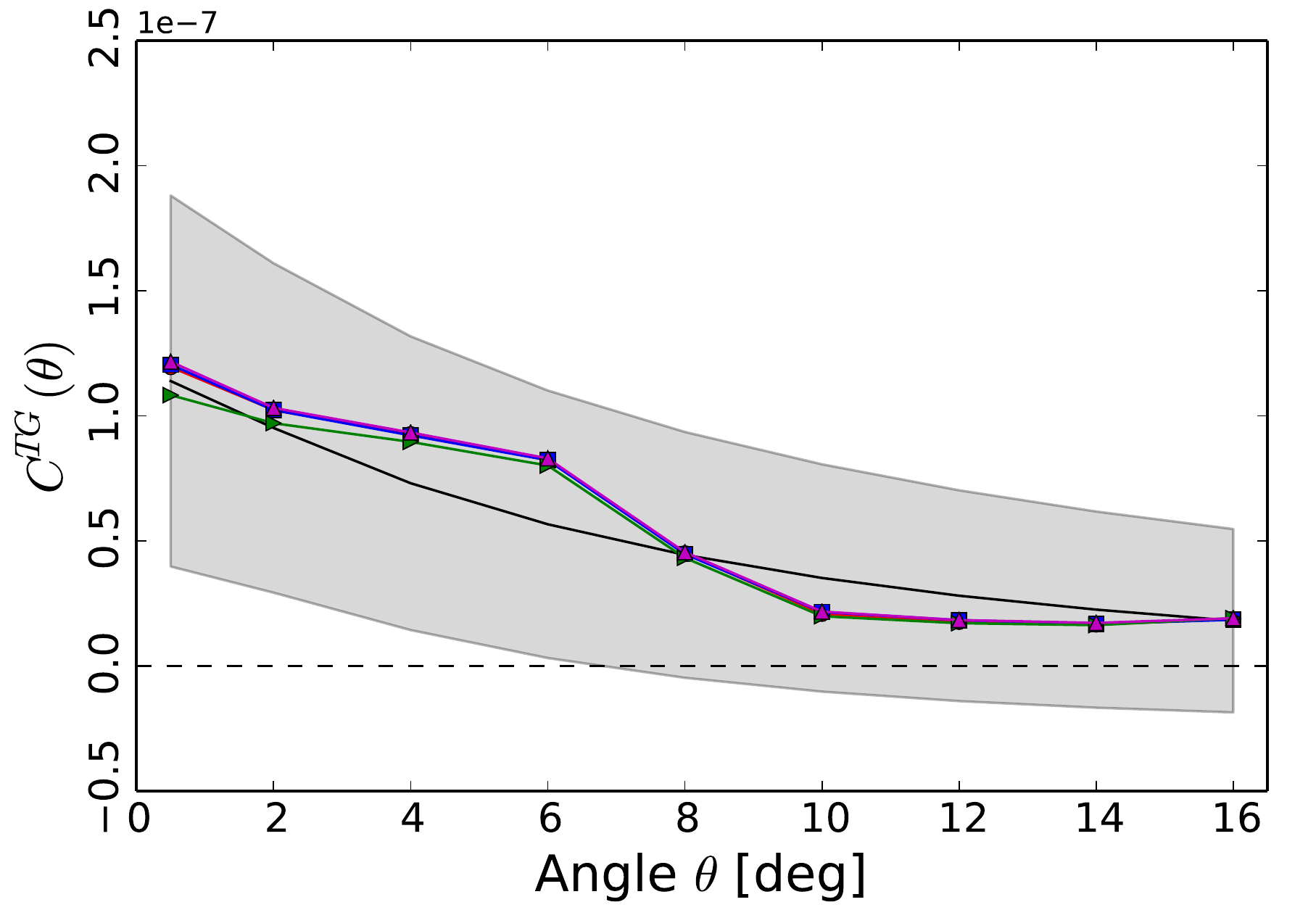}
\includegraphics[width=6cm]{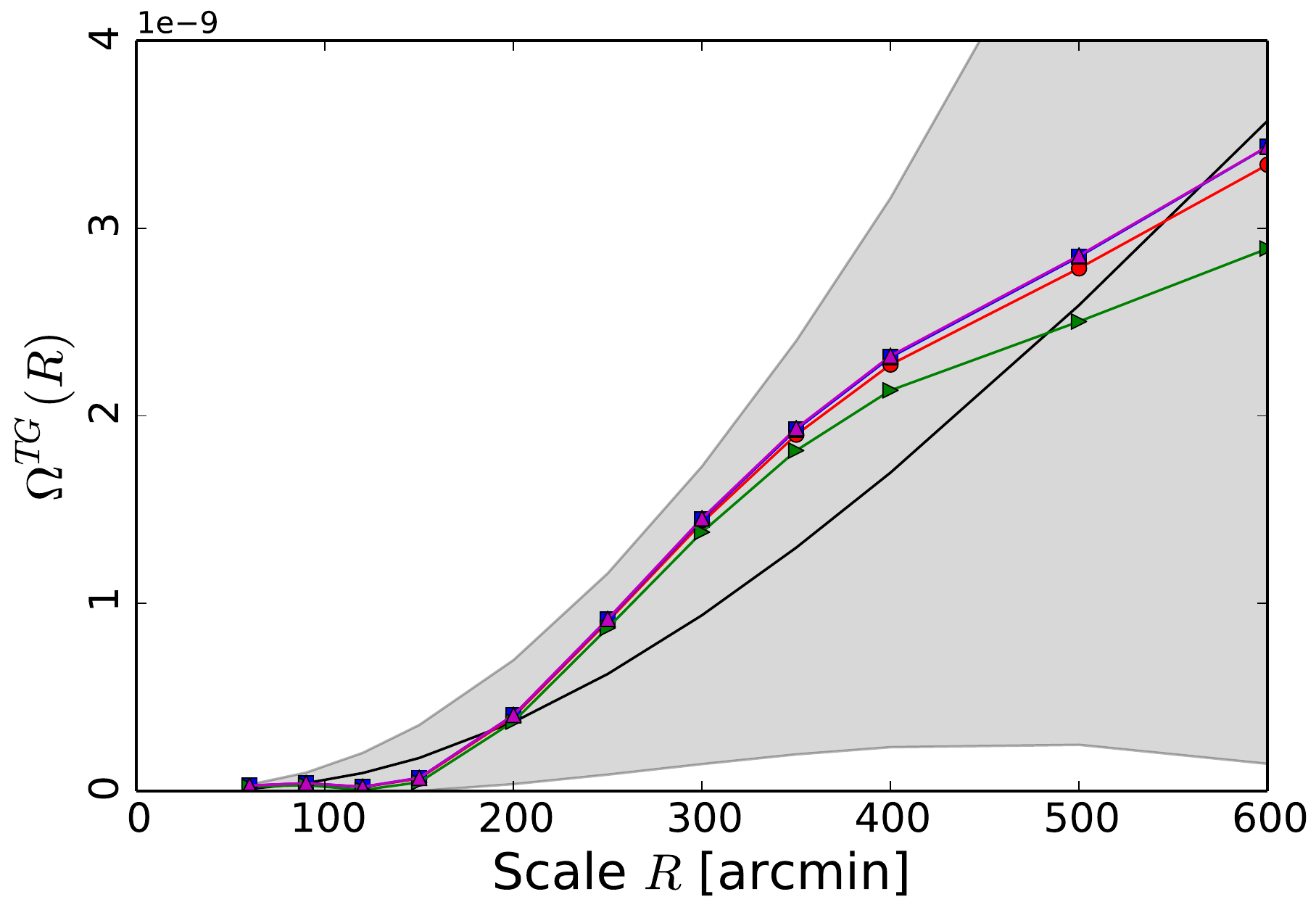}
\end{center}
\caption{\label{fig:xcorr}Observed and expected cross-correlation
signal versus multipole $\ell$, for several surveys and different
cross-correlation estimators.  Columns from left to right correspond to: CAPS;
CCF; and SMHWcov.
Rows from top to bottom represent: NVSS; \lrg; and \mg.  On each panel we
show the expected cross-correlation (black line) and the $\pm1\,\sigma$
region (grey area).} 
\end{figure*}

We have fitted the observed cross-correlations to the expected ISW signal
($C_\ell^{\rm TG}$, $C^{\rm TG}\left(\theta\right)$, and
$\Omega^{\rm TG}\left(R\right)$, see Fig.~\ref{fig:xcorr}), following
Eq.~\ref{eq:fit}, i.e., allowing for a free amplitude of the expected signal. 
Results are summarized in Table~\ref{tab:s2n_data}. 
Notice that the plots only show partial information on the total
detection level of the ISW signal, specifically coming from
the diagonal of the covariance matrix $\tens{C}_{\xi^{xy}}$.
The level of correlation among the components (i.e., the multipoles, angles
or scales) varies considerably for the different estimators, and, therefore
the whole picture is only
obtained when the full covariance is considered, as in Eq.~\ref{eq:fit}.

Overall, the ISW detection is at about the $3\,\sigma$ level and, as expected,
it is clearly dominated by the NVSS signal. There are only small differences
among estimators and CMB maps (as expected from the above discussion),
indicating that this is a robust result.
Notice that all the estimated amplitudes are compatible with unity, within the
error bars (especially for NVSS and \lrg). This is an additional validation of
how CMB and LSS are modelled. Values of $A$ deviating significantly from unity
would indicate some tension between the observed cross-correlation and the
model (in particular on the LSS modelling, which is more complex).
The CAPS-QML, applied to the \sevem\ and NVSS (i.e., the survey with the
highest signal-to-noise), yields a value of $A = 0.73\pm0.33$, which is
compatible with the CAPS, when applied to the same $N_\mathrm{side}=32$
and $\ell_\mathrm{max} = 2N_\mathrm{side}$ resolution ($A = 0.84\pm0.34$). 
Preliminary tests indicate that running the
CAPS-QML at $N_\mathrm{side}=64$ resolution could increase the sensitivity
for detecting the ISW effect with NVSS by $\approx 20\%$.

Our results indicate a somewhat smaller signal-to-noise with respect to some
previous analyses on \wmap\ data, where several (and in some case quite
similar) surveys were also considered. For instance,~\cite{Ho2008}
and~\cite{Giannantonio2012} found $3.7\,\sigma$ and $4.4\,\sigma$ detections,
respectively. Compatibility with the former is below $1\,\sigma$, whereas
there is more tension (around $1.5\,\sigma$) with the latter.
A fraction of around $0.3\,\sigma$ of these differences could be explained in terms
of the comological parameters adopted to defined the theoretical expectations.
In particular, the lower values of $H_0$ and $\Omega_\Lambda$ found by \Planck~\citep{planck2013-p11}
with respect to \wmap~\citep[e.g.,][]{Larson2011}, imply a sensitivity for the ISW $\approx 10\%$
smaller.  
The rest of the differences come either from the LSS side, or 
from the error characterization, which depends on the presence of a correlated
signal between CMB and LSS
simulations~\citep[see for instance][for a discussion]{Cabre2007}. 
Survey modelling is another important aspect: besides systematic errors
associated with the galaxy identification and redshift estimation procedures,
there are complicated aspects, such as the bias characterization. 
As was mentioned already, a strong point of our results, is the excellent
compatibility between the ISW amplitude estimates with respect to the
expected value. Whereas our estimation 
deviates by about $0.5\,\sigma$ from the expected value,
the~\cite{Giannantonio2012} result exceeds it by about $1\,\sigma$
and~\cite{Ho2008} is around $2\,\sigma$ above.

Nevertheless, the value of the ISW effect that we measure by means of NVSS
(that because of to its large sky coverage, 
redshift range, and density of galaxies is probably the best current
catalogue for studying the ISW effect) is significant, and in agreement with
previously published results using \wmap.

We have also studied the ISW signal from the point of view of its
compatibility with the null hypothesis. We have considered in this
analysis only the NVSS catalogue, since it provides the largest
detection of the ISW effect and, therefore, is the best of the existing
surveys for challenging the null hypothesis.  Probability values are
summarized in Table~\ref{tab:s2n_pvalue}. As mentioned in
Sect.~\ref{subsubsec:xcorrdef}, there is not a unique way of computing
the null hypothesis. Our approach follows Eq.~\ref{eq:xcorr_null},
where $\tens{D}_{\xi^{xy}}$ was computed out of 90{,}000 CMB
simulations that have been cross-correlated with the LSS data.
This matrix is used to
compute $\chi_\mathrm{null}^2$ from the data. This value is then
compared to its distribution for the null hypothesis, obtained from
1{,}000 realistic CMB simulations (FFP-6) uncorrelated with NVSS,
which have been processed in the same way as the \Planck\ data set.
CAPS provides the smallest probability value, but the null hypothesis is
rejected at about 10\% only; this result is not unexpected, since
an expected result since the ISW effect is weak.

The fact that the CAPS statistic provides tighter limits with respect to the
CCF and SMHWcov could have been anticipated.
In our implementation, the CAPS explores the maximum angular range allowed
for a given map, whereas the CCF and the
SMHWcov approaches are only evaluated at certain angles/scales. This
limitation is not an issue in the analysis devoted to estimating the ISW
amplitude, since these angles/scales are suitable for detecting the ISW.
However, in order to discard the null hypothesis, the larger the number of
``distances'' the better.

The previous approach is a frequentist one.  However, there is an
alternative way of addressing the null hypothesis compatibility, in the
framework of a hypotheses test.  In particular, we can study the ratio of
Bayesian evidence for both scenarios, comparing the alternative
(there {\it is\/} ISW signal) and the null (there is no correlation
between the CMB and the LSS) hypotheses.  To compute the Bayesian evidence we
just need the likelihoods and the priors associated with each hypothesis,
which are already available.  The likelihood for the alternative hypothesis is
obtained from Eq.~\ref{eq:xcorr_fit}, peaked at the best-fit value for the ISW
amplitude, and its prior could be described as a Gaussian probability peaked
at $A=1$ and with a dispersion given by $\sigma_A$ in Eq.~\ref{eq:fit}.
Conversely, the likelihood for the null hypothesis is the function given by
Eq.~\ref{eq:xcorr_null}, and its prior would be as the one for the alternative
hypothesis, but peaked at $A\equiv0$ (notice that this is justified, since,
as explained in the previous section,
$\tens{D}_{\xi^{xy}} \approx \tens{C}_{\xi^{xy}}$). Following this approach,
we obtain a Bayesian evidence ratio of around 30, which provides strong
support for the presence of the ISW signal.

\begin{table}[tmb]
\begingroup
\newdimen\tblskip \tblskip=5pt
\caption{Probability values of the CMB-LSS cross-correlation for the
NVSS survey under the null hypothesis, for the four component separation methods
and for the different
estimators.\label{tab:s2n_pvalue}
\label{tab:surveys}}
\nointerlineskip
\vskip -3mm
\footnotesize
\setbox\tablebox=\vbox{
   \newdimen\digitwidth 
   \setbox0=\hbox{\rm 0} 
   \digitwidth=\wd0 
   \catcode`*=\active 
   \def*{\kern\digitwidth}
   \newdimen\signwidth 
   \setbox0=\hbox{+} 
   \signwidth=\wd0 
   \catcode`!=\active 
   \def!{\kern\signwidth}
\halign{#\hfil\tabskip=0.2cm& \hfil#\hfil\tabskip=0.2cm&
\hfil#\hfil\tabskip=0.2cm& \hfil#\hfil\tabskip=0.2cm& \hfil#\hfil\tabskip=0.2cm& \hfil#\hfil\tabskip=0.cm\cr 
\noalign{\doubleline}
\noalign{\vskip -2pt}
LSS data& $\hat{\xi}_a^{xy}$& \ruler& \nilc& \sevem& \smica\cr 
\noalign{\vskip 3pt\hrule\vskip 5pt}
NVSS & CAPS& 0.09& 0.10& 0.10& 0.09\cr
& CCF& 0.33& 0.34& 0.40& 0.33\cr
 & SMHWcov& 0.20& 0.23& 0.27& 0.19\cr
\noalign{\vskip 3pt\hrule\vskip 3pt}}}
\endPlancktable                    
\endgroup
\end{table}

\begin{figure*}[!ht]
\begin{center}
\includegraphics[width=17.4cm]{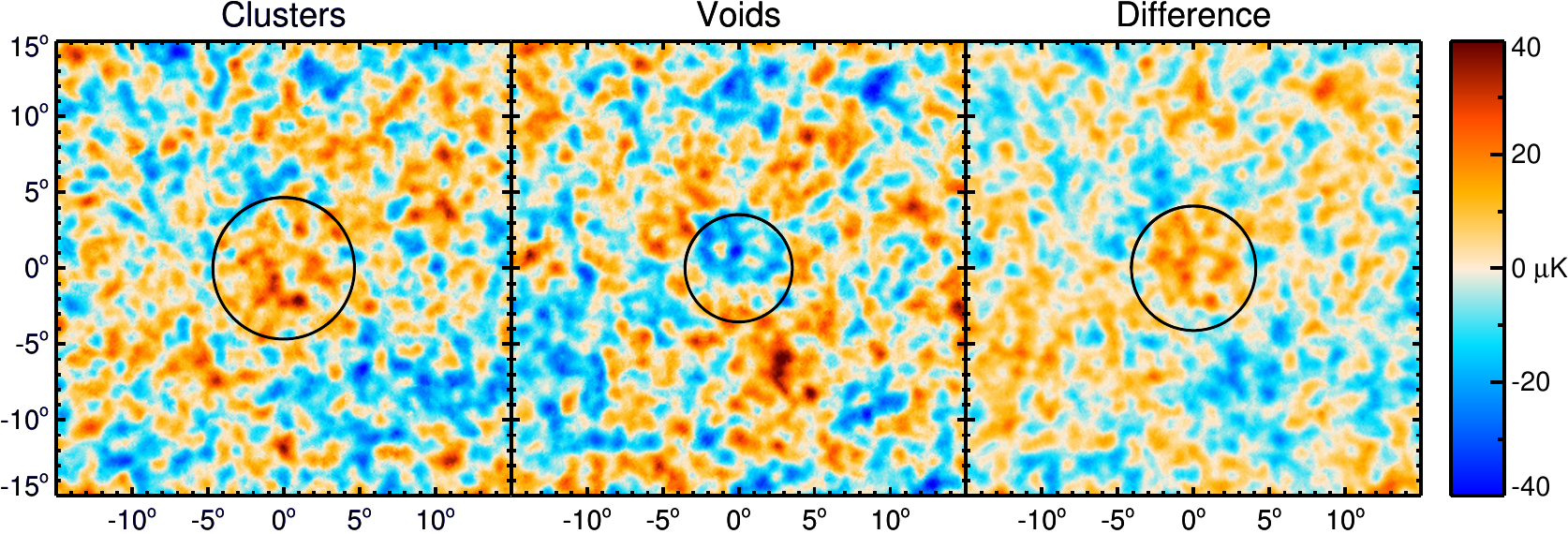}
\end{center}
\caption{\label{fig1_stack} Stacked regions of \Planck\ maps corresponding to
the locations of the superstructures identified by GR08.  
From left to right we show the images resulting from stacking of the 50
superclusters, the 50 supervoids, and the difference of both. 
The black circles superimposed indicate the angular radius at which the
signal-to-noise ratio is maximal.  See Fig.~\ref{fig2_stack} for the
corresponding temperature and photometry profiles, as well as their
statistical significance.
}
\end{figure*}

\begin{figure}[!ht]
\begin{center}
\includegraphics[width=0.95\hsize]{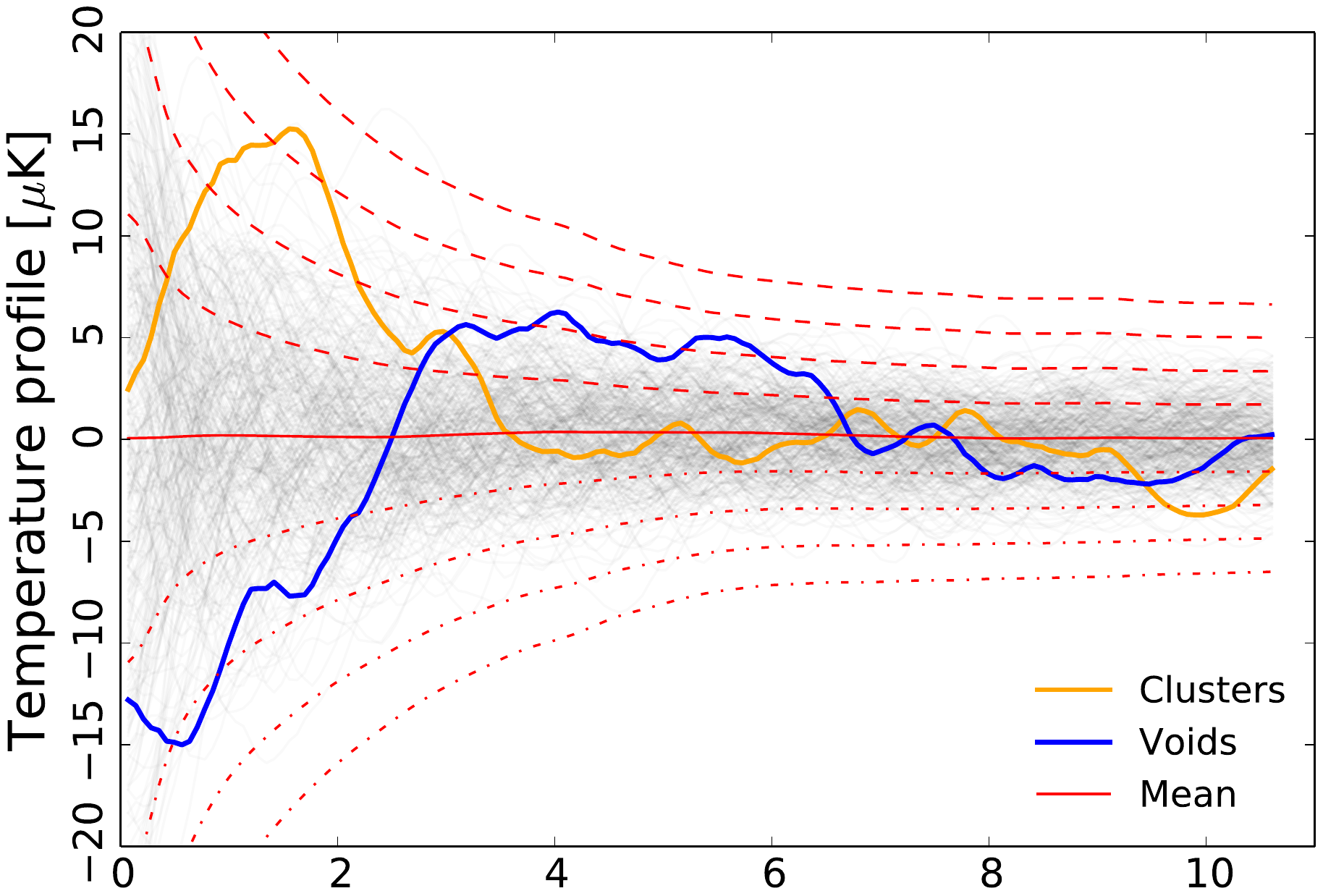}
\includegraphics[width=0.95\hsize]{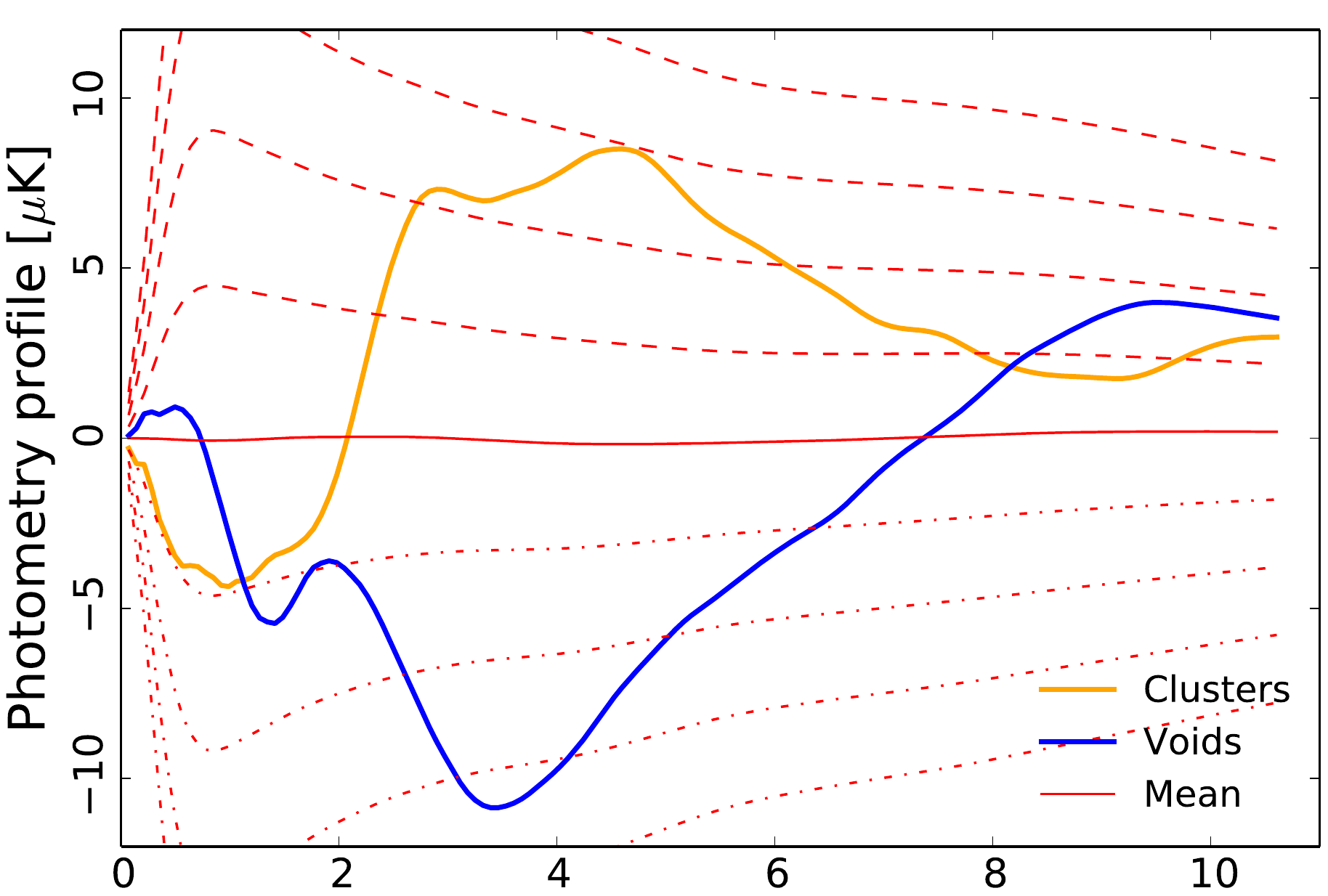}
\includegraphics[width=0.95\hsize]{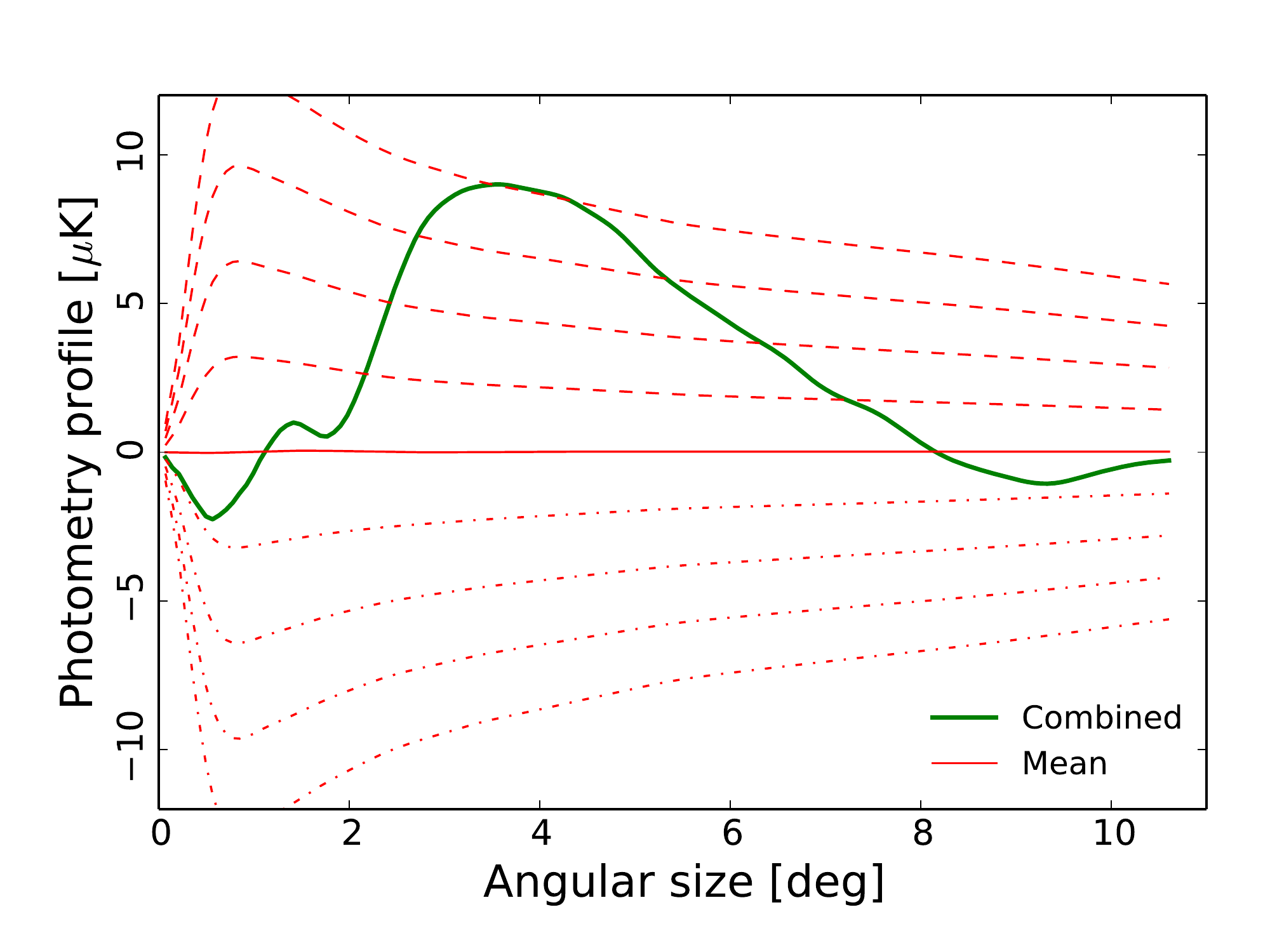}
\end{center}
\caption{\label{fig2_stack} Temperature (top) and photometry (middle and
bottom) profiles of the stacked CMB patches at the location of the 50
supervoids and 50 superclusters of GR08. The lower panel shows the combined
photometry profile (i.e., the average cluster profile minus the average
void profile). The significance is represented by 1, 2, 3, and $4\,\sigma$
level curves (dashed and dotted lines represent positive and negative error
bars, respectively). These curves represent the dispersion of the 16\,000
stacks of 50 CMB patches chosen at random positions (for illustration, on
the top panel, we represent in grey 300 of those random profiles).}
\end{figure}

\section{Stacking of large-scale structures}
\label{subsec:stack}

An alternative approach for measuring the ISW effect in \Planck\ maps is to
look for an ISW signal directly at the positions of positive and/or negative
peaks in the potential.  Since the expected (and observed) signal is very weak,
for individual structures, a stacking technique needs to be applied.
Using the {\it WMAP\/} data, it has been shown that CMB maps show hot spots
and cold spots in the direction of superclusters and supervoids, respectively
\citep[][GR08 hereafter]{Granett2008a,Granett2008b}, which appear to be barely
consistent with the predictions of standard $\Lambda$CDM
\citep[see also][]{Hernandez2012}. These structures, which are not yet
virialized, are evolving while the CMB photons travel across them and this
should contribute to the ISW effect.  We apply here the same approach to the
different \Planck\ maps, using the catalogues of superstructures introduced
in Sect.~\ref{sssect:catal}, and we test for the robustness of our findings.
We first discuss our method and the results obtained using the
catalogue provided by GR08, and then present the results obtained with the
other catalogues.

\subsection{Method}

\begin{figure}[!ht]
\begin{center}
\includegraphics[width=0.95\hsize]{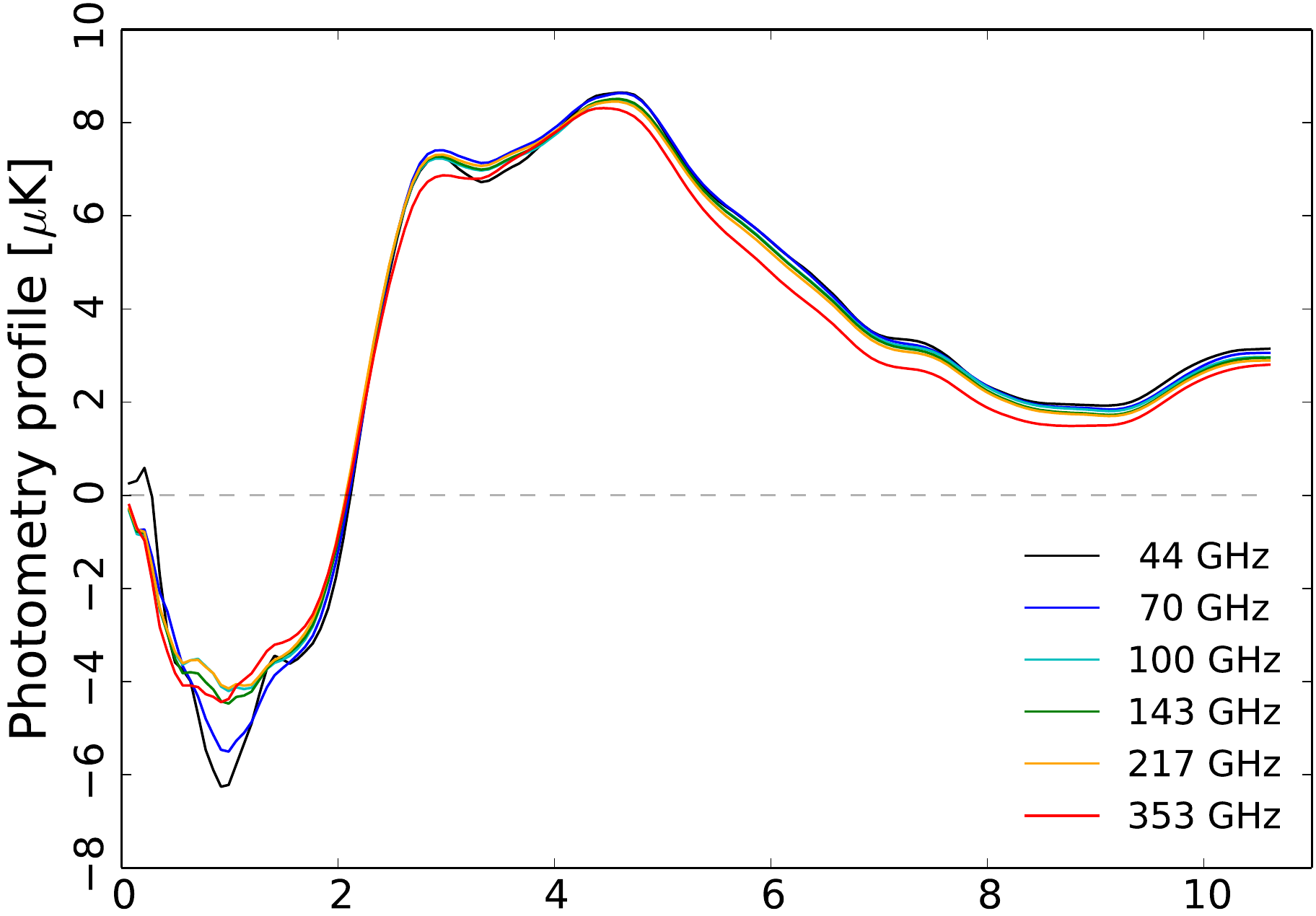}
\includegraphics[width=0.95\hsize]{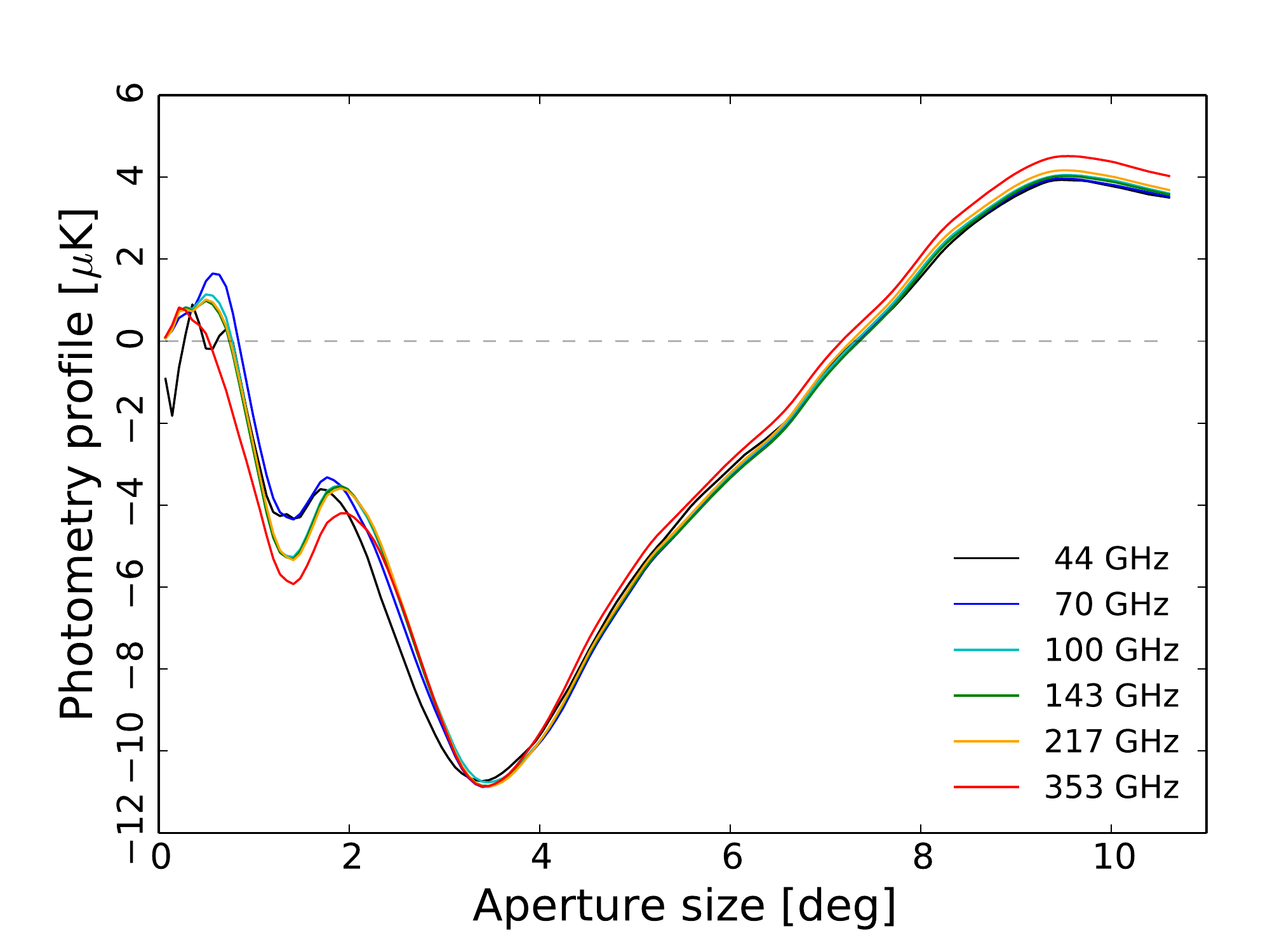}
\end{center}
\caption{\label{fig3_stack} Independence of the signal on the \sevem\
frequency channels.
The aperture photometry profiles measured in the stacked patches centred
on superclusters (top) and supervoids (bottom) are virtually identical for all
frequencies.}
\end{figure}

Our analysis is performed on the \smica\ CMB map, although we have checked
that results are compatible for the other three \Planck\ maps.
We have also used the cleaned frequency maps from \sevem\
(see Sect.~\ref{subsubsec:cmbmaps}) for some of the tests.
We first remove the monopole and dipole of the maps (outside the U73 mask), and
then apply a compact source mask based on the Planck Legacy Point Source
Catalogue~\citep{planck2013-p06} to remove the contamination from individual
point sources. 

For the purpose of comparison with the results of GR08, we smooth the
CMB maps with a common Gaussian kernel of 30\,\arcm\ FWHM. We then project them
onto patches around each position in the supervoid and supercluster catalogue.
The GR08 structures have a relatively small size on the sky (a few degrees),
but the other two catalogues considered here contain many larger and closer
voids, covering larger angular sizes. Thus we work with $30\deg\times30\deg$
CMB patches and choose the pixel size to be 6\,\arcm, so that all voids
considered are fully enclosed. We then co-add (stack) the maps, taking into
account the mask used.
On the stacked images, we calculate both the radial temperature profile and
the aperture photometry, to characterize the signal around density structures.
The temperature profile is obtained by computing the mean of the pixels in
rings of fixed width and increasing angular radius; in practice, it is
calculated for 150 radii between $0\deg$ and $15\deg$, with a width of
$\Delta \theta = 0\pdeg1$. The photometry profile is obtained by applying a
compensated filter that subtracts the average temperature of a ring from
the average temperature within the disk whose radius $\theta$ is the inner
radius ring, and where the outer radius is chosen to be $\theta\sqrt{2}$, so
that the disc and ring have the same area.  This should enhance
fluctuations of typical angular size $\theta$ against
fluctuations at smaller or larger scales.  Aperture photometry results are
also provide for at 150 angles, this time
between $0\deg$ and $15/\sqrt{2} \approx 10\pdeg6$.
In addition to the monopole and dipole, we also removed from the CMB maps
the contribution of large scale angular modes, namely $\ell=2$--10.
These modes correspond to angular scales much larger than those of the
structures under investigation, and for our purposes their only effect
is to introduce gradients in the stacked images; the high-pass filter
essentially stops such gradients getting into the stacked map (which is
equivalent to removing gradients at the end).
The contribution of the large-scale angular modes has no impact on the aperture
photometry profiles, and introduces only an offset in the temperature
profiles \citep{Ilic2013}. 

In order to estimate the significance of the resulting photometry and
temperature profiles, we follow a Monte Carlo approach based on stacked CMB
images chosen at random positions. In detail, we compute the photometry and
the temperature profiles for 16\,000 sets of 50 CMB patches randomly
distributed over the SDSS area. We then compare the profiles obtained from
the stacking at the location of the GR08 superstructures to
these random profiles, in order to compute their signal-to-noise ratio.

\subsection{Results}

\begin{figure}[!ht]
\begin{center}
\includegraphics[width=\hsize]{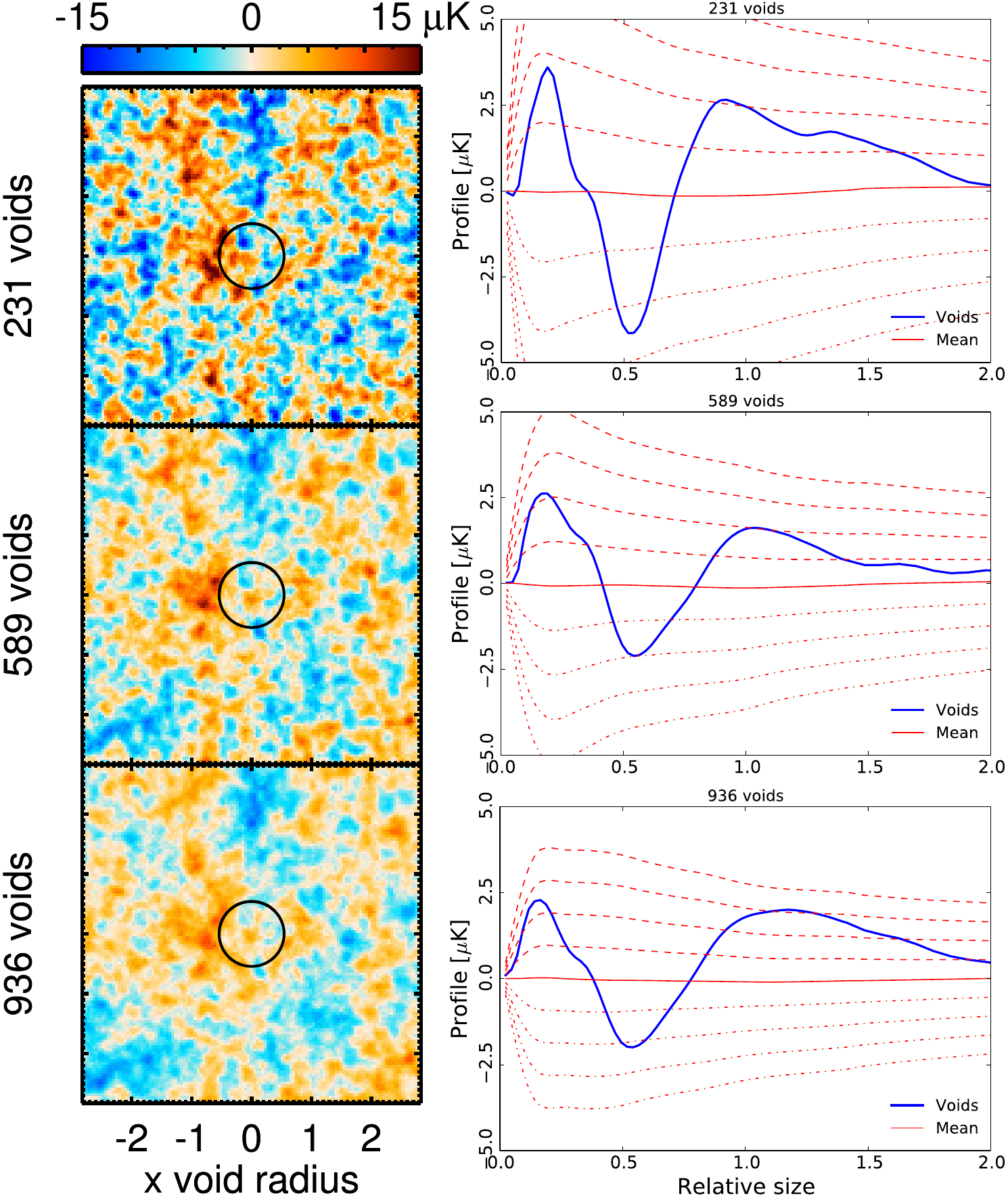}
\end{center}
\caption{\label{fig4_stack} Stacked images (left) and photometry profiles
(right) obtained for the voids of~\cite{Sutter2012} after rescaling. From top to bottom, we
show results obtained from stacking the largest 231, 589, and 936 voids.
The black circles superimposed indicate the angular radius, in terms of the effective radius after rescaling, at which the signal
is maximal. The statistical dispersion is reduced as we stack more voids.
However, the amplitude of the ``cold spot'' at about 0.5 times the effective
void radius is mostly due to the surrounding ``hot shell'', which is easiest
to see in the bottom panel.  This is further demonstrated by the $3\,\sigma$
signal detected using aperture photometry, seen at radii above 1.2 times the
effective radius (for the 936 void case). Dashed and dotted lines in the
right-hand panels represent positive and negative error bars, respectively,
from $1$ to $4\,\sigma$.}
\end{figure}

We show in Fig.~\ref{fig1_stack} the stacked images of the 50 supervoids and
50 superclusters of GR08 in the \Planck\ map.
The corresponding temperature and photometry profiles, along with their
significance levels, are shown in Fig.~\ref{fig2_stack}. 
The first thing to say is that, although the signatures are fairly weak,
the sign of the effect certainly seems to be correct.
Using the same catalogue and the \Planck\ CMB map, we find reasonable
agreement with GR08. The maximal photometric decrement, $ -10.8\,\mu$K
(essentially identical with the $-11.3\,\mu$K found by GR08),
induced by supervoids is obtained for a preferred scale of about $3\pdeg5$
($4\deg$ in GR08) and a signal-to-noise of 3.3 ($3.7\,\sigma$ in GR08),
as shown in Fig.~\ref{fig2_stack}. Superclusters produce a photometric
increment of about $8.5\,\mu$K (slightly above the $7\,\mu$K in GR08),
with a significance of $3.0\,\sigma$ (compared with $2.6\,\sigma$ in GR08)
at a slightly larger angle of $4\pdeg7$. Finally, the stack of the combined
sample (clusters minus voids) gives a temperature deviation of $8.7\,\mu$K,
with a signal strength of $4.0\,\sigma$ at $4\pdeg1$, which is consistent with
the values reported in GR08.  The values of statistical
significance for our aperture aperture photometry results are closely
related to those for the temperature profiles. Indeed, as shown in the top
panel of Fig.~\ref{fig2_stack}, the temperature profile for the void stack
shows a roughly $2\,\sigma$ deficit at small angular radii and a roughly
$2\,\sigma$ excess extending to large radii.  Since 
the aperture photometry is essentially an integral of the temperature profile
with a compensated filter, it picks up enhanced significance because of the
shape of the temperature profile.

As noted previously by several authors~\citep[e.g.,][]{Hernandez2012}, the 
amplitude and shape of the 
photometric profile found for voids and clusters is in tension (around $2\,\sigma)$ with the values
expected from pure ISW within $\Lambda$CDM.  However, it is not
straightforward to associate this entire signal with a pure ISW effect.
As seen in the Fig.~\ref{fig1_stack}, many small-scale structures -- both
cold and hot -- are present around the region delineated by the angular radius
at which the signal-to-noise ratio of the aperture photometry is maximal.
This small-scale structure contributes to the amplitude of the photometric
decrement, but at a few tens of $\mu$K, which is incompatible with the
$\Lambda$CDM predictions for the ISW effect. These are rather simply
background CMB fluctuations, with their lingering presence due to the small
number (50) of patches which are used to produce the stack.

It is intriguing that the angular sizes of the catalogued superstructures
are smaller than the angular sizes suggested by the photometry profiles.
This result is more apparent when we repeat the stacking analysis after
rescaling each CMB patch by the effective radius of the structure it contains.
Since the voids and super clusters identified by GR08 are roughly the same size,
the photometric results are similar after rescaling ($-10\,\mu$K for voids
and $7.9\,\mu$K for superclusters). However, the deviations have significance
levels of $3.3\,\sigma$ and $2.7\,\sigma$ for supervoids and superclusters,
respectively, at angular sizes of $0.9$ (voids) and $2.6$ (clusters) times the
effective radius of the structures. This mismatch, especially for clusters, could be a result of
underestimation of the structure extent the {\tt ZOBOV} and {\tt VOBOZ}
algorithms (as already suggested by GR08) or because larger potential hills
and valleys underlie the detected superstructures. Since structure in
the potential is related to the density field through the Poisson equation,
gravitational potential features are expected to cover larger scales than
structures in the density field.  Nevertheless, the factor of $2.6$ for
the case of superclusters seems large.  It is also true that
the GR08 superstructures were identified in the LRG subsample of the SDSS, and
LRGs are known to be biased tracers of the matter density field
\citep[e.g.,][]{Tegmark2006}.  This bias could help explain why structures
are larger than the scales identified by the {\it VOBOZ} algorithm, although
the argument would go in the opposite sense for the voids.
Another way of stating this is \citep{Hunt2010} that the relatively
large effect decrement found for the GR08 voids can be only be
attributed to the ISW effect only if the underdensities have been
significantly underestimated or if the LRGs are under-biased.

It is therefore difficult to be confident that the signal is
due entirely to the ISW effect. We know, however, that the ISW signal generated
by superstructures is expected to be achromatic, since it generates a
fractional perturbation of the CMB temperature. In order to check if the
signal we measure is indeed independent of frequency, we applied the same
technique to \Planck\ individual \sevem\ cleaned frequency maps from 44 to
353\,GHz. Lower (higher) frequency maps may be contaminated by radio (IR)
signals coming from our Galaxy and may thus introduce a bias in the
measurement. Figure~\ref{fig3_stack} shows the photometry profiles of
supercluster- and supervoid-stacked maps at 44, 70, 100, 143, 217, and
353\,GHz. The flux measured appears to be quite constant, which supports
the idea that the signal is due to the ISW effect induced by structures.
In the remainder of this section, we therefore apply our analysis only to
the \smica\ CMB map.
 
\subsection{Discussion and alternative catalogues}

It should be remembered that although the GR08 structures
are considered to be good tracers of the cosmic matter distribution on scales
larger than $10\, h^{-1}$\,Mpc, they are also known for their sparsity at
these redshifts ($z\approx0.4$--0.7). This sparsity could lead
to biased estimates of the properties of the reconstructed voids, in
particular their sizes and depths could be biased. Moreover, some of the
structures overlap on the sky, which could lead to systematic effects in the
stacking analysis.

We thus turn to other samples, for example
the catalogue of \citet{Pan2012}, introduced in Sect.~\ref{sssect:catal}.
The 1054 statistically significant voids it contains are larger than
$10\,h^{-1}$\,Mpc in radius and, with redshifts lower than 0.1, they are
much closer to us than the structures of GR08.  Direct stacking gives only a
weak signal at about the $1\,\sigma$ level, which is difficult to reconcile
with the previous results. This may be due to the inclusion of a large number
of small voids that could dilute the signal. Also, unlike the voids of GR08,
the voids of \cite{Pan2012} have a large scatter in angular sizes on the sky,
from about $2\deg$ to $20\deg$ \citep[e.g.,][]{Ilic2013}.
In order to try to enhance the signal, we repeated the stacking after rescaling
the voids to their effective sizes. We also subdivided the catalogue into
sub-samples based on redshift, radius, and/or angular size.
However, none of these attempts yielded any statistically significant result,
in agreement with \cite{Ilic2013}.

\begin{figure}
\begin{center}
\includegraphics[width=0.95\hsize]{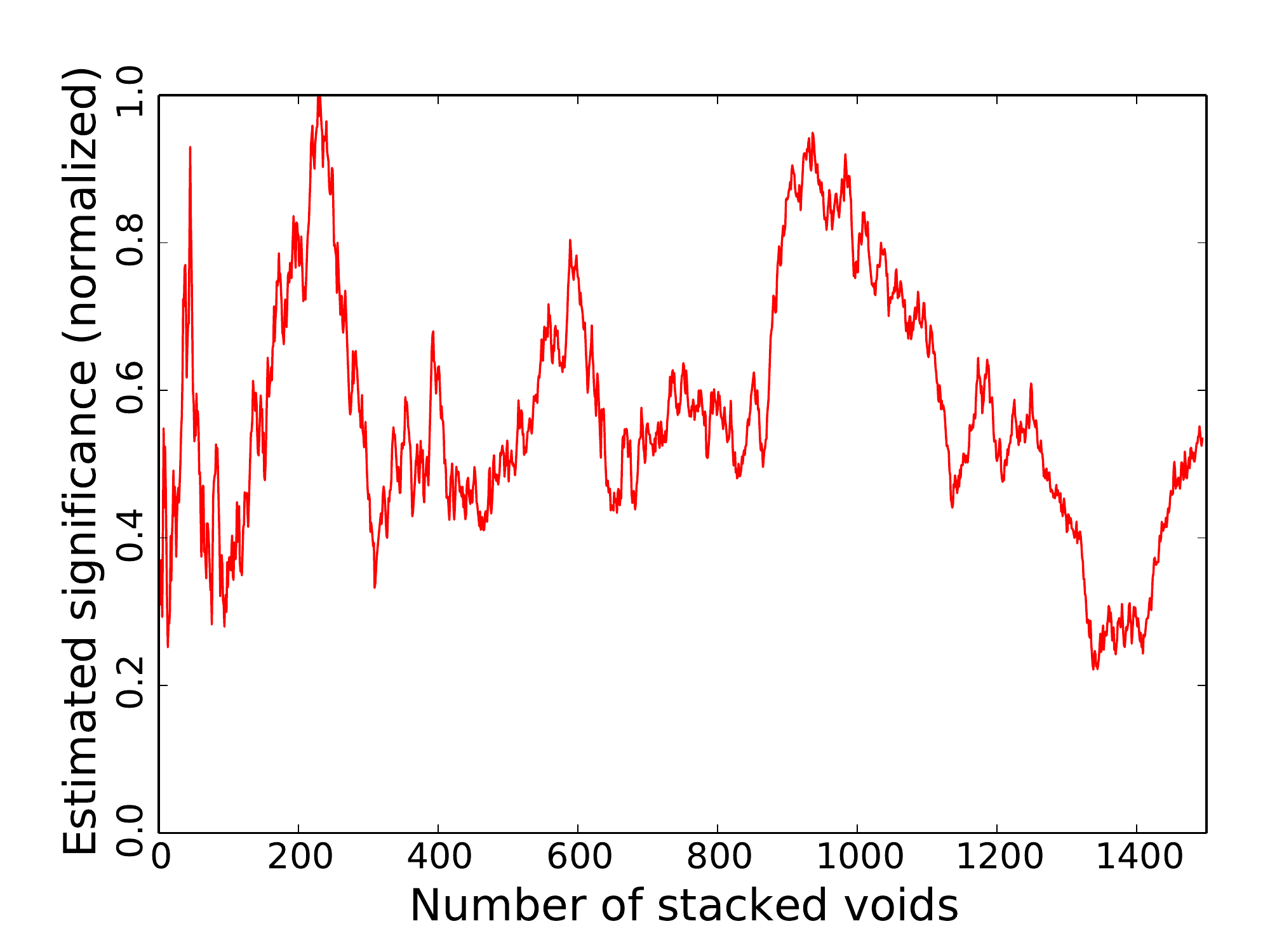}
\end{center}
\caption{\label{fig5_stack} Estimate of the optimal number of patches/voids
to stack using the catalogue of \citet{Sutter2012}. Starting from the largest
void and adding one CMB patch at a time to the stack, we find at each step
$N$ the minimum of the aperture photometry profile, and we multiply this value
by $\sqrt{N}$ to find the largest signal-to-noise, assuming that the noise
scales roughly as $1/\sqrt{N}$. The vertical axis has been normalized to the
best signal-to-noise, obtained for 231 voids.}
\end{figure}

Finally, we applied our procedure to the catalogue of voids published by
\citet[][]{Sutter2012}.  These cover a rather extended range of angular
scales (about 2--$10\deg$), and so we rescaled all the CMB patches by the
effective radius of each void.  Stacking
subsample by subsample (\textit{dim1}, \textit{dim2}, \textit{bright1},
\textit{bright2}, \textit{lrgdim}, \textit{lrgbright}), does not yield
any significant signal. Similarly, when stacked together, the entire catalogue
does not yield an ISW detection. However, since the ISW signal is expected to
be stronger for the largest and closest voids \citep[e.g.,][]{Flender2013}
we tried starting from the largest void and adding them one by one, looking
for the optimal number of voids, i.e., that for which the signal-to-noise
ratio is maximal (see Fig.~\ref{fig5_stack}). We found that stacking 231,
589 or 936 voids gives roughly the same signal-to-noise ($2.5\,\sigma$,
$2.0\,\sigma$ and $2.2\,\sigma$, respectively). However, the more voids we
stack, the smaller the amplitude of the photometry signal
(see Fig.~\ref{fig4_stack}, this being about $-2.0\,\mu$K for 936 voids,
$-2.1\,\mu$K for 589 voids and $-4.1\,\mu$K for 231 voids, at an angular size
of about 0.5 times the common rescaled radius. These amplitudes are lower than
those found with the 50 GR08 voids, although above what is expected from
numerical simulations \citep[see e.g., ][for higher
redshift and larger voids]{Hernandez2012}, but see also \cite{Cai2013}.

The apparent angular size detected (about $0.5$ times the effective void
radius) in the photometry profile is smaller than that for the GR08 voids,
but in agreement with expectations from simulations \citep{Cai2013}, and
consistent with the sizes detected using {\it WMAP\/} data \citep{Ilic2013}.

The profiles in Fig.~\ref{fig4_stack} show \modif{intriguing} hints (significance about
$2\,\sigma$) of a positive excess below about 0.2 times the effective void
radius.  \modif{This is somewhat surprising, since the centre of each void of
the \citet{Sutter2012} catalogue has been defined as the volume-weighted
barycentre of all the galaxies contained in the void volume.  This definition
in principle removes most of the bias that may arise in other methods, such as
choosing the most underdense galaxy position, which would lead to a slight
overdensity at the centre.}
Fig.~\ref{fig4_stack} also shows positive excess for larger
apertures, partly caused by the large ``hot ring'' surrounding the cold
feature in the stacked images, which raises the mean temperature of the
stacked image for discs of radii around 0.8--1.2 times the void radius.

\section{ISW map recovery}
\label{subsec:recov}

In recent years, some effort has been invested, not only to
obtain the statistical cross-correlation signal between the CMB and LSS
data, but also to recover a map of the ISW signal itself
\citep{Barreiro2008,Barreiro2013,Francis2010,Dupe2011}.
In particular, assuming the existence of a correlation between the CMB and the
gravitational potential, it is possible to recover a map of the
ISW fluctuations using a filtering method, given a tracer of the
gravitational potential (e.g., the galaxy catalogues described in
Sect.~\ref{subsec:extdata}) and the CMB fluctuations. Given the
weakness of the signal, the main objective
of this section is to provide a qualitative image of the ISW
fluctuations for visual inspection, and  an additional
consistency test of the validity of the assumed fiducial model, by
comparing the statistical properties of the recovered and expected
signals. In addition, this secondary anisotropy map could also be
used to study the large-scale properties of the CMB, and its possible
relation to some possible large-angle anomalies found in the \Planck\ data
\citep{planck2013-p09}.

\subsection{Method}

We have followed the methodology of \cite{Barreiro2008}, which applies
a linear filter to the CMB and to a gravitational potential tracer map,
in order to reconstruct an ISW map, assuming that the cross- and
auto-spectra of the signals are known. This technique has been
recently applied to reconstruct the ISW map from the {\it WMAP\/} data and
NVSS galaxy map~\citep{Barreiro2013}. The filter is implemented in
harmonic space and the estimated ISW map $\hat{s}_{\ell m}$ at each
harmonic mode is given by (see \citealt{Barreiro2008} for details)
\begin{equation}
\label{eq:rec}
\hat{s}_{\ell m}=\frac{L_{12}(\ell)}{L_{11}(\ell)}g_{\ell m} + \frac{L_{22}^2(\ell)}{L_{22}^2(\ell)+C_{\ell}^\mathrm{n}}\left( d_{\ell m}-\frac{L_{12}(\ell)}{L_{11}(\ell)}g_{\ell m}\right),
\end{equation}
where $L(\ell)$ corresponds to the Cholesky decomposition of the
covariance matrix between the considered tracer of the potential and
the ISW signal, at each multipole, which satisfies
$C(\ell)=L(\ell)L^{\rm T}(\ell)$.
Here $d_{\ell m}$ and $g_{\ell m}$ are the CMB
data and the gravitational potential tracer map, respectively, and
$C_{\ell}^\mathrm{n}$ is the power spectrum of the CMB signal without including
the ISW effect. If full-sky coverage is not available, the covariance matrix is
obtained from the corresponding pseudo-spectra. It can be shown that
the expected value of the power spectrum for the reconstructed signal
is given by
\begin{equation}
\label{eq:cl_lcb}
\left< C_\ell^{\hat{\mathrm{s}}} \right>= \frac{(C_\ell^\mathrm{gs})^2\left(\left|C(\ell)\right|+C_\ell^\mathrm{g}
C_\ell^\mathrm{n}\right)+\left|
C(\ell)\right|^2}{C_\ell^\mathrm{g}\left(\left|C(\ell)\right|+C_\ell^\mathrm{g} C_\ell^\mathrm{n}\right)},
\end{equation}
where $|C(\ell)|$ is the determinant of the tracer-ISW covariance
matrix at each multipole, and $C_\ell^\mathrm{gs}$ and
$C_\ell^\mathrm{g}$ are the
assumed cross-spectrum and gravitational potential tracer
spectra, respectively. Note that the recovered
ISW power spectrum will not contain the full ISW signal, since it can
only account for the part of the ISW signal probed by the tracer being
considered.  It is also worth noting that in detail the expected
cross-correlation depends on the assumed model.  However, in practice,
given the weakness of the signal, it would be difficult to distinguish
between two mild variants of the standard $\Lambda$CDM model.
Nevertheless this approach still provides a useful consistency check.

\subsection{Results}
\label{subsec:iswmapresults}

\begin{figure*}
\centering
\includegraphics[width=17.cm]{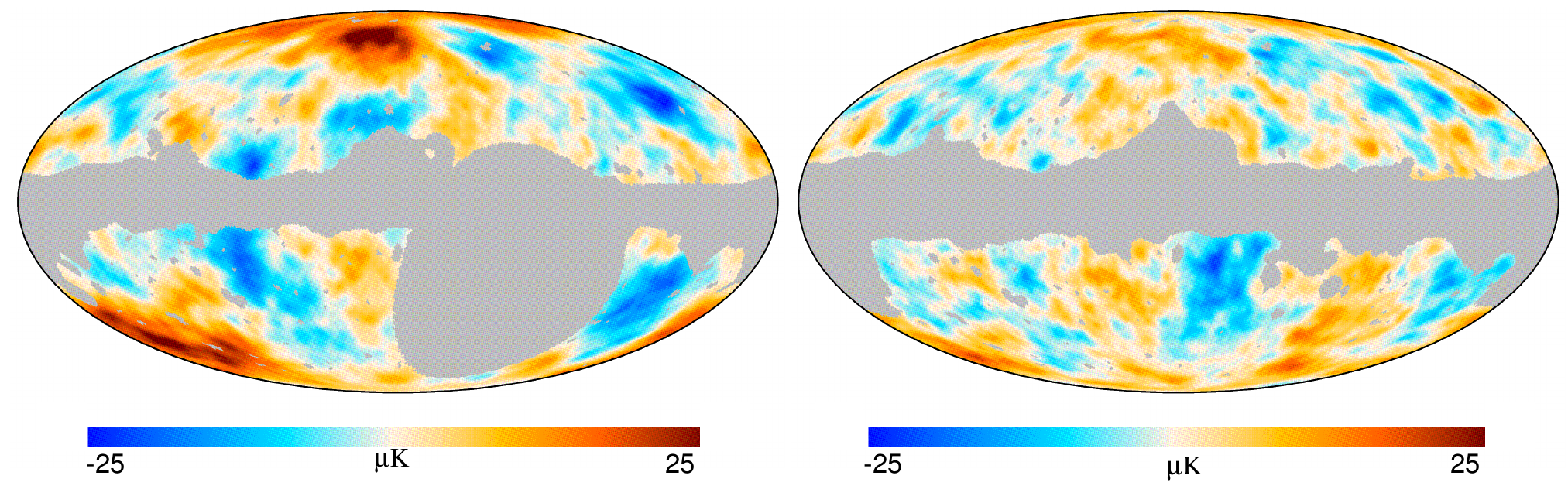}
\caption[fig:nvsssky]{Reconstructed ISW map from the \Planck\ CMB
  and NVSS data (left) and from the \Planck\ CMB and lensing potential maps
  (right). Note that the maps are not expected to look exactly the same,
  since each of them provides a partial reconstruction of the noisy ISW
  signal (see Sect.~\ref{subsec:iswmapresults} for details).  }
\label{fig:iswmap}
\end{figure*}
We have applied the filter described above to two different cases:
combining information from the CMB and the NVSS
galaxy catalogue; and applying the filter to
the CMB and the recovered lensing potential map described in
\ref{subsub:lens}. Results have been obtained for the four \Planck\ maps,
\ruler, \nilc, \sevem, and \smica. For simplicity, we show the
reconstructions only for the \sevem\
CMB map, since the four methods give very similar results. The
resolution considered for both analyses is $N_\mathrm{side}=64$.

For the first case, we are using the \Planck\ fiducial
model for the CMB and cross-power spectrum, while for the NVSS map we
assume the model described in Sect.~\ref{subsub:nvss}. We also take
into account the presence of Poissonian noise. We have excluded
the area obtained from combining the CMB mask at
$N_\mathrm{side}=64$ (described in Sect.~\ref{subsubsec:cmbmaps})
as well as the
area which has not been observed by NVSS. The final mask keeps around
62\% of the sky. Since the filter is constructed in harmonic space, we
have used an apodized version of the mask in order to reduce the
mask-induced correlations.  In any case, the degradation
introduced by the presence of a mask is small \citep{Barreiro2008}.

For the second case, the lensing map involved
applying a high-pass filter, which removed all
multipoles with $\ell < 10$. This filtering was done in harmonic
space with the presence of a mask. To take this effect into
account we used a direct estimation of the pseudo-power spectrum of these
data for the power spectrum of the lensing map,
after applying the corresponding apodized mask.
We used the \Planck\ fiducial model for the other
power spectra involved, but setting to zero the cross-power for $\ell < 10$.
A mask has been constructed by combining the CMB mask plus that
provided for the lensing potential map (described in
\citealt{planck2013-p12}), which keeps around 67\% of the sky. The
corresponding apodized version of this mask was applied before
reconstructing the ISW map. Note that the map given in
Fig.~\ref{fig:cmb_lensing_maps} (right panel) corresponds, to a good
approximation, to the first term of the right hand side of
Eq.~\ref{eq:rec}.

Figure~\ref{fig:iswmap} shows the reconstructed ISW map using the \Planck\ CMB
map and NVSS (left panel) and that obtained combining the CMB with the
lensing potential map (right panel).
There are similar structures present in both maps, but they are
not expected to look exactly the same, since each of them provides only a
partial reconstruction of the ISW signal.
This is due to the fact that the reconstruction accounts for
the part of the ISW effect probed by the considered tracer, which is
different (although correlated) for each case. Moreover,
due to the high-pass filter applied to the lensing potential map, the
power at $\ell < 10$ for this case corresponds to the Wiener-filtered
map of the CMB (to which the filter given by Eq.~\ref{eq:rec} defaults,
if the cross-correlation is set to zero, as in this case), without
additional information from the considered tracer.

For both cases, we have tested that the power spectrum of the
recovered ISW signal, as well as that of the cross-power between the
reconstructed ISW and the considered gravitational potential tracer,
are consistent with the corresponding expected values. This indicates
the compatibility between the assumed fiducial model and the
underlying statistical properties of the data.


\section{Conclusions}\label{sec:discussion}

This paper presents the first study of the ISW effect using \Planck\ data.
We derived results based on three different approaches: the detection of the
interplay between weak lensing of the CMB and the ISW effect, by looking at
non-Gaussian signatures; the conventional cross-correlations with tracers of
large-scale structure; and aperture photometry on stacks of the CMB field at
the positions of known superstructures.  A reconstruction of the ISW map
inferred from the CMB and LSS tracers was also provided.

The correlation with lensing allows, for the first time, the detection of the
ISW effect using only CMB data. This is an effective approach, because the
gravitational potential responsible for deflecting CMB photons also generates
ISW temperature perturbations. Using different estimators, we investigated
the correlation of the \Planck\ temperature map with a reconstruction of the
lensing potential on the one hand, and the estimation of the ISW-lensing
generated non-Gaussian signature on the other. 
We found that the signal strength is close to $2.5\,\sigma$, for several
combinations of estimator implementation and foreground-cleaned CMB maps.

We computed cross-correlations between the \Planck\ CMB temperature map,
and tracers of large-scale structure, namely: the NVSS survey of radio
sources; and the \lrg, and \mg\ galaxy samples. As estimators we considered
the angular cross-correlation function, the angular cross-spectra,
and the variance of wavelet coefficients as a function of angular scale. 
We performed a comparison on different component-separation maps, where we
considered \ruler, \nilc, \sevem, and \smica, and found remarkable agreement
between the results, indicating that the the low multipoles are robustly
reconstructed. Covariance matrices between the cross-correlation quantities
were estimated for a set of Gaussian realizations of the CMB for the \Planck\
fiducial model. For the ISW effect, we report detection significance levels
of $2.9\,\sigma$ (NVSS), $1.7\,\sigma$ (\lrg), and $2.0\,\sigma$ (\mg), which
are consistent among the different estimators considered.  Although these
numbers are compatible with previous claims which used \wmap\ data, they are
generally smaller. We believe that this discrepancy is mainly due to the
different characterization of the surveys and treatment of uncertainties,
since the measurement of the CMB fluctuations at the scales which contribute
to the ISW detection are very similar for \Planck\ and \wmap. Only a fraction
of these differences (around $0.3\,\sigma$) could be understood in terms of the different cosmological models used by each experiment -- in particular, the
lower values of $H_0$ and $\Omega_\Lambda$ reported by \Planck\ compared with
{\it WMAP}.

A strength of our new study lies in the fact that the amplitudes derived for
the expected signals are largely consistent with unity (i.e., the model
expectation), which indicates good modelling of the surveys.
The CMB and LSS cross-correlation has also been tested against the null
hypothesis, i.e., whether the observed signal is compatible with a null
correlation. As expected for such a weak signal, there is no strong evidence
of incompatibility with the lack of correlation. In this respect, the CAPS
approach seems to provide better constraints than the other estimators
investigated here (CCF and the SMHWcov).

We explored the aperture photometry of stacked CMB patches at the positions
of superstructures identified in the SDSS galaxy distribution. Our analysis
of the \citet{Granett2008a} catalogue (50 supervoids and 50 superstructures)
reproduced previous results, with similarly strong amplitude and significance
levels (somewhat above and below $3\,\sigma$ for voids and clusters,
respectively). While the most plausible source of this signal is the ISW
effect associated with these structures, it shows some tension with
expectations, both in terms of amplitude and scale. 
The same type of analysis was carried on the latest and much larger void
catalogues of \citet{Sutter2012} (about $1\,500$ voids) and \citet{Pan2012}
(about $1\,000$ voids). The results range from negligible to evidence at the
2--2.5\,$\sigma$ level, with a more moderate amplitude and a smaller scale,
in better agreement with theoretical predictions found in the literature.
The broad spectral coverage of \Planck\ allows us to confirm the achromatic
nature of these signals over the 44 to 353\,GHz range, supporting their
cosmological origin. 

We reconstructed maps of the ISW effect using a linear filter, by combining
the \Planck\ CMB and a gravitational potential tracers. In particular, we
considered both the NVSS catalogue and the reconstructed CMB lensing map as
LSS tracers. Again we found good agreement between different component
separation methods, as well as consistency between the expected and
reconstructed auto- and cross-power spectra for the recovered ISW map.

We conclude that the ISW effect is present in \Planck\ data
at the level expected for $\Lambda$CDM-cosmologies,
using a range of measurement methods, although there is
a possible tension with the results from stacking of CMB fields centred on
superstructures.  Generally, our results are more conservative than previous
claims using \wmap\ data, but the agreement with the expected signal is better.
Future \Planck\ data releases, including polarization information, as well as
improved understanding of foregrounds, could improve on these results, in
particular for ISW-lensing correlation and ISW-lensing map reconstruction.

\begin{acknowledgements}
The development of \Planck\ has been supported by: ESA; CNES and
CNRS/INSU-IN2P3-INP (France); ASI, CNR, and INAF (Italy); NASA and DoE (USA);
STFC and UKSA (UK); CSIC, MICINN, JA and RES (Spain); Tekes, AoF and CSC
(Finland); DLR and MPG (Germany); CSA (Canada); DTU Space (Denmark); SER/SSO
(Switzerland); RCN (Norway); SFI (Ireland); FCT/MCTES (Portugal); and PRACE
(EU). A description of the Planck Collaboration and a list of its members,
including the technical or scientific activities in which they have been
involved, can be found at \url{http://www.sciops.esa.int/index.php?project=planck&page=Planck_Collaboration}.
The modal and KSW bispectrum estimator analysis was performed on the COSMOS
supercomputer, part of the STFC DiRAC HPC Facility.
We acknowledge the computer resources, technical expertise and assistance
provided by the Spanish Supercomputing Network (RES) node at Universidad de
Cantabria, and the support provided by the Advanced Computing and e-Science team at IFCA. 
\end{acknowledgements}

\bibliographystyle{aa}
\bibliography{references_isw,Planck_bib}

\begin{thebibliography}{116}
\expandafter\ifx\csname natexlab\endcsname\relax\def\natexlab#1{#1}\fi

\bibitem[{{Adelman-McCarthy} {et~al.}(2008){Adelman-McCarthy}, {Ag{\"u}eros},
  {Allam}, {Allende Prieto}, {Anderson}, {Anderson}, {Annis}, {Bahcall},
  {Bailer-Jones}, {Baldry}, {Barentine}, {Bassett}, {Becker}, {Beers}, {Bell},
  {Berlind}, {Bernardi}, {Blanton}, {Bochanski}, {Boroski}, {Brinchmann},
  {Brinkmann}, {Brunner}, {Budav{\'a}ri}, {Carliles}, {Carr}, {Castander},
  {Cinabro}, {Cool}, {Covey}, {Csabai}, {Cunha}, {Davenport}, {Dilday}, {Doi},
  {Eisenstein}, {Evans}, {Fan}, {Finkbeiner}, {Friedman}, {Frieman},
  {Fukugita}, {G{\"a}nsicke}, {Gates}, {Gillespie}, {Glazebrook}, {Gray},
  {Grebel}, {Gunn}, {Gurbani}, {Hall}, {Harding}, {Harvanek}, {Hawley},
  {Hayes}, {Heckman}, {Hendry}, {Hindsley}, {Hirata}, {Hogan}, {Hogg}, {Hyde},
  {Ichikawa}, {Ivezi{\'c}}, {Jester}, {Johnson}, {Jorgensen}, {Juri{\'c}},
  {Kent}, {Kessler}, {Kleinman}, {Knapp}, {Kron}, {Krzesinski}, {Kuropatkin},
  {Lamb}, {Lampeitl}, {Lebedeva}, {Lee}, {Leger}, {L{\'e}pine}, {Lima}, {Lin},
  {Long}, {Loomis}, {Loveday}, {Lupton}, {Malanushenko}, {Malanushenko},
  {Mandelbaum}, {Margon}, {Marriner}, {Mart{\'{\i}}nez-Delgado}, {Matsubara},
  {McGehee}, {McKay}, {Meiksin}, {Morrison}, {Munn}, {Nakajima}, {Neilsen},
  {Newberg}, {Nichol}, {Nicinski}, {Nieto-Santisteban}, {Nitta}, {Okamura},
  {Owen}, {Oyaizu}, {Padmanabhan}, {Pan}, {Park}, {Peoples}, {Pier}, {Pope},
  {Purger}, {Raddick}, {Re Fiorentin}, {Richards}, {Richmond}, {Riess}, {Rix},
  {Rockosi}, {Sako}, {Schlegel}, {Schneider}, {Schreiber}, {Schwope}, {Seljak},
  {Sesar}, {Sheldon}, {Shimasaku}, {Sivarani}, {Smith}, {Snedden}, {Steinmetz},
  {Strauss}, {SubbaRao}, {Suto}, {Szalay}, {Szapudi}, {Szkody}, {Tegmark},
  {Thakar}, {Tremonti}, {Tucker}, {Uomoto}, {Vanden Berk}, {Vandenberg},
  {Vidrih}, {Vogeley}, {Voges}, {Vogt}, {Wadadekar}, {Weinberg}, {West},
  {White}, {Wilhite}, {Yanny}, {Yocum}, {York}, {Zehavi}, \&
  {Zucker}}]{Adelman2008}
{Adelman-McCarthy}, J.~K., {Ag{\"u}eros}, M.~A., {Allam}, S.~S., {et~al.} 2008,
  \apjs, 175, 297

\bibitem[{{Afshordi}(2004)}]{Afshordi2004b}
{Afshordi}, N. 2004, \prd, 70, 083536

\bibitem[{{Afshordi} {et~al.}(2004){Afshordi}, {Loh}, \&
  {Strauss}}]{Afshordi2004a}
{Afshordi}, N., {Loh}, Y., \& {Strauss}, M.~A. 2004, \prd, 69, 083524

\bibitem[{{Aihara} {et~al.}(2011){Aihara}, {Allende Prieto}, {An}, {Anderson},
  {Aubourg}, {Balbinot}, {Beers}, {Berlind}, {Bickerton}, {Bizyaev}, {Blanton},
  {Bochanski}, {Bolton}, {Bovy}, {Brandt}, {Brinkmann}, {Brown}, {Brownstein},
  {Busca}, {Campbell}, {Carr}, {Chen}, {Chiappini}, {Comparat}, {Connolly},
  {Cortes}, {Croft}, {Cuesta}, {da Costa}, {Davenport}, {Dawson}, {Dhital},
  {Ealet}, {Ebelke}, {Edmondson}, {Eisenstein}, {Escoffier}, {Esposito},
  {Evans}, {Fan}, {Femen{\'{\i}}a Castell{\'a}}, {Font-Ribera}, {Frinchaboy},
  {Ge}, {Gillespie}, {Gilmore}, {Gonz{\'a}lez Hern{\'a}ndez}, {Gott}, {Gould},
  {Grebel}, {Gunn}, {Hamilton}, {Harding}, {Harris}, {Hawley}, {Hearty}, {Ho},
  {Hogg}, {Holtzman}, {Honscheid}, {Inada}, {Ivans}, {Jiang}, {Johnson},
  {Jordan}, {Jordan}, {Kazin}, {Kirkby}, {Klaene}, {Knapp}, {Kneib},
  {Kochanek}, {Koesterke}, {Kollmeier}, {Kron}, {Lampeitl}, {Lang}, {Le Goff},
  {Lee}, {Lin}, {Long}, {Loomis}, {Lucatello}, {Lundgren}, {Lupton}, {Ma},
  {MacDonald}, {Mahadevan}, {Maia}, {Makler}, {Malanushenko}, {Malanushenko},
  {Mandelbaum}, {Maraston}, {Margala}, {Masters}, {McBride}, {McGehee},
  {McGreer}, {M{\'e}nard}, {Miralda-Escud{\'e}}, {Morrison}, {Mullally},
  {Muna}, {Munn}, {Murayama}, {Myers}, {Naugle}, {Neto}, {Nguyen}, {Nichol},
  {O'Connell}, {Ogando}, {Olmstead}, {Oravetz}, {Padmanabhan},
  {Palanque-Delabrouille}, {Pan}, {Pandey}, {P{\^a}ris}, {Percival},
  {Petitjean}, {Pfaffenberger}, {Pforr}, {Phleps}, {Pichon}, {Pieri}, {Prada},
  {Price-Whelan}, {Raddick}, {Ramos}, {Reyl{\'e}}, {Rich}, {Richards}, {Rix},
  {Robin}, {Rocha-Pinto}, {Rockosi}, {Roe}, {Rollinde}, {Ross}, {Ross},
  {Rossetto}, {S{\'a}nchez}, {Sayres}, {Schlegel}, {Schlesinger}, {Schmidt},
  {Schneider}, {Sheldon}, {Shu}, {Simmerer}, {Simmons}, {Sivarani}, {Snedden},
  {Sobeck}, {Steinmetz}, {Strauss}, {Szalay}, {Tanaka}, {Thakar}, {Thomas},
  {Tinker}, {Tofflemire}, {Tojeiro}, {Tremonti}, {Vandenberg}, {Vargas
  Maga{\~n}a}, {Verde}, {Vogt}, {Wake}, {Wang}, {Weaver}, {Weinberg}, {White},
  {White}, {Yanny}, {Yasuda}, {Yeche}, \& {Zehavi}}]{Aihara2011}
{Aihara}, H., {Allende Prieto}, C., {An}, D., {et~al.} 2011, \apjs, 193, 29

\bibitem[{{Barreiro} {et~al.}(2008){Barreiro}, {Vielva},
  {Hernandez-Monteagudo}, \& {Martinez-Gonzalez}}]{Barreiro2008}
{Barreiro}, R.~B., {Vielva}, P., {Hernandez-Monteagudo}, C., \&
  {Martinez-Gonzalez}, E. 2008, IEEE Journal of Selected Topics in Signal
  Processing, 2, 747

\bibitem[{{Barreiro} {et~al.}(2013){Barreiro}, {Vielva}, {Marcos-Caballero}, \&
  {Mart{\'{\i}}nez-Gonz{\'a}lez}}]{Barreiro2013}
{Barreiro}, R.~B., {Vielva}, P., {Marcos-Caballero}, A., \&
  {Mart{\'{\i}}nez-Gonz{\'a}lez}, E. 2013, \mnras, 430, 259

\bibitem[{{Bielby} {et~al.}(2010){Bielby}, {Shanks}, {Sawangwit}, {Croom},
  {Ross}, \& {Wake}}]{Bielby2010}
{Bielby}, R., {Shanks}, T., {Sawangwit}, U., {et~al.} 2010, \mnras, 403, 1261

\bibitem[{{Blake} \& {Wall}(2002)}]{Blake2002}
{Blake}, C. \& {Wall}, J. 2002, \mnras, 329, L37

\bibitem[{{Boughn} \& {Crittenden}(2004)}]{Boughn2004}
{Boughn}, S. \& {Crittenden}, R. 2004, \nat, 427, 45

\bibitem[{{Boughn} \& {Crittenden}(2002)}]{Boughn2002b}
{Boughn}, S.~P. \& {Crittenden}, R.~G. 2002, Physical Review Letters, 88,
  021302

\bibitem[{{Boughn} \& {Crittenden}(2005)}]{Boughn2005b}
{Boughn}, S.~P. \& {Crittenden}, R.~G. 2005, New Astronomy Review, 49, 75

\bibitem[{{Brookes} {et~al.}(2008){Brookes}, {Best}, {Peacock},
  {R{\"o}ttgering}, \& {Dunlop}}]{Brookes2008}
{Brookes}, M.~H., {Best}, P.~N., {Peacock}, J.~A., {R{\"o}ttgering}, H.~J.~A.,
  \& {Dunlop}, J.~S. 2008, \mnras, 385, 1297

\bibitem[{{Bucher} {et~al.}(2010){Bucher}, {van Tent}, \&
  {Carvalho}}]{Bucher2010}
{Bucher}, M., {van Tent}, B., \& {Carvalho}, C.~S. 2010, \mnras, 407, 2193

\bibitem[{{Cabr{\'e}} {et~al.}(2007){Cabr{\'e}}, {Fosalba}, {Gazta{\~n}aga}, \&
  {Manera}}]{Cabre2007}
{Cabr{\'e}}, A., {Fosalba}, P., {Gazta{\~n}aga}, E., \& {Manera}, M. 2007,
  \mnras, 381, 1347

\bibitem[{{Cabr{\'e}} {et~al.}(2006){Cabr{\'e}}, {Gazta{\~n}aga}, {Manera},
  {Fosalba}, \& {Castander}}]{Cabre2006}
{Cabr{\'e}}, A., {Gazta{\~n}aga}, E., {Manera}, M., {Fosalba}, P., \&
  {Castander}, F. 2006, \mnras, 372, L23

\bibitem[{{Cai} {et~al.}(2013){Cai}, {Neyrinck}, {Szapudi}, {Cole}, \&
  {Frenk}}]{Cai2013}
{Cai}, Y.-C., {Neyrinck}, M.~C., {Szapudi}, I., {Cole}, S., \& {Frenk}, C.~S.
  2013, ArXiv e-prints 1301.6136

\bibitem[{{Condon} {et~al.}(1998){Condon}, {Cotton}, {Greisen}, {Yin},
  {Perley}, {Taylor}, \& {Broderick}}]{Condon1998}
{Condon}, J.~J., {Cotton}, W.~D., {Greisen}, E.~W., {et~al.} 1998, \aj, 115,
  1693

\bibitem[{{Corasaniti} {et~al.}(2005){Corasaniti}, {Giannantonio}, \&
  {Melchiorri}}]{Corasaniti2005}
{Corasaniti}, P.-S., {Giannantonio}, T., \& {Melchiorri}, A. 2005, \prd, 71,
  123521

\bibitem[{{Creminelli} {et~al.}(2006){Creminelli}, {Nicolis}, {Senatore},
  {Tegmark}, \& {Zaldarriaga}}]{Creminelli2006}
{Creminelli}, P., {Nicolis}, A., {Senatore}, L., {Tegmark}, M., \&
  {Zaldarriaga}, M. 2006, \jcap, 5, 4

\bibitem[{{Crittenden} \& {Turok}(1996)}]{Crittenden1996}
{Crittenden}, R.~G. \& {Turok}, N. 1996, Physical Review Letters, 76, 575

\bibitem[{{Dup{\'e}} {et~al.}(2011){Dup{\'e}}, {Rassat}, {Starck}, \&
  {Fadili}}]{Dupe2011}
{Dup{\'e}}, F.-X., {Rassat}, A., {Starck}, J.-L., \& {Fadili}, M.~J. 2011,
  \aap, 534, A51

\bibitem[{{Fergusson} {et~al.}(2010){Fergusson}, {Liguori}, \&
  {Shellard}}]{Fergusson2010}
{Fergusson}, J.~R., {Liguori}, M., \& {Shellard}, E.~P.~S. 2010, \prd, 82,
  023502

\bibitem[{Flender {et~al.}(2013)Flender, Hotchkiss, \& Nadathur}]{Flender2013}
Flender, S., Hotchkiss, S., \& Nadathur, S. 2013, Journal of Cosmology and
  Astroparticle Physics, 2013, 013

\bibitem[{{Fosalba} \& {Gazta{\~n}aga}(2004)}]{Fosalba2004}
{Fosalba}, P. \& {Gazta{\~n}aga}, E. 2004, \mnras, 350, L37

\bibitem[{{Fosalba} {et~al.}(2003){Fosalba}, {Gazta{\~n}aga}, \&
  {Castander}}]{Fosalba2003}
{Fosalba}, P., {Gazta{\~n}aga}, E., \& {Castander}, F.~J. 2003, \apjl, 597, L89

\bibitem[{{Francis} \& {Peacock}(2009)}]{Francis2009a}
{Francis}, C.~L. \& {Peacock}, J.~A. 2009, ArXiv e-prints 0909.2494

\bibitem[{{Francis} \& {Peacock}(2010)}]{Francis2010}
{Francis}, C.~L. \& {Peacock}, J.~A. 2010, \mnras, 406, 14

\bibitem[{{Gazta{\~n}aga} {et~al.}(2006){Gazta{\~n}aga}, {Manera}, \&
  {Multam{\"a}ki}}]{Gaztanaga2006}
{Gazta{\~n}aga}, E., {Manera}, M., \& {Multam{\"a}ki}, T. 2006, \mnras, 365,
  171

\bibitem[{{Giannantonio} {et~al.}(2012){Giannantonio}, {Crittenden}, {Nichol},
  \& {Ross}}]{Giannantonio2012}
{Giannantonio}, T., {Crittenden}, R., {Nichol}, R., \& {Ross}, A.~J. 2012,
  \mnras, 426, 2581

\bibitem[{{Giannantonio} {et~al.}(2006){Giannantonio}, {Crittenden}, {Nichol},
  {Scranton}, {Richards}, {Myers}, {Brunner}, {Gray}, {Connolly}, \&
  {Schneider}}]{Giannantonio2006b}
{Giannantonio}, T., {Crittenden}, R.~G., {Nichol}, R.~C., {et~al.} 2006, \prd,
  74, 063520

\bibitem[{{Giannantonio} {et~al.}(2008){Giannantonio}, {Scranton},
  {Crittenden}, {Nichol}, {Boughn}, {Myers}, \& {Richards}}]{Giannantonio2008b}
{Giannantonio}, T., {Scranton}, R., {Crittenden}, R.~G., {et~al.} 2008, \prd,
  77, 123520

\bibitem[{{Giovi} \& {Baccigalupi}(2005)}]{Giovi2005a}
{Giovi}, F. \& {Baccigalupi}, C. 2005, in IAU Symposium, Vol. 225,
  Gravitational Lensing Impact on Cosmology, ed. Y.~{Mellier} \& G.~{Meylan},
  117--122

\bibitem[{{Giovi} {et~al.}(2003){Giovi}, {Baccigalupi}, \&
  {Perrotta}}]{Giovi2003}
{Giovi}, F., {Baccigalupi}, C., \& {Perrotta}, F. 2003, \prd, 68, 123002

\bibitem[{{Goldberg} \& {Spergel}(1999)}]{Goldberg1999}
{Goldberg}, D.~M. \& {Spergel}, D.~N. 1999, \prd, 59, 103002

\bibitem[{{G{\'o}rski} {et~al.}(2005){G{\'o}rski}, {Hivon}, {Banday},
  {Wandelt}, {Hansen}, {Reinecke}, \& {Bartelmann}}]{Gorski2005}
{G{\'o}rski}, K.~M., {Hivon}, E., {Banday}, A.~J., {et~al.} 2005, \apj, 622,
  759

\bibitem[{{Granett} {et~al.}(2008{\natexlab{a}}){Granett}, {Neyrinck}, \&
  {Szapudi}}]{Granett2008a}
{Granett}, B.~R., {Neyrinck}, M.~C., \& {Szapudi}, I. 2008{\natexlab{a}},
  \apjl, 683, L99

\bibitem[{{Granett} {et~al.}(2008{\natexlab{b}}){Granett}, {Neyrinck}, \&
  {Szapudi}}]{Granett2008b}
{Granett}, B.~R., {Neyrinck}, M.~C., \& {Szapudi}, I. 2008{\natexlab{b}}, ArXiv
  e-prints 0805.2974

\bibitem[{{Granett} {et~al.}(2009){Granett}, {Neyrinck}, \&
  {Szapudi}}]{Granett2009a}
{Granett}, B.~R., {Neyrinck}, M.~C., \& {Szapudi}, I. 2009, \apj, 701, 414

\bibitem[{{Gruppuso} {et~al.}(2009){Gruppuso}, {de Rosa}, {Cabella}, {Paci},
  {Finelli}, {Natoli}, {de Gasperis}, \& {Mandolesi}}]{Gruppuso2009}
{Gruppuso}, A., {de Rosa}, A., {Cabella}, P., {et~al.} 2009, \mnras, 400, 463

\bibitem[{{Hanson} {et~al.}(2010){Hanson}, {Challinor}, \&
  {Lewis}}]{Hanson2010}
{Hanson}, D., {Challinor}, A., \& {Lewis}, A. 2010, General Relativity and
  Gravitation, 42, 2197

\bibitem[{{Hanson} {et~al.}(2009){Hanson}, {Smith}, {Challinor}, \&
  {Liguori}}]{Hanson2009a}
{Hanson}, D., {Smith}, K.~M., {Challinor}, A., \& {Liguori}, M. 2009, \prd, 80,
  083004

\bibitem[{{Hern{\'a}ndez-Monteagudo}(2008)}]{Hernandez2008}
{Hern{\'a}ndez-Monteagudo}, C. 2008, \aap, 490, 15

\bibitem[{{Hern{\'a}ndez-Monteagudo}(2010)}]{Hernandez2010}
{Hern{\'a}ndez-Monteagudo}, C. 2010, \aap, 520, A101

\bibitem[{{Hernandez-Monteagudo} {et~al.}(2013){Hernandez-Monteagudo}, {Ross},
  {Cuesta}, {Genova-Santos}, {Prada}, {Rossi}, {Neyrinck}, {Viel},
  {Rubino-Martin}, {Scoccola}, {Zhao}, {Schneider}, {Brownstein}, {Thomas}, \&
  {Brinkmann}}]{Hernandez2013}
{Hernandez-Monteagudo}, C., {Ross}, A., {Cuesta}, A., {et~al.} 2013, ArXiv
  e-prints

\bibitem[{{Hernandez-Monteagudo} \& {Smith}(2012)}]{Hernandez2012}
{Hernandez-Monteagudo}, C. \& {Smith}, R.~E. 2012, ArXiv e-prints 1212.1174

\bibitem[{{Hinshaw} {et~al.}(2003){Hinshaw}, {Spergel}, {Verde}, {Hill},
  {Meyer}, {Barnes}, {Bennett}, {Halpern}, {Jarosik}, {Kogut}, {Komatsu},
  {Limon}, {Page}, {Tucker}, {Weiland}, {Wollack}, \& {Wright}}]{Hinshaw2003}
{Hinshaw}, G., {Spergel}, D.~N., {Verde}, L., {et~al.} 2003, \apjs, 148, 135

\bibitem[{{Hivon} {et~al.}(2002){Hivon}, {G{\'o}rski}, {Netterfield}, {Crill},
  {Prunet}, \& {Hansen}}]{Hivon2002}
{Hivon}, E., {G{\'o}rski}, K.~M., {Netterfield}, C.~B., {et~al.} 2002, \apj,
  567, 2

\bibitem[{{Ho} {et~al.}(2012){Ho}, {Cuesta}, {Seo}, {de Putter}, {Ross},
  {White}, {Padmanabhan}, {Saito}, {Schlegel}, {Schlafly}, {Seljak},
  {Hernandez-Monteagudo}, {Sanchez}, {Percival}, {Blanton}, {Skibba},
  {Schneider}, {Reid}, {Mena}, {Viel}, {Eisenstein}, {Prada}, {Weaver},
  {Bahcall}, {Bizyaev}, {Brewinton}, {Brinkman}, {Nicolaci da Costa}, {Gott},
  {Malanushenko}, {Malanushenko}, {Nichol}, {Oravetz}, {Pan},
  {Palanque-Delabrouille}, {Ross}, {Simmons}, {de Simoni}, {Snedden}, \&
  {Yeche}}]{Ho2012}
{Ho}, S., {Cuesta}, A., {Seo}, H.-J., {et~al.} 2012, ArXiv e-prints 1201.2137

\bibitem[{{Ho} {et~al.}(2008){Ho}, {Hirata}, {Padmanabhan}, {Seljak}, \&
  {Bahcall}}]{Ho2008}
{Ho}, S., {Hirata}, C., {Padmanabhan}, N., {Seljak}, U., \& {Bahcall}, N. 2008,
  \prd, 78, 043519

\bibitem[{{Hoyle} \& {Vogeley}(2002)}]{Hoyle2002}
{Hoyle}, F. \& {Vogeley}, M.~S. 2002, \apj, 566, 641

\bibitem[{{Hu}(2000)}]{Hu2000}
{Hu}, W. 2000, \prd, 62, 043007

\bibitem[{{Hu}(2002)}]{Hu2002a}
{Hu}, W. 2002, \prd, 65, 023003

\bibitem[{{Hu} \& {Okamoto}(2002)}]{Hu2002b}
{Hu}, W. \& {Okamoto}, T. 2002, \apj, 574, 566

\bibitem[{{Hu} \& {Sugiyama}(1994)}]{Hu1994}
{Hu}, W. \& {Sugiyama}, N. 1994, \prd, 50, 627

\bibitem[{{Hunt} \& {Sarkar}(2010)}]{Hunt2010}
{Hunt}, P. \& {Sarkar}, S. 2010, \mnras, 401, 547

\bibitem[{{Ili\'c} {et~al.}(2013){Ili\'c}, {Langer}, \& {Douspis}}]{Ilic2013}
{Ili\'c}, S., {Langer}, M., \& {Douspis}, M. 2013, arXiv:1301.5849

\bibitem[{{Kamionkowski} \& {Spergel}(1994)}]{Kamionkowski1994}
{Kamionkowski}, M. \& {Spergel}, D.~N. 1994, \apj, 432, 7

\bibitem[{{Komatsu} {et~al.}(2005){Komatsu}, {Spergel}, \&
  {Wandelt}}]{Komatsu2005}
{Komatsu}, E., {Spergel}, D.~N., \& {Wandelt}, B.~D. 2005, \apj, 634, 14

\bibitem[{Komatsu {et~al.}(2005)Komatsu, Spergel, \& Wandelt}]{Komatsu2003}
Komatsu, E., Spergel, D.~N., \& Wandelt, B.~D. 2005, Astrophys.J., 634, 14

\bibitem[{{Larson} {et~al.}(2011){Larson}, {Dunkley}, {Hinshaw}, {Komatsu},
  {Nolta}, {Bennett}, {Gold}, {Halpern}, {Hill}, {Jarosik}, {Kogut}, {Limon},
  {Meyer}, {Odegard}, {Page}, {Smith}, {Spergel}, {Tucker}, {Weiland},
  {Wollack}, \& {Wright}}]{Larson2011}
{Larson}, D., {Dunkley}, J., {Hinshaw}, G., {et~al.} 2011, \apjs, 192, 16

\bibitem[{{Lewis} \& {Challinor}(2006)}]{Lewis2006}
{Lewis}, A. \& {Challinor}, A. 2006, \physrep, 429, 1

\bibitem[{{Lewis} {et~al.}(2011){Lewis}, {Challinor}, \& {Hanson}}]{Lewis2011}
{Lewis}, A., {Challinor}, A., \& {Hanson}, D. 2011, \jcap, 3, 18

\bibitem[{{Li} \& {Xia}(2010)}]{Li2010}
{Li}, H. \& {Xia}, J. 2010, \jcap, 4, 26

\bibitem[{{L{\'o}pez-Corredoira} {et~al.}(2010){L{\'o}pez-Corredoira}, {Sylos
  Labini}, \& {Betancort-Rijo}}]{Lopez2010}
{L{\'o}pez-Corredoira}, M., {Sylos Labini}, F., \& {Betancort-Rijo}, J. 2010,
  \aap, 513, A3

\bibitem[{{Mangilli} \& {Verde}(2009)}]{Mangilli2009a}
{Mangilli}, A. \& {Verde}, L. 2009, \prd, 80, 123007

\bibitem[{Mangilli {et~al.}(2013)Mangilli, Wandelt, Elsner, \&
  Liguori}]{Mangilli2013}
Mangilli, A., Wandelt, B., Elsner, F., \& Liguori, M. 2013

\bibitem[{{Marcos-Caballero et al.}(2013)}]{MarcosCaballero2013}
{Marcos-Caballero et al.} 2013, in preparation

\bibitem[{{Mart{\'{\i}}nez-Gonz{\'a}lez}
  {et~al.}(2002){Mart{\'{\i}}nez-Gonz{\'a}lez}, {Gallegos}, {Arg{\"u}eso},
  {Cay{\'o}n}, \& {Sanz}}]{Martinez2002}
{Mart{\'{\i}}nez-Gonz{\'a}lez}, E., {Gallegos}, J.~E., {Arg{\"u}eso}, F.,
  {Cay{\'o}n}, L., \& {Sanz}, J.~L. 2002, \mnras, 336, 22

\bibitem[{{Martinez-Gonzalez} {et~al.}(1990){Martinez-Gonzalez}, {Sanz}, \&
  {Silk}}]{Martinez1990b}
{Martinez-Gonzalez}, E., {Sanz}, J.~L., \& {Silk}, J. 1990, \apjl, 355, L5

\bibitem[{{Massardi} {et~al.}(2010){Massardi}, {Bonaldi}, {Negrello},
  {Ricciardi}, {Raccanelli}, \& {de Zotti}}]{Massardi2010}
{Massardi}, M., {Bonaldi}, A., {Negrello}, M., {et~al.} 2010, \mnras, 404, 532

\bibitem[{{McEwen} {et~al.}(2007){McEwen}, {Vielva}, {Hobson},
  {Mart{\'{\i}}nez-Gonz{\'a}lez}, \& {Lasenby}}]{Mcewen2007}
{McEwen}, J.~D., {Vielva}, P., {Hobson}, M.~P., {Mart{\'{\i}}nez-Gonz{\'a}lez},
  E., \& {Lasenby}, A.~N. 2007, \mnras, 376, 1211

\bibitem[{{McEwen} {et~al.}(2008){McEwen}, {Wiaux}, {Hobson}, {Vandergheynst},
  \& {Lasenby}}]{McEwen2008b}
{McEwen}, J.~D., {Wiaux}, Y., {Hobson}, M.~P., {Vandergheynst}, P., \&
  {Lasenby}, A.~N. 2008, \mnras, 384, 1289

\bibitem[{Munshi \& Heavens(2010)}]{Munshi2009}
Munshi, D. \& Heavens, A. 2010, Mon.Not.Roy.Astron.Soc., 401, 2406

\bibitem[{{Neyrinck}(2008)}]{Neyrinck2008}
{Neyrinck}, M.~C. 2008, \mnras, 386, 2101

\bibitem[{{Neyrinck} {et~al.}(2005){Neyrinck}, {Gnedin}, \&
  {Hamilton}}]{Neyrinck2005}
{Neyrinck}, M.~C., {Gnedin}, N.~Y., \& {Hamilton}, A.~J.~S. 2005, \mnras, 356,
  1222

\bibitem[{{Nolta} {et~al.}(2004){Nolta}, {Wright}, {Page}, {Bennett},
  {Halpern}, {Hinshaw}, {Jarosik}, {Kogut}, {Limon}, {Meyer}, {Spergel},
  {Tucker}, \& {Wollack}}]{Nolta2004}
{Nolta}, M.~R., {Wright}, E.~L., {Page}, L., {et~al.} 2004, \apj, 608, 10

\bibitem[{{Okamoto} \& {Hu}(2003)}]{Okamoto2003}
{Okamoto}, T. \& {Hu}, W. 2003, \prd, 67, 083002

\bibitem[{{Paci} {et~al.}(2013){Paci}, {Gruppuso}, {Finelli}, {De Rosa},
  {Mandolesi}, \& {Natoli}}]{Paci2013}
{Paci}, F., {Gruppuso}, A., {Finelli}, F., {et~al.} 2013, ArXiv e-prints
  1301.5195

\bibitem[{{Padmanabhan} {et~al.}(2005){Padmanabhan}, {Hirata}, {Seljak},
  {Schlegel}, {Brinkmann}, \& {Schneider}}]{Padmanabhan2005}
{Padmanabhan}, N., {Hirata}, C.~M., {Seljak}, U., {et~al.} 2005, \prd, 72,
  043525

\bibitem[{{Pan} {et~al.}(2012){Pan}, {Vogeley}, {Hoyle}, {Choi}, \&
  {Park}}]{Pan2012}
{Pan}, D.~C., {Vogeley}, M.~S., {Hoyle}, F., {Choi}, Y.-Y., \& {Park}, C. 2012,
  \mnras, 421, 926

\bibitem[{{P{\'a}pai} \& {Szapudi}(2010)}]{Papai2010b}
{P{\'a}pai}, P. \& {Szapudi}, I. 2010, ArXiv e-prints 1009.0754

\bibitem[{{Parejko} {et~al.}(2013){Parejko}, {Sunayama}, {Padmanabhan}, {Wake},
  {Berlind}, {Bizyaev}, {Blanton}, {Bolton}, {van den Bosch}, {Brinkmann},
  {Brownstein}, {da Costa}, {Eisenstein}, {Guo}, {Kazin}, {Maia},
  {Malanushenko}, {Maraston}, {McBride}, {Nichol}, {Oravetz}, {Pan},
  {Percival}, {Prada}, {Ross}, {Ross}, {Schlegel}, {Schneider}, {Simmons},
  {Skibba}, {Tinker}, {Tojeiro}, {Weaver}, {Wetzel}, {White}, {Weinberg},
  {Thomas}, {Zehavi}, \& {Zheng}}]{Parejkoetal2012}
{Parejko}, J.~K., {Sunayama}, T., {Padmanabhan}, N., {et~al.} 2013, \mnras,
  429, 98

\bibitem[{{Pietrobon} {et~al.}(2006{\natexlab{a}}){Pietrobon}, {Balbi}, \&
  {Marinucci}}]{Pietrobon2006a}
{Pietrobon}, D., {Balbi}, A., \& {Marinucci}, D. 2006{\natexlab{a}},
  arXiv:astro-ph/0611797

\bibitem[{{Pietrobon} {et~al.}(2006{\natexlab{b}}){Pietrobon}, {Balbi}, \&
  {Marinucci}}]{Pietrobon2006b}
{Pietrobon}, D., {Balbi}, A., \& {Marinucci}, D. 2006{\natexlab{b}}, \prd, 74,
  043524

\bibitem[{{Planck Collaboration I}(2013)}]{planck2013-p01}
{Planck Collaboration I}. 2013, Submitted to \aap

\bibitem[{{Planck Collaboration II}(2013)}]{planck2013-p02}
{Planck Collaboration II}. 2013, Submitted to \aap

\bibitem[{{Planck Collaboration VI}(2013)}]{planck2013-p03}
{Planck Collaboration VI}. 2013, Submitted to \aap

\bibitem[{{Planck Collaboration XII}(2013)}]{planck2013-p06}
{Planck Collaboration XII}. 2013, Submitted to \aap

\bibitem[{{Planck Collaboration XV}(2013)}]{planck2013-p08}
{Planck Collaboration XV}. 2013, Submitted to \aap

\bibitem[{{Planck Collaboration XVI}(2013)}]{planck2013-p11}
{Planck Collaboration XVI}. 2013, Submitted to \aap

\bibitem[{{Planck Collaboration XVII}(2013)}]{planck2013-p12}
{Planck Collaboration XVII}. 2013, Submitted to \aap

\bibitem[{{Planck Collaboration XXIII}(2013)}]{planck2013-p09}
{Planck Collaboration XXIII}. 2013, Submitted to \aap

\bibitem[{{Planck Collaboration XXIV}(2013)}]{planck2013-p09a}
{Planck Collaboration XXIV}. 2013, Submitted to \aap

\bibitem[{{Raccanelli} {et~al.}(2008){Raccanelli}, {Bonaldi}, {Negrello},
  {Matarrese}, {Tormen}, \& {de Zotti}}]{Raccanelli2008}
{Raccanelli}, A., {Bonaldi}, A., {Negrello}, M., {et~al.} 2008, \mnras, 386,
  2161

\bibitem[{{Rassat} {et~al.}(2006){Rassat}, {Land}, {Lahav}, \&
  {Abdalla}}]{Rassat2006}
{Rassat}, A., {Land}, K., {Lahav}, O., \& {Abdalla}, F.~B. 2006,
  arXiv:astro-ph/0610911

\bibitem[{{Rees} \& {Sciama}(1968)}]{Rees1968}
{Rees}, M.~J. \& {Sciama}, D.~W. 1968, \nat, 217, 511

\bibitem[{{Regan} {et~al.}(2013){Regan}, {Mukherjee}, \& {Seery}}]{Regan2013}
{Regan}, D., {Mukherjee}, P., \& {Seery}, D. 2013, ArXiv e-prints 1302.5631

\bibitem[{{Ross} {et~al.}(2011){Ross}, {Ho}, {Cuesta}, {Tojeiro}, {Percival},
  {Wake}, {Masters}, {Nichol}, {Myers}, {de Simoni}, {Seo},
  {Hern{\'a}ndez-Monteagudo}, {Crittenden}, {Blanton}, {Brinkmann}, {da Costa},
  {Guo}, {Kazin}, {Maia}, {Maraston}, {Padmanabhan}, {Prada}, {Ramos},
  {Sanchez}, {Schlafly}, {Schlegel}, {Schneider}, {Skibba}, {Thomas}, {Weaver},
  {White}, \& {Zehavi}}]{Ross2011}
{Ross}, A.~J., {Ho}, S., {Cuesta}, A.~J., {et~al.} 2011, \mnras, 417, 1350

\bibitem[{{Sachs} \& {Wolfe}(1967)}]{Sachs1967}
{Sachs}, R.~K. \& {Wolfe}, A.~M. 1967, \apj, 147, 73

\bibitem[{{Sawangwit} {et~al.}(2010){Sawangwit}, {Shanks}, {Cannon}, {Croom},
  {Ross}, \& {Wake}}]{Sawangwit2010}
{Sawangwit}, U., {Shanks}, T., {Cannon}, R.~D., {et~al.} 2010, \mnras, 402,
  2228

\bibitem[{{Schiavon} {et~al.}(2012){Schiavon}, {Finelli}, {Gruppuso},
  {Marcos-Caballero}, {Vielva}, {Crittenden}, {Barreiro}, \&
  {Mart{\'{\i}}nez-Gonz{\'a}lez}}]{Schiavon2012}
{Schiavon}, F., {Finelli}, F., {Gruppuso}, A., {et~al.} 2012, \mnras, 427, 3044

\bibitem[{{Schlegel} {et~al.}(1998){Schlegel}, {Finkbeiner}, \&
  {Davis}}]{Schlegel1998}
{Schlegel}, D.~J., {Finkbeiner}, D.~P., \& {Davis}, M. 1998, \apj, 500, 525

\bibitem[{{Scranton} {et~al.}(2003){Scranton}, {Connolly}, {Nichol},
  {Stebbins}, {Szapudi}, {Eisenstein}, {Afshordi}, {Budavari}, {Csabai},
  {Frieman}, {Gunn}, {Johnson}, {Loh}, {Lupton}, {Miller}, {Sheldon}, {Sheth},
  {Szalay}, {Tegmark}, \& {Xu}}]{Scranton2003}
{Scranton}, R., {Connolly}, A.~J., {Nichol}, R.~C., {et~al.} 2003,
  ArXiv:astro-ph/0307335

\bibitem[{{Seljak} \& {Zaldarriaga}(1999)}]{Seljak1999}
{Seljak}, U. \& {Zaldarriaga}, M. 1999, \prd, 60, 043504

\bibitem[{{Serra} \& {Cooray}(2008)}]{Serra2008}
{Serra}, P. \& {Cooray}, A. 2008, \prd, 77, 107305

\bibitem[{{Sheth} \& {Tormen}(1999)}]{Sheth1999}
{Sheth}, R.~K. \& {Tormen}, G. 1999, \mnras, 308, 119

\bibitem[{{Smith} \& {Zaldarriaga}(2011)}]{Smith2011}
{Smith}, K.~M. \& {Zaldarriaga}, M. 2011, \mnras, 417, 2

\bibitem[{{Sutter} {et~al.}(2012){Sutter}, {Lavaux}, {Wandelt}, \&
  {Weinberg}}]{Sutter2012}
{Sutter}, P.~M., {Lavaux}, G., {Wandelt}, B.~D., \& {Weinberg}, D.~H. 2012,
  arXiv e-prints 1207.2524

\bibitem[{{Tegmark}(1997)}]{Tegmark1997c}
{Tegmark}, M. 1997, \prd, 55, 5895

\bibitem[{{Tegmark} \& {de Oliveira-Costa}(2001)}]{Tegmark2001}
{Tegmark}, M. \& {de Oliveira-Costa}, A. 2001, \prd, 64, 063001

\bibitem[{{Tegmark} {et~al.}(2006){Tegmark}, {Eisenstein}, {Strauss},
  {Weinberg}, {Blanton}, {Frieman}, {Fukugita}, {Gunn}, {Hamilton}, {Knapp},
  {Nichol}, {Ostriker}, {Padmanabhan}, {Percival}, {Schlegel}, {Schneider},
  {Scoccimarro}, {Seljak}, {Seo}, {Swanson}, {Szalay}, {Vogeley}, {Yoo},
  {Zehavi}, {Abazajian}, {Anderson}, {Annis}, {Bahcall}, {Bassett}, {Berlind},
  {Brinkmann}, {Budavari}, {Castander}, {Connolly}, {Csabai}, {Doi},
  {Finkbeiner}, {Gillespie}, {Glazebrook}, {Hennessy}, {Hogg}, {Ivezi{\'c}},
  {Jain}, {Johnston}, {Kent}, {Lamb}, {Lee}, {Lin}, {Loveday}, {Lupton},
  {Munn}, {Pan}, {Park}, {Peoples}, {Pier}, {Pope}, {Richmond}, {Rockosi},
  {Scranton}, {Sheth}, {Stebbins}, {Stoughton}, {Szapudi}, {Tucker}, {vanden
  Berk}, {Yanny}, \& {York}}]{Tegmark2006}
{Tegmark}, M., {Eisenstein}, D.~J., {Strauss}, M.~A., {et~al.} 2006, \prd, 74,
  123507

\bibitem[{{Verde} \& {Spergel}(2002)}]{Verde2002}
{Verde}, L. \& {Spergel}, D.~N. 2002, \prd, 65, 043007

\bibitem[{{Vielva} {et~al.}(2006){Vielva}, {Mart{\'{\i}}nez-Gonz{\'a}lez}, \&
  {Tucci}}]{Vielva2006}
{Vielva}, P., {Mart{\'{\i}}nez-Gonz{\'a}lez}, E., \& {Tucci}, M. 2006, \mnras,
  365, 891

\bibitem[{{Xia}(2009)}]{Xia2009}
{Xia}, J. 2009, \prd, 80, 103514

\bibitem[{{Xia} {et~al.}(2011){Xia}, {Baccigalupi}, {Matarrese}, {Verde}, \&
  {Viel}}]{Xia2011}
{Xia}, J.-Q., {Baccigalupi}, C., {Matarrese}, S., {Verde}, L., \& {Viel}, M.
  2011, \jcap, 8, 33

\bibitem[{{Xia} {et~al.}(2009){Xia}, {Viel}, {Baccigalupi}, \&
  {Matarrese}}]{Xia2009b}
{Xia}, J.-Q., {Viel}, M., {Baccigalupi}, C., \& {Matarrese}, S. 2009, ArXiv
  e-prints 0907.4753

\end{thebibliography}

\raggedright
\end{document}